\documentclass[12pt,a4paper]{article}
\usepackage{jheppub}

\usepackage{epsfig}
\usepackage{amsfonts}
\usepackage{amssymb}
\usepackage{amscd}
\usepackage{amsmath}
\usepackage{mathrsfs}
\usepackage{xspace}
\usepackage{mathrsfs}
\usepackage{amsthm}
\usepackage{dsfont}
\usepackage{bbm}
\usepackage{slashed}

\usepackage{array}

\usepackage{hyperref}

\newcommand{\be}{\begin{equation}}
\newcommand{\ee}{\end{equation}}
\newcommand{\ben}{\begin{eqnarray}\displaystyle}
\newcommand{\een}{\end{eqnarray}}
\newcommand{\bea}{\begin{eqnarray}}
\newcommand{\eea}{\end{eqnarray}}

\newcommand{\refb}[1]{(\ref{#1})}

\def\ZZZ{{\hbox{ Z\kern-1.6mm Z}}}
\def\RRR{{\hbox{ R\kern-2.4mm R}}}
\def\CCC{{\hbox{ C\kern-2.0mm C}}}
\def\zzz{{\hbox{z\kern-1mm z}}}

\def\ZZZ{{\mathbb Z} }
\def\RRR{{\mathbb R} }
\def\CCC{{\mathbb C} }

\newcommand{\IR}{\mathds{R}}
\newcommand{\IC}{\mathds{C}}
\newcommand{\IZ}{\mathds{Z}}

\newcommand{\p}{\partial}
\newcommand{\I}{\mathrm{i}}

\def\Tr{{\rm Tr} }

\def\Im{\,{\rm Im}\,}
\def\Re{\,{\rm Re}\,}

\def\({\left(}
\def\){\right)}
\def\[{\left[}
\def\]{\right]}
\def\hf{{1\over 2}}

\newcommand{\non}{\nonumber}

\newcommand{\vt}{\vartheta}

\newcommand{\qeq}{{\hbox{=\kern-2.3mm ? \kern.5mm }}}
\renewcommand{\qeq}{=}

\newcommand{\bG}{{\bf G}}

\newcommand{\eps}{\epsilon}

\newcommand{\vp}{\varphi}
\newcommand{\ve}{\varepsilon}

\newcommand{\bw}{\bar w}

\newcommand{\VV}{{\cal V}}

\newcommand{\AAA}{{\cal A}}

\newcommand{\OO}{{\cal O}}

\newcommand{\XX}{{\cal X}}
\newcommand{\YY}{{\cal Y}}

\newcommand{\cK}{\mathcal{K}}
\newcommand{\cL}{\mathcal{L}}
\newcommand{\cA}{\mathcal{A}}

\newcommand{\cR}{\mathcal{R}}
\newcommand{\cT}{\mathcal{T}}

\newcommand{\cZ}{\mathcal{Z}}

\newcommand{\wt}{\widetilde}

\newcommand{\NN}{{\cal N}}
\newcommand{\TT}{{\cal T}}

\newcommand{\bc}{\bar c}

\def\bw{\bar w}

\def\bz{\bar z}

\def\bW{ \bar W}
\def\bG{ \bar G }

\def\bJ{\bar J}

\def\btau{\bar \tau}
\def\bvp{\bar \vp}
\def\bnabla{\bar \nabla}

\def\tc{\tilde c}
\def\tn{n'}

\def\cl0{\tilde c_0}

\def\one{{\hbox{ 1\kern-.8mm l}}}
\def\zero{{\hbox{ 0\kern-1.5mm 0}}}

\newcommand{\dalpha}{\beta}

\def\CY{\mathfrak{Y}}

\def\gr{g^{(r)}}
\def\ghQ{g^{(h,Q)}}
\def\gi#1{g^{(#1)}}
\def\Zr{Z^{(r)}}
\def\ZhQ{Z^{(h,Q)}}
\def\Zi#1{Z^{(#1)}}
\def\cZr{\cZ^{(r)}}

\def\cZi#1{\cZ^{(#1)}}

\def\cTR{\cT^{\rm R}}

\def\KKZ{K_0}

\def\dalpha{{\dot\alpha}}
\def\dbeta{{\dot\beta}}
\def\dgamma{{\dot\gamma}}
\def\ddelta{{\dot\delta}}

\def\cKsk{\cK_{\rm sk}}
\def\cKq{\cK_{\rm q}}

\def\sp{\not  \hskip -.04in p}

\def\I{i}

\def\xxi{\rho}

\def\cpi{u}
\def\cfi{\zeta}

\def\hq{\hat q}

\def\cq{\check q}
\def\htau{\hat\tau}
\def\ttau{\tilde\tau}
\def\ctau{\check\tau}

\newcommand{\Chi}{\sigma}

\newcommand{\oo}{{$\Omega_g$}}

\title{D-instanton Induced Superpotential}

\author[1]{Sergei Alexandrov,}
\author[2]{Atakan Hilmi F{\i}rat,}
\author[2]{\newline Manki Kim,}
\author[3]{Ashoke Sen,}
\author[4]{and Bogdan Stefa\'nski, jr.}
\affiliation[1]{Laboratoire Charles Coulomb (L2C), Universit\'e de Montpellier,
CNRS, F-34095, Montpellier, France\\}
\affiliation[2]{Center for Theoretical Physics, Massachusetts Institute of Technology, Cambridge, MA 02139, USA\\}
\affiliation[3]{International Centre for Theoretical Sciences - TIFR, Bengaluru - 560089, India\\}
\affiliation[4]{Centre for Mathematical Science, City, University of London, Northampton Square, London EC1V 0HB, UK\\}

\emailAdd{sergey.alexandrov@umontpellier.fr}
\emailAdd{firat@mit.edu}
\emailAdd{mk2427@mit.edu}
\emailAdd{ashoke.sen@icts.res.in}
\emailAdd{Bogdan.Stefanski.1@city.ac.uk}

\abstract{We use string field theory to fix
the normalization of the D-instanton corrections to the superpotential
involving  the moduli fields of type II string theory compactified on an orientifold of
a Calabi-Yau threefold in the absence of fluxes.
We focus on $O(1)$ instantons whose only zero modes are the
four bosonic modes associated with translation of the instanton in non-compact
directions and a pair of fermionic zero modes associated with the two supercharges broken by the
instanton. We work with a generic superconformal field theory and express our answer in
terms of the spectrum of open strings on the instanton.
We analyse the contribution of multi-instantons of this kind to the superpotential
and argue that it vanishes when background fluxes are absent.
}

\preprint{MIT-CTP/5416}

\begin{document}

\maketitle

\section{Introduction} \label{sintro}

D-instantons generate non-perturbative contributions to string theory
amplitudes \cite{Polchinski:1994fq}. Computing these contributions is of significant importance not only
because they encode non-perturbative effects in string theory but also because
they play a prominent role in moduli stabilization \cite{Kachru:2003aw,Balasubramanian:2005zx}.
As in all instanton contributions, the D-instanton amplitude carries an
overall factor of $e^{-\cT_\Gamma}$ where $-\cT_\Gamma$ is the action of the instanton
characterized by some `charge' $\Gamma$.\footnote{We use the convention that the path
integral is weighted by $e^S$ where $S$ is the action.}
However, even at leading order,
this needs to be accompanied by
the one-loop determinant of fluctuations of open strings around the instanton since this
determines the overall normalization of the amplitude. Initial studies
involving D-instantons bypassed this problem and instead determined the D-instanton
contribution to the amplitude using a combination of S-duality, supersymmetry and
mirror symmetry / T-duality \cite{Green:1997tv,Green:1998by,Green:1997di,RoblesLlana:2006is,
RoblesLlana:2007ae,Alexandrov:2008gh,Alexandrov:2009zh,Pioline:2009ia,Alexandrov:2010ca,Alexandrov:2011va,
Alexandrov:2013yva,Alexandrov:2014rca,Balthazar:2019rnh,Balthazar:2019ypi}.

More recently, the problem of computing the one-loop determinant of fluctuations around
D-instantons was resolved by combining insights from open string field theory with explicit
world-sheet expressions for the same quantity. Even though both expressions are formal, combining
the two leads to an unambiguous result. This approach has been applied to ten-dimensional type IIB string
theory \cite{Sen:2021tpp,Sen:2021jbr}, compactification of type II string theories on
Calabi-Yau threefolds \cite{Alexandrov:2021shf,Alexandrov:2021dyl} as well as two-dimensional string theory \cite{Sen:2021qdk}
and minimal string theory \cite{Eniceicu:2022nay}. In each case the result agrees with earlier predictions based
on symmetries.

Encouraged by this success, in this paper we analyze the D-instanton generated superpotential in
type II string theory on Calabi-Yau orientifolds. Since these theories have $\NN=1$ supersymmetry
in four dimensions, dualities and symmetries are not powerful enough to determine the overall
normalization of instanton corrections. Therefore, we cannot compare our results with any
known result. Nevertheless, since we follow the same procedure that was used in the analysis
of \cite{Sen:2021tpp,Sen:2021jbr,Alexandrov:2021shf,Alexandrov:2021dyl,Sen:2021qdk,
Eniceicu:2022nay},
we expect our analysis to produce the correct result.

D-instanton contributions to the superpotential in type II string theory on Calabi-Yau orientifolds have
been extensively studied earlier in the literature, see {\it e.g.}~\cite{Witten:1996bn,Harvey:1999as,Abel:2006yk,Akerblom:2006hx,Blumenhagen:2006xt,
Ibanez:2006da,Akerblom:2007uc,Cvetic:2007ku,Billo:2007py,Blumenhagen:2009qh,Schmidt-Sommerfeld:2009ere,Grimm:2011dj,Bianchi:2011qh,Kerstan:2012cy,Cvetic:2012ts}.
Our results agree with the results in these papers. However, to the best of
our knowledge, the overall normalization of the amplitude has not been determined unambiguously
in any of the earlier papers, although some attempts were made in \cite{Blumenhagen:2006xt}.
Our main contribution is towards fixing the overall normalization.

Before presenting our final result, let us make a few remarks on the limitations of our analysis:
\begin{enumerate}
\item
We work in backgrounds containing type II string theory on Calabi-Yau threefolds with D-branes
and orientifold planes, but ignore the presence of any flux in the background. It may be possible to take into
account the effect of fluxes by treating them as
perturbations \cite{Billo:2008pg,Billo:2008sp,Cho:2018nfn},
but this has not been done.

\item
We focus on the computation of the superpotential involving only the
closed string moduli fields that parametrize the bulk world-sheet CFT and the string
coupling, and the open string moduli fields that characterize the boundary CFT associated
with the background space-filling D-branes and orientifold planes. It
may be possible to combine our approach with those in \cite{Blumenhagen:2006xt,Akerblom:2007uc,
Cvetic:2007ku,Blumenhagen:2009qh} to determine the superpotental involving
massless open string fields that live on space-filling D-branes but are not moduli,
but we have not done this.

\item
We focus on rigid instantons whose only bosonic moduli are those associated with translation
of the instanton along the four non-compact space-time directions. We also assume that the only fermion
zero modes on the instanton are the ones associated with broken supersymmetry.
\end{enumerate}

Our main result on the one-instanton induced superpotential, given in \refb{edefam}, \refb{e3.43},
takes the form:
\be\label{emain}
|W_\Gamma| \, e^{\cK/2}  = \frac{\kappa_4^3}{16\pi^{2}} \, \Re(\cT_\Gamma(\vp))\, K_0\,  |e^{-\cT_\Gamma(\vp)}|
\, ,
\ee
where
\be\label{edefamintro}
\KKZ= \lim_{\eps\to 0}\lim_{\eps'\to 0}
\exp\[\int_{\eps'}^{1/\eps}  {dt\over 2t} Z_A+\int_{\eps'/4}^{1/\eps}  {dt\over 2t}
Z_M+ 3\int_{0}^{1/\eps} {dt\over 2t} \(e^{-2\pi t} -1 \) \] .
\ee
Here $W_\Gamma$ is the superpotential induced by an instanton of charge $\Gamma$,
$\cK$ is the K\"ahler potential,
$\kappa_4$ is the four-dimensional gravitational constant, $-\cT_\Gamma(\vp)$ is the action of the instanton expressed
as a function of the moduli fields $\vp$,
$Z_A$ is the sum of annulus partition functions, either with
both ends on the D-instanton or with one end on the D-instanton and the other end on a space-filling D-brane,
and $Z_M$ is the M\"{o}bius strip contribution to the partition function
with its boundary on the instanton. $K_0$ has implicit moduli dependence
through $Z_A$ and $Z_M$. Note that due to the possibility of
K\"ahler transformations $\cK\to \cK + f +\bar f$,
$W_\Gamma\to e^{-f}W_\Gamma$, where $f$ is any holomorphic function of the moduli,
neither $\cK$  nor $W_\Gamma$ can be
defined unambiguously, but $|W_\Gamma|e^{\cK/2}$ can. This is the combination
that appears in  \refb{emain}.

Some other results obtained in this paper, within the limitations stated above, are the following:
\begin{enumerate}
\item
We argue that the multi-instanton contributions do not correct the superpotential for the
moduli fields
in the absence of fluxes. This appears to be consistent with the results of \cite{Witten:1999eg}
on worldsheet instanton corrections to the superpotential in heterotic string theory.\footnote{We remark
that in heterotic string theory, worldsheet instantons can generate two different classes of corrections
to the superpotential: terms that involve only the moduli fields characterizing the
K\"ahler class and complex structure
of the Calabi-Yau threefold and the $E_8\times E_8$
or $SO(32)$ vector bundle, and terms that
also involve the matter fields. The second class of terms is often called Yukawa couplings,
and known to receive non-trivial multi-instanton contributions \cite{Candelas:1990rm}.
Via the F-theory/heterotic duality, the first class of corrections are mapped
to D-instanton corrections that only involve the  moduli fields and
the second class of corrections are mapped to D-instanton corrections
that also involve the matter fields living on D-branes~\cite{Witten:1996bn}.
We stress that our claim on the absence of the multi-instanton contributions does
not apply to the terms involving the matter fields.}

\item
We generalize the analysis of \cite{Abel:2006yk,Akerblom:2006hx} (see also \cite{Billo:2007sw})
to show that for any compactification of type II string theory on
Calabi-Yau orientifolds preserving $\NN=1$ supersymmetry, the partition functions $Z_A$ and $Z_M$ associated with
D-instanton corrections are equal to the threshold correction to the gauge coupling when we replace the
D-instanton by a space-filling D-brane. This result is required for proving the holomorphy of the superpotential.

\item
In any solvable string compactification, {\it e.g.} orbifold models discussed in
\cite{Akerblom:2007uc}, we can explicitly
compute $Z_A$ and $Z_M$ and hence $\KKZ$ appearing in \refb{edefamintro}. For a generic compactification
we do not have such a closed form expression. Nevertheless, we show how we can organize the
contributions to $Z_A$ and $Z_M$,
both from the open string channel and from the closed string channel, using character formulas for the
representations of the extended $\NN=2$ superconformal algebra that underlies the theory associated with Calabi-Yau
compactifications~\cite{Eguchi:1988vra,Odake:1988bh,Odake:1989dm,Odake:1989ev}.
The relevant formulae can be found in \refb{emobfinal},
\refb{endfinal}, \refb{e43sum} and \refb{emobclosed}.
This can facilitate computation of $Z_A$ and $Z_M$ in theories that are obtained from solvable
models by small deformations.
\end{enumerate}

The rest of the paper is organized as follows. In \S\ref{sworld} we introduce our conventions
for the world-sheet theory, D-instantons and orientifolds. We use the language of
two-dimensional superconformal field theory (SCFT) instead of the language of Calabi-Yau
geometry so that our results are applicable even for stringy compactifications. In
\S\ref{s1} we compute the normalization of the instanton induced amplitude for a single
D-instanton by carefully taking into account the effect of zero modes.
In \S\ref{s1a} we combine the result of \S\ref{s1} with some disk amplitude results
to compute certain instanton induced amplitudes.
In \S\ref{ssugra} we compare this result with the result expected from a
supergravity theory with instanton induced superpotential to actually compute the superpotential.
This leads to the results \refb{emain}, \refb{edefamintro} quoted above.
In \S\ref{smulti} we analyze the contribution to the superpotential from multi-instanton
configurations and argue that it vanishes. In \S\ref{conclusion} we conclude the paper.

The appendices contain various technical
results. In appendix \ref{eappc} we use string field
theory action to fix the normalization of the fermion kinetic terms.  In appendix
\ref{sdeform} we discuss some aspects of the procedure for fixing the integration measure
over the open string zero modes.
Appendix \ref{sc}
contains a list of various theta function identities.  In appendix \ref{saE}
we use the formul\ae\ for the characters of the
extended $\NN=2$ superconformal algebra on the world-sheet
to organize the annulus and M\"{o}bius strip partition functions $Z_A$ and $Z_M$
that enter  \refb{edefamintro}. In appendix
\ref{sF} we establish the relation between D-instanton partition function and threshold
correction on space-filling D-branes which is used in \S\ref{ssugra}
to establish holomorphy of the
superpotential.

\section{Worldsheet theory} \label{sworld}

We shall consider type IIA or IIB string theory compactified on a Calabi-Yau threefold $\CY$. The
world-sheet theory contains the usual $b,c,\beta,\gamma$ ghost system, the
fields $\xi,\eta,\phi$ obtained by bosonizing the $\beta,\gamma$ system, the anti-holomorphic
counterpart of these fields and the matter superconformal field theory (SCFT). The latter consists
of two parts: the four space-time coordinates $X^\mu$ and their fermionic partners
$\psi^\mu,\bar\psi^\mu$ describing free field theory,
and an interacting (2,2) supersymmetric SCFT with central charge 9 describing
the non-linear sigma model with target space $\CY$.

The matter sector of the theory has four sets of spin fields,
$S_\alpha,S_{\dalpha}, \bar S_\alpha,\bar S_{\dalpha}$.
The fields $S_\alpha$, $S_\dalpha$ are holomorphic vertex operators  of dimension $5/8$, each given by the
product of a holomorphic
dimension $1/4$ spin field associated with the non-compact directions
and a holomorphic dimension $3/8$ spin field associated with the internal SCFT.
Here $\alpha,\dalpha$ denote respectively chiral undotted index
and chiral dotted index of the 3+1-dimensional Lorentz group.
$\bar S_\alpha,
\bar S_{\dalpha}$ are the anti-holomorphic counterparts of these fields. These operators are
used to construct the GSO even vertex operators in the $-1/2$ picture, i.e.
$e^{-\phi/2}S_\alpha,e^{-\phi/2}S_{\dalpha}, e^{-\bar\phi/2}\bar S_\alpha,e^{-\bar\phi/2}
\bar S_{\dalpha}$ are the GSO even dimension one vertex operators. We also introduce the matter
spin fields $\Sigma_\alpha,\Sigma_{\dalpha}, \bar \Sigma_\alpha,
\bar \Sigma_{\dalpha}$ of the opposite GSO parity so that $e^{-3\phi/2}\Sigma_\alpha$,
$e^{-3\phi/2}\Sigma_{\dalpha}$, $e^{-3\bar\phi/2}\bar \Sigma_\alpha$,
$e^{-3\bar\phi/2}\bar \Sigma_{\dalpha}$ are the GSO even  vertex operators in the $-3/2$ picture.
$\Sigma_\alpha,\Sigma_\dalpha$ differ from $S_\dalpha,S_\alpha$ respectively only in the four-dimensional
sector and similar relations hold between $\bar\Sigma_\alpha,\bar\Sigma_\dalpha$ and $\bar S_\dalpha,\bar
S_\alpha$.
In the $\alpha'=1$ unit, the singular operator product
expansions involving these fields are:
\begin{subequations}
\be \label{ematterope}
\begin{split}
&\qquad\qquad \p X^\mu(z) \p X^\nu(w) = -{\eta^{\mu\nu}\over 2 (z-w)^2}+\cdots,
\\
c(z) b(w) =& \, (z-w)^{-1}+\cdots,
\qquad\qquad\qquad\qquad
\xi(z)\eta(w) = (z-w)^{-1}+\cdots,
\\
e^{q_1\phi(z)} e^{q_2\phi(w)} =&\,  (z-w)^{-q_1q_2} e^{(q_1+q_2)\phi(w)}+ \cdots \, ,
\quad
\p \phi (z)\,  \p\phi(w) = -(z-w)^{-2} +\cdots \, ,
\end{split}
\ee
\be\label{epsiope}
\psi^\mu (z) \psi^\nu(w) = -{\eta^{\mu\nu}\over 2} \, (z-w)^{-1}+\cdots\, ,
\ee
\be\label{espinope}
\begin{split}
& e^{-\phi}\psi^\mu(z) e^{-\phi/2} S_\alpha(w) = {i\over 2}\, (z-w)^{-1} (\gamma^\mu)_\alpha^{~\dbeta}
e^{-3\phi/2} \Sigma_\dbeta + \cdots \, ,
\\
& e^{-\phi}\psi^\mu(z) e^{-\phi/2} S_\dalpha(w) = {i\over 2} \, (z-w)^{-1} (\gamma^\mu)_\dalpha^{~\beta}
e^{-3\phi/2} \Sigma_\beta + \cdots\, ,
\end{split}
\ee
\be
\begin{split}
& e^{-\phi/2} S_\beta(z) e^{-3\phi/2} \Sigma_\alpha(w)  = (z-w)^{-2} \ve_{\alpha\beta} e^{-2\phi}(w)
+\cdots \\
& e^{-\phi/2} S_\dbeta(z) \, e^{-3\phi/2} \Sigma_\dalpha(w)  = (z-w)^{-2} \ve_{\dalpha\dbeta} e^{-2\phi}(w)
+\cdots\, ,
\end{split}
\ee
\be \label{espinlast}
e^{-\phi/2} S_\alpha(z) \, e^{-\phi/2} S_\dbeta(w) = \I\,
(z-w)^{-1} (\gamma_\mu)_{\alpha\dbeta}
e^{-\phi} \psi^\mu(w) +\cdots\, ,
\ee
\end{subequations}
where $\cdots$ denotes terms that are suppressed by integer powers of $(z-w)$ compared
to the leading term and
\be
(\gamma^\mu)_{\alpha\dbeta}= (\gamma^\mu)_{\alpha}^{~\dgamma} \ve_{\dgamma\dbeta},
\quad
(\gamma^\mu)_{\dalpha\beta}= (\gamma^\mu)_{\dalpha}^{~\gamma} \ve_{\gamma\beta},
\quad
(\gamma^\mu)_{\alpha\dbeta}=(\gamma^\mu)_{\dbeta\alpha},
\quad
\{\gamma^\mu, \gamma^\nu\}=2\, \eta^{\mu\nu} {\mathds 1} .
\ee
There are also similar operator product expansions involving the anti-holomorphic fields.
Following \cite{Sen:2021tpp,Alexandrov:2021shf,Alexandrov:2021dyl},
we shall normalize the vacua for closed strings such that
\be\label{eclosednorm}
\langle k| c_{-1}\bar c_{-1} c_0\bar c_0 c_1 \bar c_1\, e^{-2\phi}(0) e^{-2\bar\phi}(0)|k'\rangle=
- (2\pi)^{4}\delta^{(4)}(k+k')\, ,
\ee
where $k$ denotes the momentum along the non-compact directions.
Several other operator products  have been described in
appendix \ref{eappc}.

The $(2,2)$ world-sheet superconformal algebra has a pair of $U(1)$ currents $J(z)$ and
$\bar J(\bz)$. It is useful to characterize the spin fields introduced above by their $(\bar J, J)$
charge. The fields $S_\alpha$ and $\Sigma_\dalpha$ carry $J$ charge $3/2$, whereas the fields
$S_\dalpha$ and $\Sigma_\alpha$ carry $J$ charge $-3/2$. A similar charge assignment holds
in the anti-holomorphic sector.

\subsection{Closed string moduli fields}

We now turn to the vertex operators associated with the closed string moduli fields. We shall
use the type IIB viewpoint for which we have identical GSO projection rules in the left and
right-moving sectors, and focus on  K\"ahler moduli and their superpartners since the
D-instanton action $-\cT_\Gamma$ is expected to depend only on these moduli.\footnote{The type
IIA viewpoint corresponds to redefining $\bar J$ as $-\bar J$ and the associated redefinition
of the other generators of the left-moving superconformal algebra.}
If we denote by $\cpi_m$ the canonically normalized
fluctuation of one such K\"ahler modulus
field around some background $\cpi^{(0)}_m$, then its vertex operator
$V^m$  in the $(-1,-1)$ picture will be given by
\be
V^{m} = c\bar c e^{-\phi} e^{-\bar\phi} W^{m}\, ,
\label{e1.6b}
\ee
where $W^{m}$ is a dimension $(1/2,1/2)$ superconformal primary of the internal SCFT with
$(\bar J, J)$
charge $(-1,1)$. The vertex operator $U^{m}$ for the complex conjugate field $(\cpi_m)^*$ has the
form
\be
U^{m} = c\bar c e^{-\phi} e^{-\bar\phi} Y^{m}\, ,
\ee
where $Y^{m}$ is a dimension $(1/2,1/2)$ superconformal primary of the internal
SCFT with $(\bar J, J)$
charge $(1,-1)$. We shall assume that $W^{m}$ and $Y^{m}$ are normalized so that
\be\label{e1.8}
W^{m}(z,\bz) Y^{n}(w,\bw)= -{\delta^{mn}\over |z-w|^2}\, .
\ee
In the language of \cite{Lerche:1989uy},
$W^m$ represents an (anti-chiral,~chiral) operator and $Y^m$ represents a
(chiral,~anti-chiral) operator.
The unusual minus sign on the right hand side of \refb{e1.8} ensures that the
operator product of $e^{-\phi}e^{-\bar\phi}W^m(z,\bz)$ and $e^{-\phi}e^{-\bar\phi}
Y^n(w,\bw)$ is given by $e^{-2\phi}e^{-2\bar\phi}\delta^{mn}|z-w|^{-4}$ without any
extra sign.

The field $\cpi_m$ has a pair of superpartner fermions $\psi_m^\alpha$ and
$\psi^{\prime \dalpha}_{m}$
from the NSR and RNS sectors, whose vertex operators can be written as
\be\label{efermuvert}
V^{m}_\alpha=c\bar c e^{-\phi/2} e^{-\bar\phi}
W^{m}_\alpha\,,\qquad\qquad V^{\prime m}_{\dalpha}=c\bar c e^{-\phi} e^{-\bar\phi/2}
W^{\prime m}_{\dalpha}\,,
\ee
in the $(-1,-1/2)$ and $(-1/2,-1)$ picture, respectively. Then we have
the operator product expansions:
\be \label{e1.9a}
\begin{split}
& e^{-\phi/2} S_\alpha (z) \ e^{-\phi/2}e^{-\bar\phi}
W_\beta^{m} (w,\bw) = (z-w)^{-1} \, \ve_{\alpha\beta}
e^{-\phi}e^{-\bar\phi}W^{m}(w,\bw)+ \cdots\, ,
\\
& e^{-\bar\phi/2}\bar S_\dalpha (\bar z) \ e^{-\phi} e^{-\bar\phi/2} W^{\prime m}_{\dbeta}(w,\bw) =
(\bz-\bw)^{-1} \, \ve_{\dalpha\dbeta} \,  e^{-\phi}e^{-\bar\phi}\, W^{m}(w,\bw) +\cdots \, .
\end{split}
\ee
$W_\alpha^{m}$ and $W^{\prime m}_{\dalpha}$ carry $(\bar J,J)$ charges $(-1,-1/2)$ and $(1/2,1)$,
respectively.

\begin{table}
\centering
{\setlength\extrarowheight{2pt}

   \begin{tabular}{|c|c|c|c|}
\hline
  Field & Vertex operator & Picture number & $(\bar J,J)$ charge  \\
    \hline
   $\cpi^m$ & $V^m=e^{-\phi}e^{-\bar\phi}W^m$ & $(-1,-1)$ & $(-1,1)$ \\
   \hline
 $(\cpi^{m})^\ast $ & $U^m=e^{-\phi}e^{-\bar\phi}Y^m$ & $(-1,-1)$ & $(1,-1)$ \\
   \hline
   $\psi^m_\alpha $&$V^m_\alpha=e^{-\phi/2}e^{-\bar\phi}W^{m}_\alpha$ & $(-1,-1/2)$ & $(-1,-1/2)$ \\
   \hline
   $\psi'^m_\dalpha $& $V^{\prime m}_{\dalpha}=
   e^{-\phi}e^{-\bar\phi/2}W^{\prime m}_{\dalpha}$ & $(-1/2,-1)$ & $(1/2,1)$ \\
   \hline
  $\cfi^m_\dalpha$&$U_\dalpha^{m}=c\bar c e^{-\phi/2} e^{-\bar\phi}Y^{m}_\dalpha$ & $(-1,-1/2)$ & $(1,1/2)$ \\
   \hline
   $\cfi'^m_\alpha$& $U^{\prime m}_{\alpha}=c\bar c e^{-\phi} e^{-\bar\phi/2}
Y^{\prime m}_\alpha$ & $(-1/2,-1)$ & $(-1/2,-1)$ \\
    \hline   \end{tabular}}
    \caption{ A list of closed string fields and vertex operators that are relevant for our analysis,
    their picture numbers and $(\bar J,J)$ charges.}
    \label{t1}
\end{table}

Similarly, the field $(\cpi_m)^*$ has a pair of superpartner fermions
$\cfi_m^\dalpha$ and $\cfi^{\prime \alpha}_m$
from the NSR and RNS sectors, with vertex operators
\be\label{efermustarvert}
U_\dalpha^{m}=c\bar c e^{-\phi/2} e^{-\bar\phi}Y^{m}_\dalpha
\,,\qquad\qquad
U^{\prime m}_{\alpha}=c\bar c e^{-\phi} e^{-\bar\phi/2}
Y^{\prime m}_\alpha\,,
\ee
in the $(-1,-1/2)$ and $(-1/2,-1)$ picture, respectively. Then we have
the operator product expansions:
\be\label{e1.10a}
\begin{split}
& e^{-\phi/2}S_\dalpha (z) e^{-\phi/2}e^{-\bar\phi}Y^{m}_\dbeta(w,\bar w) = (z-w)^{-1} \, \ve_{\dalpha\dbeta} \,
e^{-\phi}e^{-\bar\phi} Y^{m}(w,\bar w)+\cdots ,
\\
& e^{-\bar\phi/2}\bar S_\alpha (\bar z) e^{-\phi}e^{-\bar\phi/2}Y^{\prime m}_{\beta}(w,\bar w)
= (\bar z-\bar
w)^{-1} \, \ve_{\alpha\beta} \, e^{-\phi}e^{-\bar\phi}\, Y^{m}(w,\bar w)+\cdots \, .
\end{split}
\ee
$Y_\dalpha^{m}$ and $Y^{\prime m}_{\alpha}$ carry $(\bar J,J)$ charges
$(1,1/2)$ and $(-1/2,-1)$,
respectively. Since the operators at $z$ or $\bar z$ on the lhs of~\refb{e1.9a}
and~\refb{e1.10a} are supercurrents, the above OPEs encode the space-time
supersymmetry transformations.
$W_\alpha^{m}$, $W^{\prime m}_{\dalpha}$, $Y^{m}_\dalpha$ and $Y^{\prime m}_\alpha$ are
obtained by taking the product of four-dimensional spin fields with
Ramond sector ground states of the SCFT associated with $\CY$ in left/right moving part,
while keeping the right/left moving part the same \cite{Lerche:1989uy}.
For convenience we have listed in table \ref{t1} the picture numbers and $(\bar J,J)$ charges
of various vertex operators.

\subsection{Orientifold operation}
\label{subsec-orientifold}

We now consider an orientifold of this theory, obtained by taking its quotient by the
orientation reversal on the world-sheet, together with a possible $\IZ_2$ transformation $g$
that acts on the SCFT associated with $\CY$.\footnote{Early papers investigating orientifolds of type II strings include~\cite{Sagnotti:1987tw,Polchinski:1987tu,Pradisi:1988xd,Horava:1989vt} and modern reviews of the subject can be found in~\cite{Angelantonj:2002ct,Blumenhagen:2006ci}.} We shall call the combined transformation \oo.
If the transformation includes $(-1)^{F_L}$ that changes the sign of RNS and RR states,
we include it in $g$ since this can be regarded as an action on the internal parts of $\bar S_\alpha$, $\bar S_\dalpha$.
On ghost fields as well as on matter field $X^\mu$ and $\psi^\mu$ associated with the non-compact
space-time directions, the orientifold transformation
acts by exchanging the left and right-moving sectors and exchanging the
argument $z$ with $-\bz$.\footnote{Defining $z=e^{\tau+i\sigma}$, orientation reversal corresponds to the
transformation $\sigma\leftrightarrow
\pi-\sigma$, with fixed points at $\sigma=\pi/2$ and $3\pi/2$. The
fixed points may be shifted using $\sigma$ translation. This corresponds to a redefinition
$z\to e^{i\alpha}z$, $\bz\to e^{-i\alpha}\bar z$, and makes the orientifold transformation
$z\leftrightarrow -e^{-2\I\alpha}\bz$. In particular, by choosing $\alpha=\pi/2$, we can make
the orentifold transformation act as $z\to \bz$, i.e.\ $\sigma\to 2\pi-\sigma$.
However, since changing the phase of $z$ by $e^{i\alpha}$
is generated by $e^{i\alpha(L_0-\bar L_0)}$, this does not affect the transformation
laws of level matched
closed string states.}
Therefore, a holomorphic field of weight $h$ constructed from $X^\mu$, $\psi^\mu$
and the ghost fields transforms to its
anti-holomorphic counterpart with the argument replaced by $-\bz$ and an
overall multiplicative factor
of $(-1)^h$. In particular, under \oo\ we have
\ben \label{eortr}
&& \p X^\mu (z)\to - \p X^\mu(-\bz), \quad c(z)\to -\bar c(-\bz), \quad
b(z)\to \bar b(-\bz), \quad e^{-\phi}\psi^\mu(z)\to -e^{-\bar\phi}\bar\psi^\mu(-\bz),
\non\\
&&
\p\xi(z)\to -\bar\p\bar\xi(-\bz), \qquad \eta(z)\to -\bar\eta(-\bz),
\qquad e^{-2\phi}(z)\to e^{-2\bar\phi}(-\bz)\, .
\een
Some caution is needed in interpreting these transformation laws. They clearly do not
represent symmetries of a general correlation function in the conformal field theory but
preserve the operator product expansion on the real axis. The
correct way to interpret these transformation laws is to use them to find the transformation
of a state $|V\rangle=V(0)|0\rangle$ from the transformation properties of the local
operator $V$ and the $SL(2,\IR)$ invariant vacuum $|0\rangle$.

On the graviton and 2-form vertex operators \oo\ acts as
\be
e_{\mu\nu} c\bar c e^{-\phi}\psi^\mu e^{-\bar\phi}\bar\psi^\nu(z,\bz) \to
e_{\mu\nu} \bar c c e^{-\bar\phi}\bar\psi^\mu e^{-\phi}\psi^\nu(-\bz,- z) =-
e_{\mu\nu} c\bar c e^{-\phi}\psi^\nu e^{-\bar\phi}\bar\psi^\mu(-\bz,- z)\, .
\ee
Thus, the graviton vertex operator at $z=0$
is odd under \oo. In order to keep the graviton in the spectrum, we
need to declare that the $SL(2,\IC)$ invariant vacuum $|0\rangle$ is also odd.
Therefore, in the orientifold we only keep the vertex operators
that are odd under \oo. This is consistent with the fact that the
sphere three-point function of three graviton
vertex operators is non-zero even though
each of these vertex operators is odd under \oo.

Next, we consider the vertex operators associated with the compact directions. In order
to make sure that the orientifold preserves $\NN=1$ space-time supersymmetry in four
dimensions, the action of $\Omega_g$ on the extended superconformal algebra must again take
a weight $h$ holomorphic current of the algebra to $(-1)^h$ times
the anti-holomorphic current with argument
$z$ replaced be $-\bz$. Under
\oo\, we take $J(z)\rightarrow -\bar J(-\bar z)$ and we can choose
\be \label{e1.12a}
e^{-\phi/2}S_\alpha(z)\to -e^{-\bar\phi/2}\bar S_\alpha(-\bz), \qquad
e^{-\phi/2}S_\dalpha(z)\to -e^{-\bar\phi/2}\bar S_\dalpha(-\bz)\, .
\ee
The minus signs in these equations could be changed to phases by redefining
$S_\alpha$, $S_\dalpha$, $\Sigma_\alpha$, $\Sigma_\dalpha$
as $e^{i\beta}S_\alpha$, $e^{-i\beta}S_\dalpha$, $e^{-i\beta}\Sigma_\alpha$,
$e^{i\beta}\Sigma_\dalpha$
without affecting the
operator product expansion \refb{espinope}-\refb{espinlast},
but we shall stick to the convention in which
\refb{e1.12a} holds.
Under \oo,
$W^{m}$ and
$Y^{m}$ also get exchanged up to a phase. By multiplying $W^m$ and $Y^m$ by equal and opposite
phases so as not to affect \refb{e1.8}, we shall ensure that
\be \label{e1.13a}
e^{-\phi}e^{-\bar\phi}W^{m}(z,\bz)\rightarrow e^{-\phi}e^{-\bar\phi}Y^{m}(-\bz,- z)
\qquad \Rightarrow \qquad
V^{m}(z,\bz)\rightarrow -U^{m}(-\bz,- z)
\ee
under \oo.
It also follows from \refb{e1.9a}, \refb{e1.10a}, \refb{e1.12a}, \refb{e1.13a} on the real
axis that under \oo\
\be
V^{m}_\beta(z,\bar z)  \rightarrow  -U^{\prime m}_{\beta}(-\bar z,- z), \qquad
 V^{\prime m}_{\dbeta}(z,\bar z)  \rightarrow  -U^{m}_{\dbeta}(-\bar z,- z)\, .
\ee
Therefore, taking into account the fact that $|0\rangle$ is odd,
the states invariant under \oo\ are:
\be\label{evertexor}
{1\over \sqrt 2} \, [V^{m}(0)+U^{m}(0)]|0\rangle, \quad
{1\over \sqrt 2} \, [V^{m}_\beta(0) + U^{\prime m}_{\beta}(0)]|0\rangle, \quad
{1\over \sqrt 2} \, [V^{\prime m}_{\dbeta}(0) + U^{m}_{\dbeta}(0)]|0\rangle\, ,
\ee
corresponding to the fields
\be \label{einvfield}
\hat\phi_m={1 \over \sqrt 2}(\cpi_m+(\cpi_m)^*), \quad \xxi_m^\beta={1 \over \sqrt 2}(\psi_m^\beta+
\cfi^{\prime \beta}_m), \quad \xxi_m^\dbeta= {1 \over \sqrt 2}(\cfi_m^\dbeta +
\psi^{\prime \dbeta}_m)\, ,
\ee
respectively. Note that $\hat\phi_m/\sqrt 2$
becomes the real part of the scalar component  $\phi_m$ of
a chiral superfield, the imaginary part of $\phi_m$
being given by a field from the RR sector.

\subsection{D-instantons}

We now consider a D-instanton in this theory. We shall use the upper half plane
description of the open string world-sheet. First, we shall analyze the
system before taking the orientifold.
We choose Neumann boundary condition on the ghosts, Dirichlet boundary condition
on $X^\mu,\psi^\mu,\bar\psi^\mu$ and
\be\label{espininst}
J(z)=\bar J(\bar z),
\qquad
c \, e^{-\phi/2} S_\alpha(z) = -\bar c \, e^{-\bar\phi/2} \bar S_\alpha(\bz),
\qquad
c \, e^{-\phi/2} S_\dalpha(z) = \bar c \, e^{-\bar\phi/2} \bar S_\dalpha(\bz) \, ,
\ee
on the real axis. Also, we normalize the vacuum of the open strings on the
D-instanton as
\be\label{eopennorm}
\langle 0| c_{-1} c_0 c_1 \, e^{-2\phi}(0)|0\rangle=1\, .
\ee
Here, as well as in the rest of the equations in this section, $|0\rangle$ denotes the
$SL(2,\IR)$ invariant vacuum of the open string with both ends on the D-instanton.
Note that the
relative minus sign between the last two equations in \refb{espininst} is forced on us by
the consistency with the operator product expansion \refb{espinlast} and its
anti-holomorphic counterpart and the Dirichlet boundary
condition on $\psi^\mu$, but which of the two equations carries the minus sign is a matter of
choice and distinguishes a D-instanton from an anti-D-instanton. These equations are also
consistent with the fact that the $J$ charge carried by $S_\alpha$ ($S_\dalpha$) is equal
to the $\bar J$ charge carried by $\bar S_\alpha$ ($\bar S_\dalpha$). In particular, the boundary
condition $J(z)=-\bar J(\bz)$ would not be consistent with the last two boundary conditions
in \refb{espininst}.

With this choice of boundary conditions, the vertex operators $c_0^-V^{m}(i)$ and $c_0^-U^{m}(i)$
are allowed to have a non-zero one-point function
on the upper half plane.  We shall illustrate this using the one-point function
of $c_0^-V^{m}$. Since it carries
$(\bar J, J)$ charge $(-1,1)$, we can replace $c_0^-V^{m}(i)$ either by $\ointop dz J(z)
c_0^-V^{m}(i)/(2\pi \I)$ or $-\ointop' d\bar z \bar J(\bar z) c_0^-V^{m}(i)/(2\pi \I)$
where $\ointop$ and $\ointop'$ are integrals along
anti-clockwise and clockwise contours around $i$, respectively. We can now deform both
contours and make them lie along the real axis, but in opposite directions. This compensates the
minus sign due to the negative $\bar J$ charge of $c_0^-V^{m}$ and, using the $J=\bar J$ boundary
condition on the real line, we get identical results. Therefore, we do not encounter any
contradiction. Similar result holds for $c_0^-U^{m}$. In contrast, if we had vertex operators
carrying equal $J$ and $\bar J$ charges, then we would get equal and opposite results from
the $J$ and $\bar J$ contours, leading to a contradiction. Therefore, the one-point function of
such vertex operators must vanish. If we consider the compactification of type IIB string theory
on a Calabi-Yau manifold, then the vertex operators with $(\bar J, J)$ charges
$(\mp 1,\pm 1)$ represent K\"ahler moduli and the vertex operators with $(\bar J, J)$ charges
$(\pm 1,\pm 1)$ represent complex structure moduli. This is consistent with the fact that the
instanton action depends only on the K\"ahler moduli.
For type IIA compactification the two sets of moduli get exchanged,
but we can repeat the analysis presented here by reversing the sign of $\bar J$ in all equations. This would be consistent as well.

The spectrum of open strings with both ends on the D-instanton
consists of a set of massive modes and a set of zero modes --- those with vanishing $L_0$ eigenvalue.
We shall analyze the states in Siegel gauge, satisfying the constraint that $b_0$ annihilates the state.
Some of the zero modes are physical, describing the collective coordinates of the instanton. These
include the four fermionic zero modes associated with broken supersymmetry since the
instanton breaks four out of eight supersymmetries, and bosonic zero modes describing the
position of the instanton in non-compact directions.
In particular, the four zero modes associated with broken supersymmetry are described
by the vertex operators $c \, e^{-\phi/2} S_\alpha(z) = -\bar c \, e^{-\bar\phi/2} \bar S_\alpha(\bz)$ and
$c \, e^{-\phi/2} S_\dalpha(z) = \bar c \, e^{-\bar\phi/2} \bar S_\dalpha(\bz)$
on the real line.
However, there are also ghost zero modes in the NS sector corresponding to the states:
\be\label{eghost}
c_1 \beta_{-1/2} |-1\rangle=c\p\xi e^{-2\phi}(0)|0\rangle, \qquad c_1 \gamma_{-1/2} |-1\rangle
= c\eta(0)|0\rangle\, .
\ee
The presence of these ghost zero modes in the spectrum reflects
the breakdown of Siegel gauge choice \cite{Sen:2021qdk}.

The zero modes described
above are universal, but there could be additional non-universal zero modes associated with
the deformation of the instanton along the Calabi-Yau threefold $\CY$ and their fermionic partners.
We shall comment on these after discussing the effect of the orientifold projection.

\subsection{D-instantons in orientifold}

Let us now analyze the same D-instanton in the orientifold
considered above
following \cite{Gimon:1996rq,Gaberdiel:1997ia,Gaberdiel:1997mg,Ibanez:2007rs}.
We shall assume that the resulting instanton is of $O(1)$ type
\cite{Blumenhagen:2009qh,Ibanez:2007rs}
so that the D-instanton is invariant under
the orientifold projection and we do not need to add its image.
The open string spectrum on the resulting D-instanton is supposed to include the four translation
zero modes along the non-compact space-time
directions and the two fermion zero modes associated with broken supersymmetry, since the
D-instanton now breaks two out of four supersymmetries. The translation zero modes are
associated with the NS sector states
\be\label{etranslationa}
i\sqrt 2\, c_1 \psi^\mu_{-1/2} |{-1}\rangle=-i\sqrt 2\, ce^{-\phi}\psi^\mu(0)|0\rangle\, .
\ee
It follows from \refb{eortr} that under \oo, $c\psi^\mu e^{-\phi}(0)$ transforms to
$\bar c\bar\psi^\mu e^{-\bar\phi}(0)$ which can be equated to $-c\psi^\mu e^{-\phi}(0)$
using the boundary conditions~\refb{eortr}.
Therefore, in order that \refb{etranslationa} be even under
this transformation, we must declare $|0\rangle$ to be odd under \oo\, or
equivalently declare the Chan-Paton factor, which in this case is a $1\times 1$ matrix,
to be odd under \oo\ \cite{Gaberdiel:1997ia,Gaberdiel:1997mg}.
This in turn implies that the states \refb{eghost} are odd under \oo\ since $\beta$,
$\gamma$, $\xi$, $\eta$, $\phi$ satisfy Neumann boundary conditions.
Therefore, in the orientifold the ghost zero modes~\refb{eghost} are absent.
Once the action of the orientifold transformation on the vacuum $|0\rangle$
has been determined, we can also determine the spectrum of massive modes that survive
the orientifold projection.

Let us now turn to the Ramond sector of the open string on the D-instanton. Using
\refb{eortr}, \refb{e1.12a}
and \refb{espininst}, we see that under \oo:
\be
\begin{split}
& ce^{-\phi/2}S_\alpha(0)\to \bar c e^{-\bar\phi/2}\bar S_\alpha(0)
= -ce^{-\phi/2}S_\alpha(0), \\
& ce^{-\phi/2}S_\dalpha(0)\to \bar c e^{-\bar\phi/2}\bar S_\dalpha(0)
= ce^{-\phi/2}S_\dalpha(0)\, .
\end{split}
\ee
Taking into account the fact that $|0\rangle$ is odd under \oo, we see that
the zero modes that are even under \oo\  in Ramond sector are
\be\label{efermionzero}
ce^{-\phi/2}S_\alpha(0)|0\rangle\, .
\ee

Note that once we consider an orientifold, we need to have a certain number of
space-filling D-branes\footnote{We use the term `space-filling D-brane' for any
D-brane that extends along all the non-compact directions and `D-instanton' for any D-brane
that is localized at a point along the non-compact directions, irrespective of how they
extend along the internal directions.}
to cancel any RR charge carried by orientifold planes. As a result, we shall get additional
open string states with one end on the instanton and the other end on the space filling D-brane.
There could in principle be additional zero modes in this sector. Furthermore, some of the non-universal
zero modes associated with the deformation of the instanton inside $\CY$ may also survive the orientifold
projection. In our analysis we shall assume
that the only zero modes on the D-instanton are the four bosonic modes and two fermionic modes
associated with broken translation symmetry and supersymmetry.

\subsection{Normalization conventions for orientifolds} \label{snorm}

In the orientifold that involves projection into $\Omega_g$ invariant states,
the various parameters of the parent theory are related to those
in the daughter theory by powers of $2$. We shall now briefly discuss the origin of these
factors.

Since in the computation of the one-loop open string partition function in the
orientifold theory we need to average over the annulus and the M\"{o}bius strip
contribution, the contribution from the annulus carries a
multiplicative factor of $1/2$ compared to the annulus amplitude in the parent theory before
taking the orientifold projection. Since the annulus amplitude can also be viewed in the closed
string channel as the exchange of a single closed string between the D-instanton with itself, it is
proportional to the square of the action of the D-instanton. Therefore, the D-instanton action
in the orientifold is $1/\sqrt 2$ times the D-instanton action in the parent theory, provided the
two theories share the same value of the four-dimensional gravitational coupling $\kappa_4$.

More generally, the rules for computing a general amplitude in an orientifold is that for an
amplitude with $N$ external closed strings, $M$ external open strings, $b$ boundaries,
$c$ crosscaps and $g$ handles, we have an extra factor \cite{FarooghMoosavian:2019yke}
\be
2^{-g - (b+c)/2 + M/4}\, ,
\ee
compared to that in the parent theory. Therefore, closed string one-point function on the disk,
having $b=1$, carries an extra factor of $1/\sqrt 2$. Since this is proportional to
the action of the instanton, it confirms that the action $-\cT_\Gamma$
is multiplied by a factor of $1/\sqrt 2$.
On the other hand, since each external open string carries a factor of the open string
coupling constant $g_o$, the $2^{M/4}$
factor shows that $g_o$ is effectively multiplied by $2^{1/4}$. Therefore, the relation
between the real part of the action $-\cT_\Gamma^R$ of the D-instanton and the open string
coupling constant $g_o$ on the D-instanton~\cite{Sen:1999xm}
\be \label{etgo}
\cT_\Gamma^R = {1\over 2\pi^2 g_o^2}\, ,
\ee
is not affected. Similarly, the closed string coupling $\kappa_4$, captured {\it e.g.} by
the sphere three-point function of three external closed strings, remains the same as in the
parent theory since we have $g=b=c=M=0$. Therefore, the expression for the instanton action,
given in terms of the closed string coupling constant, acquires a factor of $1/\sqrt 2$.

 We shall
follow the convention that unless mentioned otherwise, the parameters used here are those
in the daughter theory, {\it e.g.} $\kappa_4$ and $-\cT_\Gamma$ will label
respectively the four-dimensional gravitational coupling and
D-instanton actions in the daughter theory.

\section{Normalization of rigid instanton amplitudes} \label{s1}

In this section, we shall compute the normalization of the instanton contribution to the
amplitude by evaluating
the exponential of the annulus and M\"{o}bius strip diagrams. However, we need to treat the
contributions from the zero modes carefully with the help of open string field theory.
For this, we need to first review
some aspects of free open string field theory on the D-instanton.

Since an open string with both ends on the D-instanton has states with zero $L_0$ eigenvalue
that leads to zero modes in the open string spectrum, we shall first consider a regulated
system where we put slightly different boundary conditions on the two ends of the open
string \cite{Sen:2021tpp,Alexandrov:2021shf,Alexandrov:2021dyl}.
This lifts the zero modes and helps us to make the correspondence between the worldsheet
expressions and string field theory path integral expressions for the normalization manifest.
Once this correspondence is established, we shall take the limit of zero separation and
compute the normalization by evaluating the string field theory
path integral.
Further discussion on this deformation can be found in appendix \ref{sdeform}.
For
simplicity of notation we shall continue to refer to the modes that become zero
modes in the limit of zero separation as zero modes even for non-zero
separation.

For our analysis we shall need some aspects of open string field theory that will be described
in \S\ref{sopensft}. Later, in order to
fix the normalization of the closed string fields, we shall also need some aspects of closed string
field theory which we review in appendix \ref{eappc}. The reader interested in
more details can consult
\cite{Taylor:2003gn,Okawa:2012ica,deLacroix:2017lif,Erler:2019vhl,Erler:2019loq,Erbin:2021smf}.

\subsection{Open string field theory} \label{sopensft}

The NS sector open string field in Siegel gauge corresponds to a general
GSO even state $|\Phi_{NS}\rangle$ of picture number $-1$, subject to the condition
\be
b_0|\Phi_{NS}\rangle=0\, .
\ee
Let $\{|\vp_r\rangle\}$ denote a set of basis states and $|\Phi_{NS}\rangle = \sum_r \xi_r |\vp_r\rangle$.
The quadratic part of the string field theory action has the form
\be
\label{esNS}
S_{NS}={1\over 2} \, \langle \Phi_{NS}|c_0 L_0|\Phi_{NS}\rangle \, ,
\ee
where $L_n$ are the Virasoro generators. We shall normalize the basis states $|\vp_r\rangle$
so that in any given $L_0$ eigenvalue sector, the matrix $-\langle \vp_r|c_0|\vp_s\rangle$
has determinant 1. This does not completely specify the choice of basis, but for our
purpose any basis satisfying this condition will suffice. The minus sign in the definition of the
matrix is related to our sign convention for the action --- we use $e^S$ as the weight factor in
the path integral. For classical string fields carrying ghost number 1,
such a choice of basis can be achieved by taking the basis states
$|\vp^{(1)}_r\rangle$ to satisfy $\langle\vp^{(1)}_r| c_0|\vp^{(1)}_s\rangle=-\delta_{rs}$.
For other ghost numbers we can achieve this by pairing the basis states
$|\vp^{(n)}_s\rangle$ of ghost number $n$ with the basis states $|\vp^{(2-n)}_s\rangle$
of ghost number $(2-n)$ such that
$\langle\vp^{(2-n)}_r|c_0|\vp^{(n)}_s\rangle=-\delta_{rs}$ for $n\le 0$.

For the R sector there are various formalisms, but for our analysis it will be simplest to use the
one given in \cite{Kunitomo:2015usa,Kunitomo:2016bhc}
(see also \cite{Kazama:1985hd,Kazama:1986cy,Terao:1986ex})
and restrict to string fields with $L_0>0$ using the trick of considering open strings
stretched between a separated pair of instantons. In this formalism the R sector string
field in Siegel gauge
is a general GSO even state $|\Phi_R\rangle$ of  picture number $-1/2$, subject to the conditions
\be
b_0|\Phi_R\rangle=0,
\qquad
\beta_0|\Phi_R\rangle=0\, .
\ee
Let $\{|\tilde \vp_a\rangle\}$ denote the set of basis states and
$|\Phi_{R}\rangle = \sum_a \tilde\xi_a |\tilde\vp_a\rangle$ be a general string field in Siegel gauge,
obtained by the action of $b,c,\beta,\gamma$ oscillators and matter operators
acting on $e^{-\phi/2}S_\alpha(0)|0\rangle$ or $e^{-\phi/2}S_\dalpha(0)|0\rangle$.
The quadratic part of the action has the form
\be
\label{eopenraction}
S_R= {1\over 2} \, \langle \Phi_{R}|c_0 G_0 \delta(\gamma_0)|\Phi_{R}\rangle \, ,
\ee
where $G_n$ are the modes of the super-stress tensor $-2\psi_\mu\p X^\mu+\cdots$
of the combined matter-ghost theory, normalized so that $G_0^2=L_0$.
Since the operator $\delta(\gamma_0)$ has not been introduced before, we need to describe
how to compute matrix elements of $\delta(\gamma_0)$. It is a Grassmann odd operator
with specific (anti-)commutation rules with the usual operators.
Most of the rules are obvious, {\it e.g.}
if this is multiplied by a factor of $\gamma_0$ the result will vanish and it (anti-)commutes with
all non-zero mode $\beta,\gamma$ oscillators as well as all $b,c$ and
matter operators. Since $\beta_0$ annihilates the vacuum in the $-1/2$ picture,
we do not encounter products of $\beta_0$ and $\delta(\gamma_0)$ for evaluation of \refb{eopenraction}.
Using these rules, we can manipulate the matrix element appearing in \refb{eopenraction} so that
$\delta(\gamma_0)$ acts on $e^{-\phi/2}S_\alpha(0)|0\rangle$ or
$e^{-\phi/2}S_\dalpha(0)|0\rangle$ to the right.
In these cases we follow the rules
\be
\label{e3.5new}
\delta(\gamma_0) e^{-\phi/2}S_\alpha(0)|0\rangle=e^{-3\phi/2}S_\alpha(0)|0\rangle,
\qquad \delta(\gamma_0) e^{-\phi/2}S_\dalpha(0)|0\rangle=e^{-3\phi/2}S_\dalpha(0)|0\rangle\, .
\ee
Once the factor of $\delta(\gamma_0)$ has been
removed this way, we can use \refb{eopennorm} and operator product expansions
to compute the remaining matrix element.
We shall normalize the basis states
$|\tilde\vp_a\rangle$ such that inside a subspace of states of fixed $L_0$ eigenvalue $h$,
the matrix
$ M_{ab}=\langle \tilde\vp_a|c_0 {G_0\, L_0^{-1/2}}  \delta(\gamma_0)|\tilde\vp_b\rangle$
satisfies $\det(M)=1$. As can be verified from \refb{epsiope}-\refb{espinlast},
for the zero modes associated with broken supersymmetry, this simply amounts to
choosing the zero mode vertex operators to be $ce^{-\phi/2}S_\alpha$ and
$ce^{-\phi/2}S_\dalpha$ without any further normalization.
This normalization also agrees with the one used in \cite{Sen:2021tpp,Alexandrov:2021shf,Alexandrov:2021dyl}.

We further note that in the formulation of superstring field theory, it is possible to integrate
out a set of degrees of freedom and work with the remaining fields. There is a large amount of
freedom in deciding which degrees of freedom
we integrate out. We can use this
freedom to work with a truncated version of string field theory with a finite number of
degrees of freedom that contains all the zero modes and possibly some other modes.
In this context we note that if in the R sector we integrate out all the non-zero modes,
then we can also work with the action where in the kinetic term $G_0\delta(\gamma_0)$
is replaced by a more conventional operator $\YY_0 Q_B$
where $Q_B$ is the BRST operator and $\YY_0$ is the zero mode of the inverse
picture changing operator \cite{Sen:2021tpp}. The resulting action coincides with
\refb{eopenraction} in the zero mode sector. The choice \refb{eopenraction} allows us to also
include some of the non-zero modes in the string field theory, but the final result
is not affected by this choice.

\subsection{Normalization before the orientifold projection} \label{sorientdef}

We first analyze the normalization of the instanton amplitude before the orientifold
projection. Formally this is given by the exponential of the annulus amplitude with
the boundaries lying on the D-instanton. Our strategy will be to interpret (part of)
the annulus amplitude as arising from path integral over open string modes and then
use insights from string field theory to tame the divergences associated with the
zero modes.
Even though this has been discussed in detail
elsewhere~\cite{Sen:2021qdk,Sen:2021tpp,Sen:2021jbr,Alexandrov:2021shf,
Alexandrov:2021dyl},  we shall reformulate
the earlier analysis in a language that makes the extension to the orientifold case
straightforward.

We begin by writing down the usual expression for the exponential of the annulus amplitude:
\be\label{epowerB}
e^A=
\exp\[\int_0^\infty {dt\over 2t} \, Z_A(t)\].
\ee
This has possible divergences from the $t\to\infty$ end due to massless open string states
propagating in the loop and also from the $t\to 0$ end due to massless closed string
states exchanges between the instanton and itself. Our goal will be to make sense of these
divergences using the language of string field theory and extract a finite, unambiguous
answer. For this, we first express \refb{epowerB} as
\be \label{eccd2}
e^A=
\exp\[\int_0^{\delta} {dt\over 2t} \, Z_A(t)
+ \int_{\delta}^\infty {dt\over 2t} \, Z_A(t)\] ,
\ee
where $\delta$ is some small positive constant.
The two terms appearing in this expression can be  interpreted as follows.

For the first term in the exponent we make a change of variables $\ell=1/t$ and
interpret it as the matrix element of\,\footnote{As will be discussed below \refb{eg11pm},
under the assumption that the
only fermion zero modes on the instanton are those associated with broken supersymmetry,
the contribution to $Z_A+Z_M$ from the odd spin structure sector vanishes. Otherwise
the relation between the open string channel expression and the closed string channel
expression is more complicated due to the existence of $\beta,\gamma$ zero modes.}
\be\label{eccd1}
c_0^- b_0^+ \int_{1/\delta}^\infty \, d\ell \, e^{-\pi \ell (L_0+\bar L_0)}
= {1\over \pi}\, c_0^- b_0^+ e^{-\pi (L_0+\bar L_0)/\delta}/ (L_0+\bar L_0) \, ,
\ee
between the boundary states of the D-instanton. Here $L_0$ and $\bar L_0$ are
the Virasoro generators acting on the closed string states. Physically this represents a
correction to the D-instanton action due to the emission and absorption of a
closed string, with the factor $1/(L_0+\bar L_0)$ coming from the closed string propagator
with both ends attached to the instanton. The $e^{-\pi (L_0+\bar L_0)/\delta}$ factor
can be absorbed into the definition of the two off-shell one-point functions of the closed string on the D-instanton.
Potential divergence of the left hand side of \refb{eccd1} from
the $\ell\to\infty$ region can be traced to the contribution from
closed string states with $L_0+\bar L_0=0$.
For the annulus with both boundaries on the D-instanton the
closed string carries four-dimensional momentum $k$ and we get a factor proportional to
$\int d^4 k/k^2$ which has no divergence from the small $k$ region. This implies
finiteness of the matrix element of \refb{eccd1} between the
boundary state of the D-instanton.

To analyze the second term in the exponent of \refb{eccd2}, we first express $Z_A$ as:
\be \label{egenZA}
Z_A=\Tr_{NS}  \(e^{-2\pi t L_0} (-1)^F\) + {1\over 2} \, \Tr_{R}\( e^{-2\pi t L_0} (-1)^F\)
=  \sum_r s_r e^{-2\pi t h_r} +{1\over 2}\sum_a \tilde s_a e^{-2\pi t \tilde h_a}\, ,
\ee
where $h_r$ and $\tilde h_a$ are respectively the $L_0$ eigenvalues of the basis states
$|\vp_r\rangle$ and  $|\tilde\vp_a\rangle$ in the NS and R sectors, $F$ is the space-time fermion number,
$s_r$ is 1 or $-1$ depending on whether the mode $\xi_r$ is Grassmann even or Grassmann odd, and similarly
$\tilde s_a$ is 1 or $-1$ depending on whether the mode $\tilde\xi_a$ is Grassmann even or Grassmann odd.
Both in the NS sector and the R sector the Grassmann parity is correlated with the $(-1)^F$ eigenvalue.
The extra factor of $1/2$ in front of the R-sector reflects the fact that the kinetic operator in the
R sector is $G_0$, which is the square root of $L_0$.
As long as we have $h_r>0$ and $\tilde h_a>0$, \refb{egenZA} substituted into the second term
in \refb{eccd2} gives a finite result, while divergences come from states with $h_r=0$ and/or
$\tilde h_a=0$. To deal with these cases, we shall first work with the deformed system
discussed at the beginning of this section by considering a slightly shifted boundary
condition at the two boundaries of the annulus so that
we have $h_r>0$, $\tilde h_a>0$ for all $r$ and $a$ and then recover the result for
coincident boundary condition by taking the appropriate limit.

We now rewrite \refb{egenZA} as
\be
\label{ebeginA}
Z_A=Z_A'+Z_A''\, ,
\ee
where
\be
\label{e311new}
Z_A'={\sum_r}' s_r e^{-2\pi t h_r} +{1\over 2}\,{\sum_a}' \tilde s_a e^{-2\pi t \tilde h_a},
\qquad
Z_A''={\sum_r}'' s_r e^{-2\pi t h_r} +{1\over 2}\,{\sum_a}'' \tilde s_a e^{-2\pi t \tilde h_a}\, .
\ee
Here the set labelled by $'$ gets contributions from a finite number of modes and the rest of the modes are
included in the set labelled by $''$. The set $'$ is supposed to satisfy the following restrictions:
\begin{enumerate}
\item The set $'$ includes all the $h_r$'s and $\tilde h_a$'s which will be
zero when we remove the deformation that separates the two boundaries.
\item In the quadratic part of the string field theory action given in \refb{esNS} and
\refb{eopenraction} there should not be any cross term between modes in the set
$'$ and the modes in the set $''$. Therefore, if $|\vp_r'\rangle$ is a basis state in the
set $'$ and $|\vp''_s\rangle$ is a basis state in the set $''$, then in the NS sector
$\langle \vp_r'|c_0 L_0|\vp_s''\rangle$ must vanish and in the R sector
$\langle \vp_r'|c_0 G_0\delta(\gamma_0)|\vp_s''\rangle$ must vanish.
\item The normalization condition on the basis states, as described below \refb{esNS}
and \refb{e3.5new}, must hold separately within the set $'$ (and so also within the set $''$).
\item The set $'$  should have the
property that
\be \label{eccd4}
{\sum_r}' s_r +{1\over 2} \,  {\sum_a}'\tilde s_a =0\, .
\ee
This condition may force us to include  in the set labelled by $'$
some modes which will continue to
have non-zero $h_r$ and/or
$\tilde h_a$ even in the limit when we remove the deformation.
\end{enumerate}
Clearly, there is a lot of freedom involved in this decomposition into the two sets,
but the final result will be independent of it. For example, once we have chosen the
set containing the zero modes,
we could first add to this set sufficient number of
$(-1)^F=-1$ NS sector states to make the left hand side into a negative integer.
We then add
the correct number of NS sector states of ghost number 1, each of which contributes
1 and does not need to have a pair, to cancel the negative integer that we had.

We now write
\be
\label{eccd5}
\begin{split}
\int_{\delta}^\infty {dt\over 2t} \, Z_A'(t)
=&\, \int_{\delta}^\infty {dt\over 2t} \, \[{\sum_r}' s_r e^{-2\pi t h_r} +{1\over 2} \,
{\sum_a}' \tilde s_a e^{-2\pi t \tilde h_a}\]
\\
=&\,  - {1\over 2}\, {\sum_r}' s_r \ln h_r
- {1\over 4}\, {\sum_a}' \tilde s_a \ln \tilde h_a\, ,
\end{split}
\ee
where the last equality can be seen by redefining $th_r$ or $t\tilde h_a$ as a new
variable $u$ in individual terms in the sum, carrying out the $u$ integration in the small
$\delta$ limit and using \refb{eccd4} to show that the $\delta$ dependence drops out.
More specifically, we need $h_r\delta$ and $\tilde h_a\delta$ to be small
for this manipulation to hold. Therefore, it is important that we apply this formula
only on the primed set containing a finite set of states with bounded $h_r$, $\tilde h_a$.
Using this, the contribution to the second term in the exponent in \refb{eccd2} may
be expressed as:
\ben \label{eccd6}
 \exp\[ \int_{\delta}^\infty {dt\over 2t} \, Z_A(t)\]
&=& \exp\[ \int_{\delta}^\infty {dt\over 2t} \, Z_A''(t)\]
{\prod_r}' (h_r)^{-s_r/2} {\prod_a}' (\tilde h_a)^{-\tilde s_a/4}\non \\
&=& \exp\[ \int_{\delta}^\infty {dt\over 2t} \,  Z_A''(t) \]
\, N\, \int {\prod_r}' {d\xi_r} \, {\prod_a}' d\tilde\xi_a\,  e^{S_{NS}'+S_R'} \, ,
\een
where $S_{NS}'$ and $S_R'$ are respectively the open string field theory action
\refb{esNS} and \refb{eopenraction} restricted to the primed sector and
the normalization factor $N$ is chosen such that the intergration measure is
$d\xi/\sqrt{2\pi}$ for Grassmann even modes and $d\xi$ for Grassmann odd modes, irrespective
of whether the mode arises from the NS or the R sector.
If all the states have positive $L_0$ eigenvalue, then the equality of the first and
the second line of \refb{eccd6} follows using the relation $G_0^2=L_0$ and the form of
the actions \refb{esNS} and \refb{eopenraction}.

If we now remove the deformation that puts shifted boundary conditions on the two
boundaries of the annulus, we get back the zero modes on the instanton corresponding to having some vanishing
$h_r$ and/or $\tilde h_a$ in the primed set. As a result, the integral in the exponent on
the left hand side of \refb{eccd6} diverges from the $t\to\infty$ end.
To deal with this, we use the right hand side
of \refb{eccd6} as the definition of the left hand side and then substitute this
and \refb{ebeginA} into \refb{eccd2} to compute $e^A$.
Some of the zero modes are physical, {\it e.g.} those
describing the collective coordinates of the instanton.
As mentioned earlier, for compactification on Calabi-Yau threefolds, these
include the four fermionic zero modes
associated with broken supersymmetry since the
instanton breaks four out of eight supersymmetries, and bosonic zero modes describing the
position of the instanton in space-time as well as possible (generalized) motion along $\CY$.
The procedure for integrating over these zero modes will be reviewed shortly.
However, there are also ghost zero modes in the NS sector corresponding to the states \refb{eghost}
which reflect the breakdown of Siegel gauge choice. These can be dealt with by using a
gauge invariant form of the path integral \cite{Sen:2021qdk}. As will become
clear soon, the last step can be avoided for orientifold compactifications in the case of
$O(1)$ instantons.

\subsection{Effect of the orientifold projection}
\label{subsec-effectorient}

We now consider an orientifold of this theory, obtained by taking a quotient of this theory by
the trasformation \oo\ introduced in \S\ref{subsec-orientifold}.
In this case the normalization factor \refb{epowerB} is replaced by
\be\label{epowerF}
e^A=\exp\[\int_0^\infty {dt\over 2t}  \(Z_A(t) + Z_M(t)\)\] ,
\ee
where $Z_A$ is the annulus contribution and $Z_M$ is the M\"{o}bius strip contribution.
The annulus contribution itself gets modified from the one appearing in \refb{epowerB}
in two ways. First of all, due to the orientifold projection we have an extra
factor of 1/2 in the expression for $Z_A$.
Second, once we have an orientifold, typically we need to also have a certain
set of space-filling D-branes to cancel tadpoles of massless closed string fields. Therefore,
we shall also have annuli with one end on the D-instanton and the other end on a space-filling
D-brane. $Z_A$ appearing in \refb{epowerF} includes all these contributions.

Following the logic leading to \refb{eccd2}, we again express this as
\be\label{eccd10}
e^A=\exp\[\int_0^\delta {dt\over 2t} \(Z_A(t) + Z_M(t)\) + \int_\delta^\infty {dt\over 2t}
\(Z_A(t) + Z_M(t)\)\] ,
\ee
for some small constant $\delta$. As before, the first term can be interpreted as due to
the emission and absorption of closed strings by the D-instanton or
exchange of closed strings between a D-instanton and a space-filling D-brane or
orientifold plane. There is however a small subtlety in that the Schwinger parameter
$\ell$ in the closed string channel is related to the parameter $t$ by the relation $\ell=1/t$
for the annulus, but by the relation $\ell = 1/(4t)$ for the M\"{o}bius
strip \cite{Polchinski:1998rq,Polchinski:1998rr}. Therefore, for the
annulus we have the operator \refb{eccd1} sandwiched between the D-brane boundary states,
whereas for the M\"{o}bius strip we have the operator
\be\label{eccd7}
c_0^- b_0^+ \int_{1/(4\delta)}^\infty \, d\ell \, e^{-\pi \ell (L_0+\bar L_0)}
= {1\over \pi}\, c_0^- b_0^+ e^{-\pi (L_0+\bar L_0)/(4\delta)}/ (L_0+\bar L_0) \, ,
\ee
sandwiched between the D-instanton boundary state $|I\rangle$ and the crosscap states.
In the following we shall denote by $|S\rangle$  the total
boundary state associated with all the space-filling D-branes and by  $|C\rangle$ the
total crosscap state associated with all the orientifold planes.
Individually both the M\"{o}bius strip contribution and the annulus contribution may
diverge due to the presence of states with $L_0+\bar L_0=0$ in
$|S\rangle$ and $|C\rangle$. Since $|S\rangle$ and $|C\rangle$
do not carry any momentum along the non-compact directions, these
divergences are not softened by momentum integrals of the form $\int d^4 k/k^2$ and
involve evaluation of $1/k^2$ at $k=0$. However, in a consistent theory where the tadpoles
cancel, the sum $|S\rangle+|C\rangle$ has no $L_0+\bar L_0=0$ states.
Therefore, we can organize \refb{eccd7} as
\be
c_0^- b_0^+ \int_{1/(4\delta)}^{1/\delta} \, d\ell \, e^{-\pi \ell (L_0+\bar L_0)} +
c_0^- b_0^+ \int_{1/\delta}^{\infty} \, d\ell \, e^{-\pi \ell (L_0+\bar L_0)}\, ,
\ee
and then combine the contribution of the
second term with that from \refb{eccd1} so that we get the matrix element of
\refb{eccd1} between the D-instanton boundary state $|I\rangle$ and $|S\rangle+|C\rangle$.
Thus, the first term on the right hand side of \refb{eccd10} may be written in a
manifestly divergence free form as:
\ben
&& \langle I| c_0^- b_0^+ \int_{1/\delta}^{\infty} \, d\ell \, e^{-\pi \ell (L_0+\bar L_0)}|I\rangle
+2 \, \langle I| c_0^- b_0^+ \int_{1/\delta}^{\infty} \, d\ell \, e^{-\pi \ell (L_0+\bar L_0)}(|S\rangle
+|C\rangle) \non\\
&& \hspace{1in} + 2 \, \langle I| c_0^- b_0^+ \int_{1/(4\delta)}^{1/\delta}
\, d\ell \, e^{-\pi \ell (L_0+\bar L_0)}|C\rangle\, .
\een
This is equivalent to putting a uniform upper
cut-off $1/\eps'$ on the $\ell$ integration and taking
the limit $\eps'\to 0$
after summing the contributions of the annulus and the M\"{o}bius strip. Since $t=1/\ell$ for
the annulus and $1/(4\ell)$ for the M\"{o}bius strip, this translates to the following replacement in \refb{eccd10}:
\be\label{eccd9}
\int_0^\delta {dt\over 2t} \(Z_A(t) + Z_M(t)\)
\quad \Rightarrow \quad \lim_{\eps'\to 0}
\[\int_{\eps'}^\delta {dt\over 2t} \, Z_A(t) + \int_{\eps'/4}^\delta {dt\over 2t} \, Z_M(t)\] .
\ee

We now turn to the second term on the right hand side of \refb{eccd10}.
We express $Z_M+Z_A$ as in \refb{egenZA}, with the only difference that
the sum on the right hand side runs only over those open string states that are
invariant under the orientifold projection and we include states with both ends on the
D-instanton as well as states with one end on the D-instanton and the other end on the
space-filling D-brane.
We now analyze the resulting expression exactly as in \refb{ebeginA}-\refb{eccd5}, arriving at
an expression similar to \refb{eccd6}.

As stated before, we shall assume that in the orientifold, the only zero modes
the D-instanton has are the four translation zero modes along the non-compact space-time
directions associated with NS sector states \refb{etranslationa}
and the two fermion zero modes associated with broken supersymmetry, described by the
R sector states \refb{efermionzero}.
As discussed below \refb{etranslationa},
if the translation zero modes
are invariant under orientifold transformation, the ghost zero modes \refb{eghost}
are odd and are projected out. This is related to the fact that before the orientifold projection
these ghosts arise out of an attempt to gauge fix the rigid $U(1)$ transformation on the D-instanton.
Since $U(1)$  breaks to $O(1)$ in the orientifold, we do not have the gauge fixing problem that we
encounter before the orientifold projection.
Therefore, \refb{eccd6} does not contain any ghost zero mode contribution and
does not need to be reexpressed in a gauge invariant form.
There is however one subtlety. Since $O(1)$ is a group of
two elements, 1 and $-1$, the gauge invariant form of the path integral should contain a
division by 2 representing the volume of the gauge group. This effect is not included in the
perturbatively gauge fixed form of the path integral given in \refb{eccd6} and must be
included as an extra factor of $1/2$ multiplying the expression for $e^A$. Thus we get
\be
\label{eccd21}
\begin{split}
e^A =&\, {1\over 2} \,\lim_{\eps'\to 0}\exp\bigg[\int_{\eps'}^\delta {dt\over 2t}
Z_A(t) + \int_{\eps'/4}^\delta {dt\over 2t} \, Z_M(t)
\bigg]
\\
&\, \times \exp\[ \int_{\delta}^\infty {dt\over 2t} \(Z_A''(t) +Z_M''(t)\) \]
\, N\, \int {\prod_r}' {d\xi_r} \, {\prod_a}' d\tilde\xi_a\,  e^{S_{NS}'+S_R'}\, .
\end{split}
\ee

Since the primed set includes the four bosonic zero modes $\xi^\mu$ from the NS sector
and the two fermionic zero modes $\chi^\alpha$ from the R sector,
they give a net contribution of 3 to the left hand side of
\refb{eccd4}. Therefore, to satisfy this condition, we must include some non-zero mode states in the primed set.
Since the final result is independent of how we choose them, let us choose $n$ modes with $L_0$ eigenvalues
$y_1,\cdots,y_n$ and weights $w_1,\cdots,w_n$,
with $w_i$ being given by $s_r$ for NS sector modes
and $\tilde s_a/2$ for R-sector modes, subject to the condition $\sum_{i=1}^n w_i =-3$.
Then according to the logic leading to \refb{e311new}, we have
\be
\begin{split}
Z_A'(t)+Z_M'(t) =&\,  3  + \sum_{i=1}^n w_i \, e^{-2\pi y_i t} ,
\\
Z_A''(t)+Z_M''(t) = &\,
Z_A(t)+Z_M(t) - 3 -  \sum_{i=1}^n w_i \, e^{-2\pi y_i t}\, .
\end{split}
\ee
Eq.\refb{eccd21} can now be written as
\ben\label{eccd23}
e^A &=& {1\over 2} \, \lim_{\eps'\to 0}\exp\bigg[\int_{\eps'}^\delta {dt\over 2t}
Z_A(t) + \int_{\eps'/4}^\delta {dt\over 2t} \, Z_M(t)
\bigg]
\\
& \times & \exp\[ \int_{\delta}^\infty {dt\over 2t}  \(Z_A(t)+Z_M(t) - 3
-\sum_{i=1}^n w_i \, e^{-2\pi y_i t}\) \]
\int \prod_{\mu=0}^3 {d\xi_\mu\over \sqrt{2\pi}} \prod_{\alpha=1}^2 d\chi_\alpha
\, \prod_{i=1}^n y_i^{-w_i/2}\, ,
\non
\een
where the last factor $\prod_{i=1}^n y_i^{-w_i/2}$ comes from explicitly
performing the integration over the non-zero modes in the primed set.
Even though we arrived at this result by taking the $y_i$'s to be among
the set of $L_0$ eigenvalues that appear in the spectrum of open string modes,
one can check, by taking a derivative of \refb{eccd23} with respect to $y_i$,
that the right hand side of \refb{eccd23} is independent of the choice of the
$y_i$'s as $\delta \to 0$, which we will eventually consider for the integral under consideration.
Therefore, we shall choose $y_i=h$ for some constant $h$. Furthermore, although the
integration over $t$ in the second line of \refb{eccd23} has no divergence from the
$t\to\infty$ end due to subtraction of the zero mode contribution reflected in the $-3$
term, we shall find it more convenient to put an explicit upper cut-off $1/\eps$ and take
$\eps\to 0$ limit at the end so that we can separate the different terms in the intermediate
steps. This gives, using $\sum_{i=1}^n w_i=-3$,
\be \label{epowerC}
e^A= {1\over 2} \int \prod_{\mu=0}^3 {d\xi_\mu\over \sqrt{2\pi}} \prod_{\alpha=1}^2 d\chi_\alpha\, \KKZ \, ,
\ee
where
\be\label{edefam}
\KKZ= h^{3/2} \lim_{\delta\to 0}\lim_{\eps\to 0}\lim_{\eps'\to 0}
\exp\[\int_{\eps'}^{1/\eps}  {dt\over 2t}\, Z_A+\int_{\eps'/4}^{1/\eps}  {dt\over 2t}\,Z_M
+3 \int_{\delta}^{1/\eps} {dt\over 2t} \(e^{-2\pi h t}-1  \) \] .
\ee
By splitting each of the integrals at some finite point in the middle one can verify that the term
in the exponent can be written as a sum of terms, each of which depends at most on one of
the small parameters $\eps,\eps'$ and $\delta$. Therefore, the order of limits in \refb{edefam}
is not important. Furthermore, since the last integral has no divergence from the $t\to 0$
end, we can set $\delta=0$.
Setting $h=1$ and $\delta=0$ in \refb{edefam}, we recover \refb{edefamintro}.

\subsection{Integration over the zero modes}

The integration over the zero modes appearing in \refb{epowerC} can be carried out as follows.
Bosonic zero modes can be related to the D-instanton position $\wt\xi^\mu$ by comparing (a)
the coupling of the bosonic zero modes $\xi^\mu$
and (b) an amplitude of closed strings carrying momentum $p$ to the expected coupling of $\wt\xi^\mu$ via
the $e^{ip.\wt\xi}$ factor. This gives \cite{Sen:2021tpp,Alexandrov:2021shf,Alexandrov:2021dyl}
\be
\label{e2.13}
\xi^\mu = \frac{\wt\xi^\mu}{\sqrt{2}\,\pi g_o}\, ,
\ee
where $g_o$ is the open string coupling constant on the D-instanton,
arising in \refb{e2.13} due to the fact that the coupling of an open string to a closed
string amplitude is proportional to $g_o$.
The open string coupling is related to
the real part $\cTR_\Gamma$
of the instanton action via \refb{etgo}. The integration over $\wt\xi^\mu$ is to be
done at the end, producing the momentum conserving delta function $(2\pi)^4
\delta^{(4)}(p)$ where $p$ is the total momentum carried by all the closed strings.
On the other hand, the integration over the fermion zero modes requires that in the
rest of the amplitude we insert a pair of $\chi^\alpha$ vertex operators. These are
given by $ce^{-\phi/2}S_\alpha$, each accompanied by a factor of $g_o$. If we denote by
$\AAA_{\alpha\beta}$ the rest of the amplitude with insertion of the open string vertex
operators $ce^{-\phi/2}S_\alpha$ and $ce^{-\phi/2}S_\beta$ in all possible ways,
then the full D-instanton induced amplitude
will be given by
\be \label{enorm}
{1\over 2!} \, e^{-\cT_\Gamma}\, (2\pi)^4 \delta^{(4)}(p)\,  (2\pi)^{-2} g_o^{-4}2^{-3} \pi^{-4}\,
g_o^2 \, \ve^{\gamma\delta}
\KKZ \, \AAA_{\gamma\delta}\, ,
\ee
Here $1/2!$ is a combinatorial factor that cancels the factor of 2 coming from $\ve^{\gamma\delta}A_{\gamma\delta}=2 A_{12}$,
the factor $(2\pi)^4 \delta^{(4)}(p)\,  (2\pi)^{-2} g_o^{-4}2^{-3} \pi^{-4}$
comes from $\xi^\mu$ integration via \refb{e2.13} and the factor of $1/2$ in
\refb{epowerC}, the factor $g_o^2 \, \ve^{\gamma\delta}$ comes from the integration over
the open string fermion zero modes, $K_0$ comes from the massive modes
as given in \refb{edefam}, and $\AAA_{\gamma\delta}$
comes from rest of the amplitude.

If we have a solvable world-sheet conformal field theory, {\it e.g.} an orbifold compactification,
then \refb{edefam} can be evaluated explicitly, leading to a
finite answer. For general compactification we cannot get such explicit results.
Nevertheless, the contributions to $Z_M$ and $Z_A$ can be organized using the
extended superconformal algebra underlying the world-sheet theory~\cite{Eguchi:1988vra,Odake:1988bh,Odake:1989dm,Odake:1989ev}.
This is discussed in appendix \ref{saE}, with the final formul\ae\ given in
\refb{emobfinal}, \refb{endfinal}, \refb{e43sum} and \refb{emobclosed}.
This is useful for proving general results {\it e.g.} vanishing
of $Z_A$ with the D-instanton boundary conditions at both boundaries, and the equality of
the D-instanton partition function and threshold corrections to gauge coupling on a space-filing D-brane, required
for proving holomorphy of the superpotential, as discussed
in appendix \ref{sF}.
This may also be useful if the theory under consideration
is a deformation of a solvable theory and one attempts to compute $Z_A$ and $Z_M$ using
perturbation expansion in the deformation parameter.

\section{Instanton induced amplitudes and effective action} \label{s1a}

In this section we shall use the results of \S\ref{s1}
to compute some amplitudes which will eventually lead to the determination of the holomorphic
superpotential defining the effective action in four dimensions.

Let us denote by $\Phi_m$ the chiral superfields whose scalar components represent the K\"ahler moduli of $\CY$.
We choose a background value $\phi^{(0)}_{m}$ of $\Phi_m$ and denote
by $\phi_m$ and $\xxi_m^\alpha$
the bosonic and fermionic fluctuations around this background, as introduced in \refb{einvfield}.
Due to the reality condition \refb{ereality}, $\xxi_m^\dalpha$ introduced in
\refb{einvfield} is related to the hermitian conjugate of $\xxi_m^\alpha$.
Also note that \refb{einvfield} only defines $\hat\phi_m=(\phi_m+\phi_m^*)/\sqrt 2$,
the imaginary part of $\phi_m$ being a field in the RR sector.
The background $\phi^{(0)}_{m}$ will be taken to be real so that it represents an NSNS background.
We shall denote by $\kappa_4 $ the four-dimensional gravitational coupling and
by $g^{(s)}_{\mu\nu}$ the four-dimensional string metric.
Covariantizing the kinetic terms of string field theory found in
\refb{ensns4} and \refb{ec.26}, which are written in the background
$g^{(s)}_{\mu\nu}=\eta_{\mu\nu}$, we can express the kinetic terms of the various fields as:
\be\label{estringaction}
\int d^4 x \sqrt{-  g^{(s)}} \[ {1\over 2\, \kappa_4 ^2} \,\cR^{(s)}
-{1\over 2}\,g^{(s)\mu\nu} \p_\mu\hat\phi_{m} \p_\nu\hat\phi_m
 - 2\,\bar {\xxi}_m E^{(s)\mu}_a \gamma^a \p_\mu {\xxi}_m\] ,
\ee
where $E^{(s)\mu}_a$ is the inverse vierbein.

Our goal will be to compute the leading instanton contribution
to the amplitude involving moduli fields $\hat\phi_{m_3},\cdots,\hat\phi_{m_n}$
labelling the real parts of the
scalar components of chiral superfields and a pair of chiral fermions $\xxi_{m_1}^\alpha$,
$\xxi_{m_2}^\beta$, and use this to determine the instanton correction to the effective action.
The relevant contribution comes from the product of $k$ disk / upper half plane amplitudes, of which $(k-2)$ disks
carry the vertex operators of the scalars $\hat\phi_{m_3},
\cdots \hat\phi_{m_k}$ and each of the remaining two disks
carries the vertex operator of a closed string fermion $\xxi_{m_1}^\alpha$ or
$\xxi_{m_2}^\beta$ and an open string fermion zero mode.

The disk one-point function of the corresponding vertex
operator of $\hat\phi_m$,  given in \refb{evertexor}, takes
the form:
\be\label{e117a}
{1\over 2} \,\kappa_4 \cT^R_\Gamma\, {1\over \sqrt 2}\, \langle c_0^- (V^{m}(i)+U^{m}(i))\rangle
=  -{\p \cT_\Gamma\over \p \hat\phi_m} \, , \qquad c_0^-\equiv {1\over 2} (c_0-\bar c_0)\, .
\ee
The left hand side follows from the rules for computing disk one-point functions
in the normalization convention of \cite{Alexandrov:2021shf},
and the right hand side follows from the standard
rule that the disk one-point function of a field is given by the derivative of the
instanton action with respect to that field.

On the other hand, the disk two-point function of the field $\xxi_m^\alpha$,
whose vertex operator is also
given in \refb{evertexor}, and the fermion zero mode described by the vertex operator
$c e^{-\phi/2}S_\gamma$, is given by \cite{Alexandrov:2021shf}:
\be
\label{e121a.1}
\I\pi\kappa_4\, \cT^R_\Gamma {1\over \sqrt 2} \langle c \, e^{-\phi/2} S_\gamma(0)
\( V_\alpha^{m} (i)
+ U^{\prime m}_{\alpha}(i)\) \rangle\, .
\ee
To evaluate this amplitude, we first analyze the correlation function
$\left\langle c e^{-\phi/2} S_\gamma (z) V_\alpha^{m} (i)\right\rangle$
in the upper half plane. Due to \refb{efermuvert}, we can write
\be\label{eainitial}
\left\langle c e^{-\phi/2} S_\gamma (z) V_\alpha^{m} (i)\right\rangle
= \left\langle c e^{-\phi/2} S_\gamma (z) c\bar c e^{-\phi/2}e^{-\bar\phi}
W_\alpha^{m} (i)\right\rangle .
\ee
Using $SL(2,\IR)$ symmetry of the upper half plane, and
the fact that both vertex operators are conformally invariant, one can show that the
result does not depend on the locations of the vertex operators as long as $z$ lies on the
real axis. However, we can now use holomorphy of the operator $c e^{-\phi/2} S_\gamma (z)$
to extend the result to the case where $z$ lies anywhere in the upper half plane.
Therefore, we can
evaluate this correlation function by taking the $z\to i$ limit.
Using the operator product expansion \refb{e1.9a}
we can write:
\be \label{ea.4}
\begin{split}
 \left\langle c e^{-\phi/2} S_\gamma (z) V_\alpha^{m} (i)\right\rangle
=&\, \ve_{\gamma\alpha} \left\langle \p c \, c\, \bar c \, e^{-\phi}e^{-\bar\phi}
W^m(i) \right\rangle
\\
=&\, {1\over 2}\,
\ve_{\gamma\alpha} \left\langle (\p c-\bar\p \bar c) \, c\, \bar c \, e^{-\phi}e^{-\bar\phi}
W^m(i) \right\rangle=
\ve_{\gamma\alpha} \left\langle c_0^- V^m(i) \right\rangle .
\end{split}
\ee
In arriving at the second equality we have used the replacement rule $\bar c(\bar z) \to c(z)$ and the explicit
expression for the correlation functions of the $c$'s in the upper half plane.

Next we shall analyze the correlation function
$\left\langle c e^{-\phi/2} S_\gamma (z) U^{\prime m}_{\alpha} (i) \right\rangle$ in the
upper half plane. Using the boundary conditions~\refb{eortr} on the real axis allows us to replace $c e^{-\phi/2} S_\gamma (z)$
by $-\bar c e^{-\bar\phi/2} \bar S_\gamma (\bar z)$, so that we can express the correlation function as
$-\left\langle \bar c \bar e^{-\bar\phi/2} \bar S_\gamma (\bar z)U^{\prime m}_{\alpha} (i) \right\rangle$.
We can now take the $\bz\to -i$ limit as before
and use \refb{efermustarvert}, \refb{e1.10a}  to write
\be
\label{ea.7}
\begin{split}
 \left\langle c e^{-\phi/2} S_\gamma (z) U^{\prime m}_\alpha (i)\right\rangle
=&\, -\ve_{\gamma\alpha}  \left\langle \bar\p \bar c (i) \, c(\I) \bar c(\I)\, e^{-\phi}e^{-\bar\phi}\, Y^{m}(i)
\right\rangle
\\
= &\, {\ve_{\gamma\alpha}\over 2}   \left\langle (\p c-\bar\p \bar c) \, c \, \bar c\,
e^{-\phi}e^{-\bar\phi}\, Y^{m}(i)
\right\rangle = \ve_{\gamma\alpha} \left\langle c_0^-U^{m}(i)
\right\rangle .
\end{split}
\ee

Using \refb{ea.4}, \refb{ea.7} and \refb{e117a}, we can express \refb{e121a.1} as
\be\label{e121a}
\I\, \pi\kappa_4\, \cT^R_\Gamma \, \ve_{\gamma\alpha}
{1\over \sqrt 2}  \langle c_0^- (V^{m}(i)+U^{m}(i))\rangle
=
-2\, \I\, \pi
\ve_{\gamma\alpha} {\p \cT_\Gamma\over \p \hat\phi_m}\, .
\ee
Combining \refb{e117a} and \refb{e121a} with \refb{enorm}, we get the leading
contribution to an amplitude for
a pair of fermions $\rho_{m_1}^\alpha, \rho_{m_2}^\beta$ and $(k-2)$ scalars
$\hat\phi_{m_3},\cdots,\hat\phi_{m_k}$
\be\label{e449}
{1\over 2!} \, e^{-\cT_\Gamma} (2\pi)^4 \delta^{(4)}(p)\,
(2\pi)^{-2}\, 2^{-3}\pi^{-4} g_o^{-2} \,
\KKZ\, \ve^{\gamma\delta} (\ve_{\gamma\alpha}\ve_{\delta\beta}-
\ve_{\gamma\beta}\ve_{\delta\alpha}) (-2i\pi)^2
\prod_{i=1}^k \left(-{\p \cT_\Gamma\over \p \hat\phi_{m_i}} \right) .
\ee
Note that we have summed over the
possibility of exchanging the zero mode vertex operators
$c e^{-\phi/2}S_\gamma$ and $c e^{-\phi/2}S_\delta$ on the two disks keeping the
closed string vertex operators fixed. The result \refb{e449} translates to the following contribution
to the effective action
\be \label{einteraction}
-{2^{-3} \pi^{-4} \over (k-2)! \, 2!} \int d^4 x \sqrt{-  g^{(s)}}e^{-\cT_\Gamma}\,
\ve_{\alpha\beta} \, \xxi_{m_1}^\alpha \xxi_{m_2}^\beta
\, g_o^{-2}  \KKZ \prod_{j=3}^k \hat\phi_{m_j}
\prod_{i=1}^k \left(-{\p \cT_\Gamma\over \p \hat\phi_{m_i}}
\right) .
\ee

We now define the Einstein metric using the relation
\be
g^{(s)}_{\mu\nu}= \kappa_4^2\, g^{(E)}_{\mu\nu}.
\ee
In order to canonically normalize the kinetic terms, we define
\be
\hat\phi_m = \kappa_4 ^{-1} \tilde\phi_m, \qquad \xxi_m^\alpha = \kappa_4 ^{-3/2} \tilde\xxi_m^\alpha.
\ee
Then the  kinetic terms take the form
\be
\int d^4 x \sqrt{-  g^{(E)}} \[{1\over 2}\,R^{(E)}
- {1\over 2}\, g^{(E)\mu\nu}\p_\mu\tilde\phi_m \p_\nu\tilde\phi_m
- 2 \,\bar {\tilde\xxi}_m E^{(E)\mu}_a \gamma^a\p_\mu \tilde\xxi_m\].
\label{renorm-lagrang}
\ee
In these variables the interaction term \refb{einteraction} becomes
\be \label{einteractiontwo}
-{2^{-3} \pi^{-4} \over (k-2)! \ 2!}
\int d^4 x \sqrt{-  g^{(E)}}\, e^{-\cT_\Gamma} \, \ve_{\alpha\beta} \,
\tilde\xxi_{m_1}^\alpha\, \tilde\xxi_{m_2}^\beta\,
\kappa_4^3 \, g_o^{-2}  \, \KKZ\, \prod_{j=3}^k \tilde\phi_{m_j}
\prod_{i=1}^k \left(-{\p \cT_\Gamma\over \p \tilde\phi_{m_i}}
\right) .
\ee
In the next section we shall translate this result to a formula for the instanton generated holomorphic
superpotential.

\section{Supergravity analysis} \label{ssugra}

Eq.\refb{einteractiontwo} gives the result for the leading instanton induced
scattering amplitude of a pair of fermions and
arbitrary number of scalars. In this section we shall compute the same amplitude
in an
$\NN=1$ supergravity theory coupled to matter with a superpotential. The comparison
of the two results will then be used
to determine the explicit form of the D-instanton induced superpotential.

\subsection{General structure} \label{s4.1}

We take the following convention for $\mathcal{N}=1$ supergravity action \cite{Freedman:2012zz}
\begin{equation}
\cL\supset \frac{1}{2}\,\mathcal{R}
-\cK_{I\bJ}\left(\partial \vp^I\partial\bvp^{\bJ}+\overline{\Chi^J} \slashed{D}\Chi^{I}\right)
-V_F+\cL_m,
\label{SUGRA-lagr}
\end{equation}
where $\vp^I$ are complex scalars, $\Chi^I$ are complex left-handed Weyl fermions,
$\cK$ is the K\"ahler potential
and we define
\begin{subequations}
\begin{equation}
\cK_{I\bJ}\equiv\partial_{I}\partial_{\bJ}\cK,
\end{equation}
\begin{equation}
V_F=e^{\cK}\(\cK^{I\bJ}\nabla_I W\bnabla_{\bJ}\bW-3 |W|^2\),
\end{equation}
\be
\mathcal{L}_m\supset -\frac{1}{2}\, e^{\cK/2}(\nabla_I\nabla_J W)\,
\ve_{\alpha\beta} {\Chi}^{I\alpha}\Chi^{J\beta}+h.c.\, .
\label{fm-term}
\ee
\end{subequations}
Here $W$ is a holomorphic function of the moduli known as superpotential and
\be
\nabla_I W=\p_I W+(\p_I\cK) W.
\ee
Comparing the Einstein-Hilbert terms in \refb{renorm-lagrang} and \refb{SUGRA-lagr}, we see
that the metric in \refb{SUGRA-lagr} has been normalized in the same way as the Einstein
metric.

Typically the massless chiral multiplet fields
of $\NN=1$ supersymmetric type II string compactifications include, besides closed
string moduli, the open string moduli, {\it e.g.} the locations of space-filling D-branes inside
the Calabi-Yau space,
and other massless open string fields {\it e.g.} those living at
the intersections of space-filling D-branes. The dependence on the open string moduli is
already captured in $K_0$ appearing in \refb{einteractiontwo}, since the annuli with
one end on the instanton and the other end on a space-filling D-brane will depend on the moduli
of space-filling D-branes. The dependence on other chiral multiplet open string
fields may be obtained by
inserting their vertex operators in an amplitude, as discussed
in \cite{Akerblom:2007uc,Cvetic:2007ku,Blumenhagen:2009qh}. Since our main aim is
to describe how to fix the overall normalization of the superpotential, we shall ignore the
dependence on the open string fields in the subsequent analysis, working at their fixed values.

Even though our results are valid for general compactification, we list below as an example the specific closed string
moduli that arise in type IIB string compactification on Calabi-Yau threefolds with O7-D7-O3-D3
systems. For such compactifications the complex scalar fields $\vp^I$ comprise \cite{Grimm:2004uq,Grimm:2005fa}
\begin{itemize}
\item
the complex structure moduli $z^k$, $k=1,\dots, h_{2,1}^-$;

\item
the axio-dilaton
\be\label{eaxio}
\tau=c^0+\I e^{-\phi};
\ee

\item
moduli arising from RR and NS 2-forms ($a=1,\dots h_{1,1}^-$)
\be
G^a=c^a-\tau b^a;
\ee

\item
complexified K\"ahler moduli ($\alpha=1,\dots h_{1,1}^+$)\footnote{Comparing to \cite{Grimm:2004uq},
we renormalized $T_\alpha$ by the factor $\frac23$.}
\be\label{eTalpha}
T_\alpha=\frac{e^{-\phi}}{2}\, \kappa_{\alpha\beta\gamma}t^\beta t^\gamma +\I \tc_\alpha- \zeta_\alpha(\tau,G),
\qquad
\zeta_\alpha=-\frac{\I}{2(\tau-\btau)}\, \kappa_{\alpha bc}G^b(G-\bG)^c,
\ee
\end{itemize}
where $\kappa_{\alpha\beta\gamma},\kappa_{\alpha bc}$ encode triple intersection numbers of 4-cycles on $\CY$,
$t^\alpha$ is the volume of 2-cycles $\gamma_\alpha\in H_2^+(\CY,\IZ)$.
and $h^\pm_{p,q}$ denote the number of harmonic $(p,q)$ forms that are even / odd
under the orientifold action.

At the classical level, the K\"ahler potential is given by
\be \label{ekahlerpot}
\cK=\cKsk+\cKq,
\ee
where
\begin{subequations}
\bea \label{eKkahler}
\cKsk(z)&=&-\log\(\I \int \Omega\wedge\overline{\Omega}\),
\\
\cKq(\tau,G,T) &=&-2\log(V^{(E)})-\log\(-\I(\tau-\bar{\tau})\) ,
\label{qKkahler}
\eea
\end{subequations}
with $\Omega$ being the holomorphic 3-form on $\CY$,
and we introduced the Einstein-frame volume
\be
V^{(E)}=e^{-3\phi/2} V,
\qquad
V=\frac{1}{6}\, \kappa_{\alpha\beta\gamma}t^\alpha t^\beta t^\gamma.
\ee
In this coordinate system the D-instanton being analyzed here represents
an Euclidean D3-brane wrapped on a 4-cycle of $\CY$ and its action depends only on the coordinates
$T_\alpha$. On the other hand, the scalar vertex operators given in the first equation of
\refb{evertexor} describe fluctuations of the real parts of $i T_\alpha/\tau$ and $G^a/\tau$
since they correspond to marginal deformations of the CFT describing the
Calabi-Yau compactification.

\subsection{The superpotential}

The general contribution $W_\Gamma$ to the superpotential from an instanton of charge
$\Gamma$ is expected to have the structure
\be\label{esuperpot}
W_\Gamma =\cA_\Gamma(\vp)\, e^{-\cT_\Gamma(\vp)},
\ee
where $-\cT_\Gamma(\vp)$ is the classical instanton action and $\cA_\Gamma(\vp)$ is
the correction term that we are trying to determine.
According to \eqref{fm-term}, this superpotential gives rise to a fermion mass term in the effective action.
Below we compare this term with \refb{einteractiontwo} with $k=2$ to determine
the prefactor $\cA_\Gamma(\vp)$.

We have
\be\label{eWone}
\nabla_I W_\Gamma
=- \p_I \cT_\Gamma \, W_\Gamma + \p_I\cA_\Gamma \, e^{-\cT_\Gamma}
+ \p_I\cK \, W_\Gamma\, .
\ee
This relation is valid in any complex coordinate system in the moduli space.
Let us choose a coordinate system such that when we set the RR fields to zero, only
one of the moduli depends on the string coupling whose expectation value
has an additive term proportional to
$\ln g_s$, and the remaining moduli are $g_s$ independent functions of the K\"ahler and complex
structure moduli of $\CY$. For instance, in the example of \S\ref{s4.1} we can take the moduli
to be $\ln\tau$,  $z^k$, $T_\alpha/\tau$ and $\zeta_\alpha/\tau$.
Then $\cT_\Gamma(\vp)$ contains an explicit factor of $1/g_s$.
Therefore, in the weak coupling limit the $\p_I\cT_\Gamma(\vp)$
term in \refb{eWone}  has an extra factor of
$1/g_s$ compared to the other two terms and will dominate, giving
\be\label{eWtwo}
\nabla_I W_\Gamma\simeq - \p_I \cT_\Gamma \, W_\Gamma,
\qquad
\nabla_I\nabla_J W_\Gamma\simeq \p_I \cT_\Gamma \, \p_J \cT_\Gamma \, W_\Gamma\, .
\ee
Using this, we can express the leading contribution to \refb{SUGRA-lagr} as:
\be
\begin{split}
\cL \supset&\, \frac{1}{2}\, \mathcal{R}-\cK_{I\bJ}\left(\partial \vp^I\partial\bvp^{\bJ}+
\overline{\Chi^J} \slashed{D}\Chi^I\right)-
e^{\cK}\(\cK^{I\bJ}\p_I \cT_\Gamma\bar\p_{\bJ}\overline \cT_\Gamma-3\)|W_\Gamma|^2
\\
&\,  -\[\frac{1}{2}\, e^{\cK/2}(\p_I \cT_\Gamma)\, (\p_J \cT_\Gamma)\, W_\Gamma
\, \ve_{\alpha\beta} {\Chi}^{I\alpha}\Chi^{J\beta}+h.c\] .
\end{split}
\label{SUGRA-lagrtwo}
\ee
Note that we have kept only the $W_\Gamma$
term in the
superpotential that arises from the instanton of charge $\Gamma$.

Let us introduce the matrix $R$ such that $(R^\dagger R)_{I\bJ}=\cK_{I\bJ}(\vp^{(0)})$ where
$\vp^{(0)}$ denotes the background fields around which we carry out our computation.
Then relabelling
$R\vp$ and $\vp$ and $R\Chi$ as $\Chi$, we can express \refb{SUGRA-lagrtwo} as
\be
\begin{split}
\cL \supset&\, \frac{1}{2}\, \mathcal{R}-\delta_{I\bJ}\left(\partial \vp^I\partial\bvp^{\bJ}+\overline{\Chi^J}
\slashed{D}\Chi^I\right)-
e^{\cK}\(\delta^{I\bJ}\p_I \cT_\Gamma \, \bar\p_{\bJ} \bar\cT_\Gamma -3\)|W_\Gamma|^2
\\
&\,  - \[\frac{1}{2}\, e^{\cK/2} \, \p_I \cT_\Gamma\,\p_J \cT_\Gamma\, W_\Gamma\,
\ve_{\alpha\beta} {\Chi}^{I\alpha}\Chi^{J\beta}+h.c\],
\end{split}
\label{SUGRA-lagrthree}
\ee
up to higher order terms involving Taylor expansion of $\cK_{I\bJ}(\vp)$ around $\vp^{(0)}$
which will not be relevant for our analysis.

Since the form of \refb{SUGRA-lagrthree} remains unchanged under a unitary rotation of the
coordinates $\vp^I$, we
shall use this freedom to ensure that the fields $\tilde \phi_m$ appearing
in \refb{renorm-lagrang}, \refb{einteractiontwo} can be identified, up to normalization, as
the real part of
one set of chiral fields $\vp^m$
appearing in \refb{SUGRA-lagrthree}. The corresponding fermionic partners
$\tilde\rho_m$ appearing in \refb{renorm-lagrang}, \refb{einteractiontwo}
can then be identified, up to a normalization,
with superpartners $\Chi^m$ of $\vp^m$ appearing in
\refb{SUGRA-lagrthree}.  To determine the relative normalization factors, we
match the kinetic terms for fermions and scalars in \eqref{renorm-lagrang}
with \eqref{SUGRA-lagrthree}.
This requires the identification
\be \label{eferid}
\tilde\rho_m={1\over \sqrt 2}\, e^{i\xi/2} \, \Chi^{m}, \qquad \tilde\phi_{m}={1\over \sqrt 2} \,
\(\vp^{m} + (\vp^m)^*\).
\ee
where $e^{i\xi/2}$ is an arbitrary phase that
does not affect the fermion kinetic terms. The phase $\xi$ could depend on the moduli since
at the leading order any term that may arise from derivative of $\xi$ will be subdominant,
the dominant term coming from the same derivative acting on $\cT_\Gamma$.

We now focus on the part of the action containing the
mass term for the fermion fields $\Chi^m$ in \refb{SUGRA-lagrthree}:
\be\label{predict-sugra}
-\frac{1}{2}\, \int d^4 x\, \sqrt{-  g}\, e^{\cK/2}\, (\p_m \cT_\Gamma) (\p_n \cT_\Gamma) \,
W_\Gamma  \, \ve_{\alpha\beta} \Chi^{m\alpha}\Chi^{n\beta}\, .
\ee
The hermitian conjugate term will come from anti-D-instanton and will not be
considered here.
On the other hand, using \refb{eferid} and the fact that $\cT_\Gamma$ is a holomorphic function
of the $\vp^m$'s, we get
\be\label{erelation}
\ve_{\alpha\beta} \tilde\rho_{m_1}^\alpha \tilde\rho_{m_2}^\beta
\prod_{i=1}^2 \left.\left(-{\p \cT_\Gamma\over \p \tilde\phi_{m_i}}\right) \right|_{\phi^{(0)}}
= {1\over 4} \, \(-{\p\cT_\Gamma\over \p \vp^{m_1}}\) \(-{\p\cT_\Gamma\over \p \vp^{m_2}}\)
\ve_{\alpha\beta} \Chi^{m_1\alpha} \Chi^{m_2\beta}\, e^{i\xi}\, .
\ee
Using this, \refb{einteractiontwo} for $k=2$ may be expressed as:
\be \label{einteractionthree}
-{1 \over 16\pi^4} \int d^4 x \sqrt{-  g^{(E)}}e^{-\cT_\Gamma}\,
\ve_{\alpha\beta} \Chi^{m_1\alpha} \Chi^{m_2\beta}\,
{\kappa_4^3 \, e^{i\xi}\over 4 g_o^{2}} \, \KKZ
\prod_{i=1}^2 \left.\left(- {\p \cT_\Gamma \over \p \vp^{m_i}}
\right) \right|_{\phi^{(0)}}.
\ee
Comparing \refb{predict-sugra} and \refb{einteractionthree} and identifying
$g_{\mu\nu}$ with $g^{(E)}_{\mu\nu}$, we get
\be\label{e3.43}
W_\Gamma  =\frac{\kappa_4^3\, e^{i\xi}}{32\pi^{4}g_o^2}
\, K_0\, e^{-\cK/2}\, e^{-\cT_\Gamma(\vp)}\, .
\ee
Finally, using the relation $\Re(\cT_\Gamma)=(2\pi^2 g_o^2)^{-1}$, we recover \refb{emain}.

Even though in the analysis in this section we have used the two-point function of
the superpartners of the K\"ahler moduli to fix the normalization of the superpotential, we could
also use the two-point function of the superpartner of the four-dimensional dilaton to
fix the same normalization. The analysis will proceed in an identical manner with the vertex
operators for the K\"ahler moduli and their fermionic partners replaced by those of the
four-dimensional dilaton and its fermionic partner. At the end we shall get back the same
normalization since the same superpotential controls these different contributions to the
amplitude. A manifestation of this is already reflected in the fact that the final form
of the superpotential \refb{e3.43} does not depend on which K\"ahler modulus and its
superpartner we use for our computation.

\subsection{Holomorphy of the superpotential}

While \refb{e3.43} gives a concrete, calculable expression for $W_\Gamma$ at the leading
order in the string coupling, it is
not immediately obvious that it is a holomorphic function of the moduli. We shall now
show that this is indeed the case. Our starting point will be a result given in
\cite{Akerblom:2007uc} (see {\it e.g.} eq.(6.11))
\be\label{e3n.44}
\lim_{\eps'\to 0}\[\int_{\eps'}^{\infty} {dt\over 2t} \,  Z_A \, e^{-\mu^2 t} + \int_{\eps'/4}^{\infty} {dt\over 2t} \,  Z_M
\, e^{-\mu^2 t} \]
= -8\pi^2 \Re(f^{(1)}) +{3\over 2}\, \log{M_p^2\over \mu^2} +{1\over 2}\, \cK
-\log {V_\Gamma\over g_s}\, ,
\ee
in the $\mu\to 0^+$ limit.  Here $f^{(1)}$ is a holomorphic function of the moduli, $M_p$ is
the four-dimensional Planck mass, $V_\Gamma$ is the volume of the cycle $\Gamma\in H_4(\CY,\IZ)$
measured in the string metric and $g_s$ is the ten-dimensional string coupling.
Before we go on, let us review the origin and some of the features of this equation:
\begin{enumerate}

\item
To the best of our knowledge, \refb{e3n.44} has not been proved from first principles. The way
this equation was arrived at was by arguing that the same expression appears in the result for
the threshold correction to the gauge coupling on a space-filling brane \cite{Lust:2003ky,Berg:2004ek},
given by the one-loop correction to the gauge kinetic term,
and then using some general properties of the threshold correction \cite{Kaplunovsky:1993rd}.
This correspondence with the threshold correction
can be traced to the fact that the left hand side of \refb{e3n.44} may be regarded as a one-loop
correction to the D-instanton action. If we now regard the D-instanton as a regular instanton on
a space-filling brane obtained
by changing the Dirichlet boundary condition to Neumann boundary condition along the
non-compact directions, then the D-instanton action should be controlled by the gauge
coupling on this space-filling brane.  Therefore, the one-loop correction to the D-instanton action
should be related to the one-loop correction to the gauge coupling on the space-filling D-brane.
This correspondence was also explicitly
checked for some cases in \cite{Abel:2006yk,Akerblom:2006hx}.
We shall give a detailed derivation
of this correspondence in appendix \ref{sF} (see \refb{e3n.44thrnew}).

\item
The parameter $\mu$ serves as an infrared cut-off and makes the integral finite. The exact way of
regulating the divergence is often not stated in the literature. For example, the regularization
used here is the same one used in \cite{Berg:2004ek},
while \cite{Antoniadis:1999ge} used an upper cut-off $\mu^2$ on the
$t$ integral. In the $\mu\to 0$ limit different regulators give results that differ by a numerical constant
which can be absorbed into the definition of $f^{(1)}$.
Since the goal of this subsection is to prove the holomorphy of $W_\Gamma$, having already
obtained an unambiguous expression in \refb{e3.43}, we can make any choice of the regulator.
We shall use the one given in \refb{e3n.44} since this will make direct connection with
the expression for $K_0$ given in \refb{edefam}.

\item
To resolve all ambiguities on the right hand side of \refb{e3n.44}, we need to make a specific choice of
$\cK$, which is defined up to a K\"ahler transformation $\cK\to \cK+h(\vp) + \bar h(\bar\vp)$.
This can be absorbed into a shift of $f^{(1)}$ by $h(\vp)/(8\pi^2)$ and reflects
the usual ambiguity in the definition of the superpotential.
We choose $\cK$ to be whatever choice we have made in arriving at \refb{e3.43}.

\item
We shall fix all further ambiguities in the definitions of various quantities on the right hand
side by taking
\be\label{e424}
M_p=\kappa_4^{-1},
\qquad
{V_\Gamma\over g_s} = \cT^R_\Gamma(\vp) = {1\over 2\pi^2 g_o^2}\, .
\ee
With this \refb{e3n.44} can be viewed as a definition of $f^{(1)}$. The non-trivial content of
\refb{e3n.44} is that $f^{(1)}$ so defined is a holomorphic function.

\item
Since the left hand side of this equation is independent of the string coupling
$g_s=\langle e^\phi\rangle = \langle \tau_2^{-1}\rangle$, the right hand side should
also be independent of $g_s$.
There are however various terms on the right hand side that depend on $g_s$. In particular,
since $\kappa_4\propto g_s$ we get $-3\log g_s$ from the term with $\log M_p^2$ and
$\log g_s$ from the last term. The form of $\cK$ depends on the specific compactification
we use, but the $g_s$-dependence is expected to be the same for all compactifications.
For the class of compactifications described in \S\ref{s4.1}, $\cK$ has $3\log g_s$ from the first
term in \refb{qKkahler} and $\log g_s$ from the last term.
Putting these results together, we see that the $g_s$-dependent terms indeed all cancel.

\item
One-loop quantum effects are known to correct the definition of the
moduli fields to $\vp^I\to\vp^I+ v^I(\vp,\bar\vp)$ where $v^I$ have a power of the string coupling, see for example \cite{Haack:2018ufg}.
Due to this redefinition, we have $e^{-\cT_\Gamma(\vp)}\to e^{-\cT_\Gamma(\vp)} e^{-\p_I\cT_\Gamma(\vp) v^I(\vp,\bar\vp)}$.
Since the threshold correction to the gauge coupling is also affected by this redefinition, the
right hand side of \refb{e3n.44} is understood to include an
additive term $-\p_I\cT_\Gamma(\vp) v^I(\vp,\bar\vp)$ to compensate for this effect.
Since this term is independent of $g_s$, it does not
affect the $g_s$-independence of the right hand side of \refb{e3n.44}.
\end{enumerate}
Taking into account these effects, we now write a more accurate version of \refb{e3n.44}
\be
\label{e3n.44new}
\begin{split}
\lim_{\eps'\to 0} &\[ \int_{\eps'}^{\infty} {dt\over 2t} \,  Z_A \, e^{-\mu^2 t}
+ \int_{\eps'/4}^{\infty} {dt\over 2t} \,  Z_M
\, e^{-\mu^2 t}\]
\\
& = -8\pi^2 \Re(f^{(1)}) +{3\over 2}\, \log{1\over \kappa_4^2\mu^2} +{1\over 2}\, \cK
+\log (2\pi^2 g_o^2)-\p_I \cT_\Gamma(\vp) v^I(\vp,\bar\vp)\, .
\end{split}
\ee

We shall now use \refb{e3n.44new} to evaluate $K_0$ given in \refb{edefam}.
First, we note that we could have put an upper cut-off $1/\eps$ on the integrals over $t$ on
the left hand side of \refb{e3n.44new}. This allows us to express the left hand side of
\refb{e3n.44new} as:
\be\label{einter11}
\lim_{\delta\to 0}\lim_{\eps\to 0} \lim_{\eps'\to 0}\[ \int_{\eps'}^{1/\eps} {dt\over 2t} \,  Z_A \, e^{-\mu^2 t}
+ \int_{\eps'/4}^{1/\eps} {dt\over 2t} \,  Z_M
\, e^{-\mu^2 t} -3\int_{\delta}^{1/\eps} {dt\over 2t}   e^{-\mu^2 t}
+ 3\int_{\delta}^{1/\eps} {dt\over 2t}   e^{-\mu^2 t} \],
\ee
where we have added and subtracted the same quantity.
Noting that $Z_A+Z_M-3$ receives contributions only from massive open string
states which are exponentially suppressed for large $t$,
we can drop the factor $e^{-\mu^2 t}$ from the first three terms.
Therefore, \refb{einter11} can be replaced by
\be\label{einter12}
\lim_{\delta\to 0}\lim_{\eps\to 0} \lim_{\eps'\to 0}\[ \int_{\eps'}^{1/\eps} {dt\over 2t} \,  Z_A
+ \int_{\eps'/4}^{1/\eps} {dt\over 2t} \,  Z_M
+ 3\int_{\delta}^{1/\eps} {dt\over 2t}   \(e^{-\mu^2 t}-1\) \] \, ,
\ee
in the $\mu\to 0$ limit. We now compare this with \refb{edefam}.
Since \refb{edefam} is independent of $h$, we can take $2\pi h=\mu^{2}$. With this \refb{einter12}
may be identified as $\log \((2\pi)^{3/2}K_0/\mu^3\)$, and \refb{e3n.44new} now gives
\be
\log K_0
= -8\pi^2 \Re(f^{(1)}) -{3} \log{\kappa_4} +\frac{1}{2}\,\cK
 + \log (2\pi^2 g_o^{2}) -{3\over 2}\, \log(2\pi) - \p_I\cT_\Gamma(\vp) v^I(\vp,\bar\vp) .
\ee
Notice that the $\mu$-dependence on both sides cancelled.
We now substitute this into \refb{e3.43} and choose $\xi=-8\pi^2 \Im(f^{(1)})$. This gives
\be
\begin{split}
W_\Gamma =&\,  e^{-8\pi^2 f^{(1)}(\vp)} 2^{-11/2} \pi^{-7/2} e^{- \p_I\cT_\Gamma(\vp) v^I(\vp,\bar\vp)}e^{-\cT_\Gamma(\vp)}
\\
=&\, 2^{-11/2} \pi^{-7/2}  e^{-8\pi^2 f^{(1)}(\vp)}
e^{-\cT_\Gamma(\vp + v(\vp,\bar\vp))}\, .
\end{split}
\ee
Since $\vp^I + v^I(\vp,\bar\vp)$ give the quantum corrected holomorphic coordinates in
the moduli space, we see that $W_\Gamma$ is a holomorphic function of the moduli.

When additional zero modes are present from open strings at the intersection of the D-instanton
and space-filling D-branes, there are additional non-holomorphic terms on the right hand side of
\refb{e3n.44new}. However, it was shown in \cite{Akerblom:2007uc}
that these terms are still consistent with the holomorphy of the superpotential,
which now involves additional chiral multiplets living on the space-filling D-branes.

\section{Multi-instanton contribution} \label{smulti}

We shall now compute  the overall normalization of the
$k$-instanton amplitude. As in the case of single instanton amplitudes, we shall carry out
our analysis in the absence of background fluxes. Since the single instanton effect already
generates a superpotential and one needs to switch on fluxes to find an extremum, one might
question the usefulness of the multi-instanton results in the absence of fluxes.
Nevertheless, as long as the effects of fluxes are small, our results may be interpreted as
the analysis of the leading order contribution to the superpotential from multi-instanton effects.
We shall argue that this contribution actually vanishes for the multi-instantons whose
single-instanton constituents only have the two universal fermionic zero modes
that we have been considering throughout this paper.

Formally, the normalization factor is given by the exponential of the annulus amplitude for open
strings living on the configuration of $k$ identical instantons.
Our analysis will closely follow section 4 of \cite{Alexandrov:2021shf}, the main difference being that in the zero
mode sector we only have a subset of the degrees of freedom that were present in
\cite{Alexandrov:2021shf}.

The system of $k$ identical $O(1)$ instantons has an underlying $O(k)$
gauge symmetry and it is useful to classify the zero modes by their transformation properties
under $O(k)$.
This is done as follows:
\begin{enumerate}
\item
In the NS sector we have four bosonic zero modes $\xi^\mu_A$ in the symmetric rank 2 tensor
representation of $O(k)$, arising from a symmetric Chan-Paton matrix multiplying the
states \refb{etranslationa}. Here the index $A$ labels the gauge index and runs over
$k(k+1)/2$ values. Of these the trace part, denoted by $A=0$, gives the collective mode
associated with space-time translations, while the other values of $A$ label the elements of
a symmetric traceless rank two tensor.

\item
In the NS sector we have a pair of Grassmann odd zero modes $p_a,q_a$
in the anti-symmetric rank two tensor
representation (adjoint representation) of $O(k)$, arising out of antisymmetric Chan-Paton
matrix multiplying \refb{eghost}. The gauge index $a$ runs over $k(k-1)/2$ values.

\item
In the R-sector we have two fermionic zero modes  $\chi_A^\alpha$ in the symmetric rank two tensor
representation of $O(k)$, with $\alpha$ labelling the undotted spinor index. These arise from
symmetric Chan-Paton matrix multiplying \refb{efermionzero}.
The trace part $\chi_0^\alpha$ describes the fermionic collective coordinates associated with broken
supersymmetry up to a normalization, while the rest belongs to the symmetric traceless
rank two tensor representation of $O(k)$.

\item
In the R-sector we have two fermionic zero modes  $\chi_a^\dalpha$ in the
adjoint representation of $O(k)$, with $\dalpha$ labelling the dotted spinor index.
These arise from anti-symmetric Chan-Paton matrix multiplying the states of the
form \refb{efermionzero}, with $\alpha$ replaced by $\dalpha$.
\end{enumerate}

Let us first count the $(-1)^F$ weighted
number of zero modes in the NS and R sector since this will determine how to analyze the non-zero
mode contribution. In the NS sector the number of $(-1)^F$ weighted zero modes is
$2k(k+1)- k(k-1)=k^2+3k$. On the other hand, in the R-sector all the zero modes are fermionic
and the number of $(-1)^F$ weighted zero modes is $-k(k+1)-k(k-1)=-2k^2$.
Adding the NS sector contribution to half of the R-sector contribution, we get $3k$.
Following the discussion leading to \refb{edefam}, and using the $h$ independence of the
result to set $h=1$, we now see that the non-zero mode contributions may be
expressed as:
\be\label{edefammulti}
\KKZ^{(k)}=
\exp\[\int {dt\over 2t} \,\Bigl( Z_A(t) + Z_M(t) + 3k\( e^{-2\pi  t} -1\) \Bigr) \] ,
\ee
with appropriately chosen limits of integration as in \refb{edefam}.
We remark that \eqref{edefammulti} is independent of $g_s,$ since
the open string spectrum is independent of $g_s$ and all the divergences can be
properly subtracted by introducing cut-offs as in \refb{edefam} without introducing $g_s$ dependence.
Compared to the case of single instantons,  the annulus contribution with both
boundaries on the instanton is multiplied by
$k^2$ due to the trace over the Chan-Paton factor on its two boundaries, while the
M\"{o}bius strip contribution and the annulus contribution with one boundary on the
D-instanton and the other boundary on a space-filling D-brane
are multiplied by $k$.

We now turn to the contribution due to the zero modes. First, we consider the Grassmann
odd zero modes $p_a,q_a$ in the NS sector. These modes represent the breakdown of Siegel
gauge choice and are dealt with by using the gauge invariant formulation \cite{Sen:2021qdk}.
This introduces an integration over a set of out of Siegel gauge bosonic modes $\phi_a$ multiplying the
states $c_0 \beta_{-1/2}|-1\rangle$, with action $-\sum_a\phi_a^2/4$, and division by the
volume of the gauge group labeled by the parameters $\theta_a$.
Therefore, the integration over $p_a,q_a$ is replaced by
\be\label{e4.2}
\int \prod_{a=1}^{k(k-1)/2} d\phi_a \, e^{-\phi_a^2/4} \bigg/ \int \prod_{a=1}^{k(k-1)/2}D\theta_a\, .
\ee
The integral over $\phi_a$ gives the result $(2\sqrt\pi)^{k(k-1)/2}$, while the integral over
$\theta_a$ can be found by comparing the string field theory gauge transformation parameters
with the rigid $O(k)$ transformation parameters $\wt\theta^a$.
At the leading order in the expansion in powers of $\kappa_4$, the relation takes the
form $\theta_b=2\wt\theta_b/g_o$ \cite{Sen:2021qdk}. Therefore, we can express
\refb{e4.2} as:
\be\label{e4.3}
(\sqrt{\pi}\, g_o)^{k(k-1)/2} / (2\, V_{SO(k)})\, ,
\ee
where $V_{SO(k)}$ denotes the volume of the $SO(k)$ group in appropriate normalization and the
factor of 2 accompanying $V_{SO(k)}$
takes into account the contribution from the $\IZ_2$ subgroup of $O(k)$ that is
outside $SO(k)$.

Next we turn to the integration over the zero modes $\xi^\mu_A$. Of these $\xi^\mu_0$
is related to the center of mass coordinate $\wt\xi^\mu$ of the instanton system by an equation
analogous to \refb{e2.13}:
\be
\xi^\mu_0 = \frac{\sqrt k}{\sqrt{2}\, \pi g_o}\, \wt \xi^\mu\, ,
\ee
where the extra factor of $\sqrt k$ can be traced to the fact that the Chan-Paton
factor accompanying the correctly normalized
vertex operator for $\xi_0^\mu$ is given by
$1/\sqrt k$ times the identity matrix \cite{Sen:2021jbr}.
The rest of the $\xi^\mu_A$'s are redefined as
\be
x^\mu_A= g_o^{1/2} \, \xi^\mu_A\, , \qquad \hbox{for $1\le A\le k(k+1)/2-1$},
\ee
so that the action written in terms of $x^\mu_A$ does not have any factor of the
string coupling \cite{Sen:2021jbr,Alexandrov:2021shf}. This will be seen explicitly
in \refb{e14}.

For the fermion zero modes, we first consider the zero modes $\chi_0^\alpha$ that are
related to the collective mode associated with supersymmetry breaking. The redefinition that makes the
vertex operators of the collective modes independent of the string coupling
is \cite{Alexandrov:2021shf}:
\be
\tilde\chi^\alpha = g_o \, \chi^\alpha_0\, .
\ee
Finally, the fermion zero modes $ \chi_A^\alpha$ and $\chi_a^\dalpha$ for $A\ge 1$ are
redefined as
\be \label{e4.7}
y_A^\alpha = g_o^{1/4} \, \chi_A^\alpha, \qquad y_a^\dalpha = g_o^{1/4} \, \chi_a^\dalpha,
\ee
so that the action expressed in terms of the zero modes $x^\mu_A,y_A^\alpha,y_a^\dalpha$
does not have any explicit dependence on the string
coupling \cite{Sen:2021jbr,Alexandrov:2021shf}. In fact, by defining
\be
X^\mu = x^\mu_A \, L^A ,
\qquad
Y^\alpha = y_A^\alpha\, L^A , \qquad Y^\dalpha = y_a^\dalpha\, T^a \, ,
\label{Tvar}
\ee
where $L^A$'s give a basis of real, symmetric Chan-Paton factors and $T^a$'s
give a basis of imaginary, anti-symmetric Chan-Paton factors, one can write the action for the zero modes as
\be \label{e14}
S = {1\over 8}\, \Tr\bigl( [X_\mu, X_\nu] [X^\mu,X^\nu]\bigr) + {1\over \sqrt 2}\, \gamma^\mu_{\alpha\dbeta}
\Tr\bigl(Y^\alpha[X_\mu, Y^\dbeta]\bigr)\, .
\ee
This has the same form as described in \cite{Alexandrov:2021shf} that came from dimensional
reduction of $\NN=1$ supersymmetric Yang-Mills theory from 4 to 0 dimensions,
except that there
$X^\mu,Y^\alpha$ and $Y^\dbeta$ were all given by arbitrary linear combination of the generators of $SU(k)$.

Using \refb{e4.3}-\refb{e14}, we can now represent the exponential of the annulus
amplitude as:
\be
\begin{split}
&{1\over 2}\,(\sqrt{\pi}\, g_o)^{k(k-1)/2}
\(\frac{\sqrt k}{\sqrt{2}\, \pi g_o}\)^4 \, g_o^2\, (g_o^{-1/2})^{2k (k+1)-4}
\\
& \hskip 1in \times
(g_o^{1/4})^{k(k+1)-2 + k(k-1)} \,\KKZ^{(k)}\,
M_k \, \int\prod_{\mu=0}^3 \frac{d\wt\xi^\mu}{\sqrt{2\pi}} \prod_{\delta=1}^{2} d\tilde\chi^\delta,
\end{split}
\ee
where
\be \label{e6.11nk}
M_k ={1\over V_{SO(k)}} \int \prod_{A=1}^{k(k+1)/2 -1} \left\{\prod_{\mu=0}^3 \frac{dx^\mu_A}{\sqrt{2\pi}} \right\}
\left\{\prod_{\delta=1}^2 dy^\delta_A \right\} \int\prod_{a=1}^{k(k-1)/2}
\prod_{\ddelta=1}^2 dy^\ddelta_a\ e^S\, .
\ee
This result has to be combined with the product of the disk amplitudes as in \refb{enorm}.
Each disk gives an extra factor of $k$ from the trace over the Chan-Paton factors and the
Chan-Paton factor associated with the fermion zero mode $\tilde\chi^\alpha$ has a factor
of $1/\sqrt k$. Therefore, the product of two disk amplitudes with insertion of a pair
of fermion zero modes gives a net factor of $k$, replacing $A_{\alpha\beta}$ in
\refb{enorm} by $k A_{\alpha\beta}$. The integration over the $\wt\xi^\mu$'s produce the
momentum conserving delta functions as usual.
Thus, the net amplitude is given by:
\be \label{e4.12}
{1\over 2!}\, e^{-k\TT(\phi^{(0)})}\, (2\pi)^4 \delta^{(4)}(p)\,  2^{-5} \pi^{k(k-1)/4 -6} k^3
g_o^{-(3k+1)/2}
\, \KKZ^{(k)}\,M_k   \,  \AAA_{\gamma\delta}.
\ee

We shall now argue that $M_k$ vanishes. The point is that each term in the action
\refb{e14} has equal number of dotted and undotted spinors. Therefore, when we expand $e^S$
in powers of the second term in \refb{e14}, we have equal powers of dotted and undotted spinors in each term
in the expansion. However, the integration over the zero modes is saturated only when there
are $k(k+1)/2-1$ undotted spinors and $k(k-1)/2$ dotted spinors in the expansion. Therefore, the
result vanishes unless $k=1$.
This is in agreement with the fact that the amplitude is consistent with the holomorphy of the
superpotential only when the power of $g_o$ is $-2$, corresponding to $k=1$, in which
case we get \refb{einteractiontwo}. This however leaves
open the possibility that higher order contributions could affect the superpotential. In particular,
the counting of powers of $g_o$ indicates that the $k$-instanton amplitude can receive
contributions at order $g_o^{3(k-1)/2}$ above the leading contribution given by the product of
disk amplitudes.\footnote{In this argument we have assumed that all the $x^\mu_A$'s in
the integral \refb{e6.11nk} are of the same order, in which case the action
\refb{e14} forces them to be of order
unity. \refb{e14} also has flat directions that correspond to large commuting $X^\mu$'s.
Physically this represents the freedom of separating the instantons in space-time. In
terms of effective field theory, the contribution to an amplitude from this region of
integration corresponds to using the instanton induced superpotential
term in the action multiple times in a Feynman diagram instead of correction to the
superpotential itself.}

We can analyze this possibility as follows. In order to saturate the zero mode integrals, we need
to insert the excess $y^\alpha$ zero mode vertex operators in the amplitude. There are
$2(k-1)$ of these modes, taking into account the fact that $\alpha$ takes two values. Since $\chi^\alpha$
are canonically normalized, each $\chi^\alpha$ vertex operator
carries a factor of $g_o$. Since \refb{e4.7} gives $dy_A^\alpha = g_o^{-1/4} \, d\chi_A^\alpha$,
we see that each extra $y^\alpha_A$ integration will be accompanied by a factor of
$g_o^{3/4}$. Therefore, we get a net factor of $g_o^{3(k-1)/2}$ which can combine with the
$g_o^{-(3k+1)/2}$ factor in \refb{e4.12} to give $g_o^{-2}$. Therefore, the net power of $g_o$
we get is independent of $k$ and the same argument as in \S\ref{ssugra} will give a superpotential
independent of $g_o$.

There is however a further consistency check that we need to perform. We recall from \S\ref{sworld} that each
$S_\alpha$ that appears in the $-1/2$ picture vertex operator carries a $J$ charge of $3/2$.
Therefore, we have a net excess $J$ charge of $3(k-1)$. Since $J+\bar J$ charge is conserved
by the boundary condition \refb{espininst}, we must compensate for this excess $J$ charge.
To compute the superpotential of bulk fields we need to insert a number of scalars
from the chiral multiplet, each of which carries vanishing $J+\bJ$ charge, and a pair of
chiral fermions, each carrying $J+\bJ$ charge $-3/2$ according to the results given in
\S\ref{sworld}. Therefore, we see that the closed string vertex operators carry net $J+\bJ$ charge $-3$.
This compensates for the 3 units of $(J+\bJ)$ charge carried by the zero mode vertex
operators associated with broken supersymmetry and therefore does not help in cancelling the
excess $3(k-1)$ units of $J+\bJ$ charge. The only other
source of excess $J$ charge are the picture changing operators (PCO). Each PCO has a term
$e^\phi T_F$ where $T_F=-\psi_\mu \p X^\mu+
T_F^++T_F^-$ is the matter super-stress tensor current and $T_F^\pm$
carry $J$ charge $\pm 1$. Since we have $2(k-1)$ extra open string vertex operators in $-1/2$
picture, we need $(k-1)$ PCOs to conserve the picture number. Even if we pick $T_F^-$ from
each of these PCOs, we can at most get a net $J$ charge of $-(k-1)$. Since this cannot
compensate the $3(k-1)$ units of excess
$J$ charge carried by the additional zero mode vertex operators, the amplitude
must vanish. At higher genus we need more PCOs which could compensate the $J$ charge,
but this will involve additional power of string coupling. In particular, to compensate for the
$2(k-1)$ unit of excess $J$ charge we need to insert at least $2(k-1)$ PCOs which will take us
to genus $(k-1)/2$, producing an extra factor of $g_s^{k-1}\sim g_o^{2(k-1)}$. This
will be inconsistent with holomorphy of the superpotential.
Thus, we conclude that there are no multi-instanton contributions to the superpotential.

In the above analysis we have assumed that we are working on a fixed world-sheet,
consisting of the exponential of the annulus or M\"obius strip and a product of disks, with each disk
containing one closed string vertex operator and some open string vertex operators. But
we also have the possibility of adding more disks containing only open string vertex
operators which will take into account the effect of additional terms in the
effective action \refb{e14}. Since the counting of the powers of $g_o$ associated with the
vertex operators remains unchanged, the net effect will be to generate additional powers of
$g_o^{-2}$ --- one for every additional disk. Therefore, one could ask whether these negative
powers of $g_o$ could compensate for the the extra powers of $g_o$ that we saw in the
previous paragraph. To this end, note that on the disk the total picture number of all the operators
must add up to $-2$. Therefore, if we have $r$ extra disks, then the total number of required
PCOs will change from $(k-1)$ to $(k-1-2r)$ in the previous paragraph.  On the other hand, the
minimal number of PCOs required for $J$-charge conservation remains fixed at $3(k-1)$.
Thus, we now have to increase the genus by $(k-1+r)/2$ to get the extra $2(k-1+r)$
PCOs. This gives an additional factor of $g_o^{2(k-1+r)}$. From this we see that the extra
factors of $g_o^{-2}$, that we had gotten for each extra disk, are precisely cancelled by the
extra factor of $g_o^{2r}$ that we get from imposing $J$-charge conservation and picture
number conservation. Therefore, having extra disks does not help us generate a multi-instanton
contribution to the superpotential.

It is easy to see that the argument based on conservation of $J$-charge
generalizes in a straightforward fashion even to the cases
where we have different types of instantons instead of identical instantons. In particular, for
a  total of $k$ instantons we always have $2(k-1)$ extra Grassmann integrals over undotted
spinor zero modes. To saturate these integrals, we need to insert $2(k-1)$ vertex operators
of the form $ce^{-\phi/2} S_\alpha$ into the amplitude. We can then use  $J$-charge conservation
to argue that any contribution to the amplitude from products of disks and
annuli vanishes.

One must keep in mind however that our argument for the vanishing contribution to the superpotential does not apply to systems in which the fermion zero mode structure on
the individual instantons is different from that of the $O(1)$ instantons
considered here. In \cite{Garcia-Etxebarria:2007fvo} the authors considered a
combination of $U(1)$ instanton and $O(1)$ instanton and argued that this system can
give non-vanishing contribution to the superpotential. The $U(1)$ instanton has equal
number of chiral and anti-chiral fermion zero modes and therefore in the combined
system the difference between chiral and anti-chiral fermion zero modes remains fixed at
two, allowing it to contribute to the superpotential. 
Refs. \cite{Blumenhagen:2008ji,Blumenhagen:2012kz} also provided examples
of multi-instanton systems which could contribute to the superpotential. In this
system many of the constituent instantons had non-universal fermion zero modes for
which our computation of the $J$-charge does not hold.

\section{Conclusion} \label{conclusion}

In this paper we have described a systematic procedure for computing the D-instanton induced
superpotential in an orientifold of type IIB/IIA string theory compactified on a Calabi-Yau
threefold, preserving $\NN=1$ supersymmetry in four dimensions. Our analysis includes
all Euclidean D-branes preserving half of the space-time supersymmetry. This problem has been
addressed by many authors in the past, and our main contribution has been to fix the overall
normalization of the correction terms by drawing insights from string field theory. Our final
results are quoted in \refb{emain}, \refb{edefamintro}.

We have focussed on instantons whose only zero modes are the four bosonic zero modes
associated with translation of the instanton along non-compact  directions and the
fermionic zero modes associated with the supersymmetries that are
broken on the D-instanton world-volume. More general instantons are possible, and while in
principle our methods could be extended to these cases as well, we have not addressed
those in this paper.

\bigskip

\noindent{\bf Acknowledgement}: We wish to thank Harold Erbin for collaboration at the
initial stages of this work. We thank  Daniel Harlow, Hong Liu, Liam McAllister, Jakob Moritz,
Richard Nally, Washington Taylor, and Barton Zwiebach for discussions.
A.S. is supported by ICTS-Infosys Madhava Chair Professorship
and the J. C. Bose fellowship of the Department of Science and Technology,
India. M.K. is supported by the Pappalardo Fellowship.
B.S. acknowledges funding support from The Science and Technology Facilities Council through
a Consolidated Grant ‘Theoretical Particle Physics at City, University of London’ ST/T000716/1.
For the purpose of open access, the authors have applied a Creative Commons Attribution (CC BY)
licence to any Author Accepted Manuscript version arising.
This material is based upon work supported by the U.S. Department of Energy, Office of Science,
Office of High Energy Physics of U.S. Department of Energy under grant Contract Number DE-SC0012567.

\appendix

\section{Normalization of the fermion kinetic term} \label{eappc}

In this appendix we shall find the normalization of the fermion kinetic terms obtained from
string field theory. We begin by listing some additional operator product expansions in the
world-sheet theory beyond those listed in \S\ref{sworld}:
\begin{subequations} \label{ec}
\be\label{ec.1}
\begin{split}
e^{-\phi}\psi^\mu(z) \, e^{-\phi/2} e^{-\bar\phi} W_\alpha^{m}(w,\bw)
=&\, {i\over 2}\, (\gamma^\mu)_{\alpha}^{~\dbeta} (z-w)^{-1}
e^{-3\phi/2} e^{-\bar\phi} \, \wt W_\dbeta^{m}(w,\bw) +\cdots,
\\
e^{-\bar\phi}\bar\psi^\mu(\bz) \, e^{-\phi} e^{-\bar\phi/2} \, W^{\prime m}_\dalpha(w,\bw)
=&\, {i\over 2}\, (\gamma^\mu)_{\dalpha}^{~\beta} (\bz-\bw)^{-1}e^{-\phi} e^{-3\bar\phi/2} \,
\wt W^{\prime m}_\beta(w,\bw)+\cdots,
\\
e^{-\phi}\psi^\mu(z) \, e^{-\phi/2} e^{-\bar\phi} Y_\dalpha^{m}(w,\bw)
=&\, {i\over 2}\, (\gamma^\mu)_{\dalpha}^{~\beta} (z-w)^{-1}
e^{-3\phi/2} e^{-\bar\phi} \, \wt Y_\beta^{m}(w,\bw)+\cdots,
\\
e^{-\bar\phi}\bar\psi^\mu(\bz) \, e^{-\phi} e^{-\bar\phi/2} \, Y^{\prime m}_\alpha(w,\bw)
=&\, {i\over 2}\, (\gamma^\mu)_{\alpha}^{~\dbeta} (\bz-\bw)^{-1}e^{-\phi} e^{-3\bar\phi/2} \,
\wt Y^{\prime m}_\dbeta(w,\bw)+\cdots,
\end{split}
\ee
\be \label{ec.2}
\begin{split}
e^{-\phi/2} S_\dgamma (z) e^{-\phi/2}e^{-\bar\phi}
W_\alpha^{m} (w,\bw) = &\, \OO(1)\, ,
\\
e^{-\bar\phi/2}\bar S_\alpha (\bar z) e^{-\phi} e^{-\bar\phi/2} W^{\prime m}_\dbeta(w,\bw) =&\,\OO(1),
\\
e^{-\phi/2}S_\alpha (z) e^{-\phi/2}e^{-\bar\phi}Y^{m}_\dbeta(w,\bar w) =&\, \OO(1),
\\
e^{-\bar\phi/2}\bar S_\dalpha (\bar z) e^{-\phi}e^{-\bar\phi/2}Y^{\prime m}_{\beta}(w,\bar w)
=&\, \OO(1)\, ,
\end{split}
\ee
\be \label{ec.2.5}
\begin{split}
e^\phi T_F^\pm (z) \, \, c\, \bar c\, e^{-3\phi/2} e^{-\bar\phi} \wt W^m_\dalpha(w,\bw)
=&\, \OO(z-w),
\\
e^{\bar \phi} \bar T_F^\pm (\bz) \, \,
c\, \bar c\, e^{-\phi} e^{-3\bar\phi/2}  \wt W^{\prime m}_\alpha(w,\bw)
=&\,\OO(\bz-\bw),
\\
e^\phi T_F^\pm (z) \, \, c\, \bar c\, e^{-3\phi/2} e^{-\bar\phi} \wt Y^m_\alpha(w,\bw) =&\, \OO(z-w),
\\
e^{\bar \phi} \bar T_F^\pm (\bz) \, \,
c\, \bar c\, e^{-\phi} e^{-3\bar\phi/2} \wt Y^{\prime m}_\dalpha(w,\bw) =&\, \OO(\bz-\bw),
\end{split}
\ee
\be\label{ec.3}
\begin{split}
e^{-3\phi/2} e^{-\bar\phi} \, \wt W_\dalpha^{m}(z,\bz) \ e^{-\phi/2} e^{-\bar\phi} Y_\dbeta^n(w,\bw)
= &\, K\, |z-w|^{-4} \ e^{-2\phi} e^{-2\bar\phi} \, \ve_{\dalpha\dbeta}\, \delta_{mn} +\cdots,
\\
e^{-3\phi/2} e^{-\bar\phi} \, \wt Y_\alpha^{m}(z,\bz) \ e^{-\phi/2} e^{-\bar\phi} W_\beta^n(w,\bw)
= &\, K\, |z-w|^{-4} \ e^{-2\phi} e^{-2\bar\phi} \, \ve_{\alpha\beta}\, \delta_{mn}+\cdots,
\\
e^{-\phi} e^{-3\bar\phi/2} \, \wt W^{\prime m}_\alpha(z,\bz) \ e^{-\phi} e^{-\bar\phi/2}
Y^{\prime n}_\beta(w,\bw)
=&\, K\, |z-w|^{-4} \ e^{-2\phi} e^{-2\bar\phi} \, \ve_{\alpha\beta}\, \delta_{mn}+\cdots,
\\
e^{-\phi} e^{-3\bar\phi/2} \, \wt Y^{\prime m}_\dalpha(z,\bz) \ e^{-\phi} e^{-\bar\phi/2}
W^{\prime n}_\dbeta(w,\bw)
=&\, K\, |z-w|^{-4} \ e^{-2\phi} e^{-2\bar\phi} \, \ve_{\dalpha\dbeta}\, \delta_{mn}+\cdots,
\end{split}
\ee
\end{subequations}
where $K$ is a constant to be determined and $T_F^\pm$, $\bar T_F^\pm$ are the
superconformal generators of the $(2,2)$ superconformal algebra associated with
Calabi-Yau compactification, carrying $(\bJ,J)$ charges $(0,\pm 1)$ and $(\pm 1,0)$
respectively.
The operators $\wt W$ and $\wt Y$ appearing in \refb{ec.1} differ from the corresponding
operators $W$ and $Y$ appearing on the left hand side of the equations only in the chirality
of the four-dimensional spin fields, but the parts coming from the CFT associated with the
Calabi-Yau threefold remain unchanged and continue to represent the Ramond sector ground
states. \refb{ec.2} follows from the observation that the net $J$ ($\bar J$) charge of the
operators on the left hand side is given by $\pm 2$ and therefore the matter sector
contribution to the
$L_0$ ($\bar L_0$) eigenvalue of any operator on the right hand side is bounded from
below by $|J|/2=1$ ($|\bar J|/2=1$) \cite{Lerche:1989uy}. \refb{ec.2.5} follows from the fact that
the zero modes $G_0^\pm$ of $T_F^\pm$ ($\bar G_0^\pm$ of $\bar T_F^\pm$)
annihilate the Ramond ground states of the
CFT associated with the Calabi-Yau threefold \cite{Lerche:1989uy}.

To determine the constant $K$ appearing in \refb{ec.3}, we begin with the worldsheet
correlator:
\be
\label{ec.4}
\begin{split}
I =&\, \bigg\langle
\ointop_{z_1} {dz} \(e^{-\phi/2} S_\gamma(z)\)  \(c\, e^{-\phi/2} S_\ddelta(z_1)\)
\(\bar\p \bc c\bc\, e^{-\phi/2} e^{-\bar\phi}\, W^{m}_\alpha(z_2,\bz_2)\)
\\
& \hskip 0.5in
\times \(c\bc\, e^{-\phi/2} e^{-\bar\phi} \,Y^n_\dbeta(z_3,\bz_3)\)
\bigg\rangle \, ,
\end{split}
\ee
where $\ointop$ is normalized to include the $1/(2\pi i)$ factor so that
$\ointop dz/z=1$.
Using \refb{espinlast}, we get
\be\label{ec.5}
I = - i\, (\gamma_\mu)_{\gamma\ddelta}
\left\langle
 c \,e^{-\phi} \psi^\mu(z_1) \,
\bar\p \bc c\bc\, e^{-\phi/2} e^{-\bar\phi}\, W^{m}_\alpha(z_2,\bz_2)\,
c\bc\, e^{-\phi/2} e^{-\bar\phi}\, Y^n_\dbeta(z_3,\bz_3)
\right\rangle.
\ee
Using the normalization of the closed string vacuum given in \refb{eclosednorm}
and \refb{ec.1}, \refb{ec.3}, we can express \refb{ec.5} as
\be\label{ec.7a}
I= i\, (\gamma_\mu)_{\gamma\ddelta} \,  (2\pi)^4 \, \delta^{(4)}(0)\
{i\over 2} \gamma^\mu_{\alpha\dbeta} \, K\,  \delta_{mn}
=   (2\pi)^4 \, \delta^{(4)}(0)\, K\, \ve_{\gamma\alpha} \ve_{\ddelta\dbeta}\, \delta_{mn},
\ee
where in the last step we used
\be
(\gamma^\mu)_{\gamma\ddelta} (\gamma_\mu)_{\alpha\dbeta}
= -2\, \ve_{\gamma\alpha} \ve_{\ddelta\dbeta}\, .
\ee
On the other hand, deforming the $z$ integration contour in \refb{ec.4} away from $z_1$ to pick
the residues from $z_2$ and $z_3$ using \refb{e1.9a} and \refb{ec.2}, we get
\be \label{ec.8}
I=\ve_{\gamma\alpha} \left\langle c e^{-\phi/2} S_\ddelta(z_1)\,
\bar\p \bc c\bc \,e^{-\phi} e^{-\bar\phi}\, W^{m}(z_2,\bz_2)\,
c\bc\, e^{-\phi/2} e^{-\bar\phi}\, Y^n_\dbeta(z_3,\bz_3)
\right\rangle .
\ee
Using \refb{e1.10a} and \refb{e1.8}, we can express this as
\be\label{ec.9}
I=- (2\pi)^4 \, \delta^{(4)}(0)\ \ve_{\gamma\alpha} \ve_{\ddelta\dbeta}\, \delta_{mn}.
\ee
Comparing \refb{ec.7a} with \refb{ec.9}, we get
\be\label{ec.12}
K =-1\, .
\ee

We shall now determine the normalization of the fermion kinetic term in closed string field theory.
Since the kinetic term couples NSR sector to NSR sector and RNS sector to RNS sector,
we can analyze the two sectors independently. First, let us consider the NSR sector.
The massless string fields in Siegel gauge take the form:
\be\label{ephiRrep}
|\phi_{NSR}\rangle = \int{d^{4}p\over (2\pi)^{4}}\, \[
\psi_m^\alpha(p) \, c\, \bar c\, e^{-\phi/2} e^{-\bar\phi} W^m_\alpha
+ \cfi_m^\dbeta(p) \, c\, \bar c\, e^{-\phi/2} e^{-\bar\phi} Y^m_\dbeta\] e^{\I p.X}(0)|0\rangle \, ,
\ee
in the $(-1,-1/2)$ picture, and
\be \label{efg1rep}
|\wt\phi_{NSR}\rangle = \int{d^{4}p\over (2\pi)^{4}}\, \[
\wt\psi_m^\dalpha(p) \, c\, \bar c\, e^{-3\phi/2} e^{-\bar\phi} \wt W^m_\dalpha
+ \wt \cfi_m^\beta(p) \, c\, \bar c\, e^{-3\phi/2} e^{-\bar\phi} \wt Y^m_\beta\] e^{\I p.X}(0)|0\rangle \, ,
\ee
in the $(-1,-3/2)$ picture. Note that the fields $\psi_m^\alpha$, $\cfi_m^\dbeta$,
$\wt\psi_m^\dalpha$ and $\wt \cfi_m^\beta$ are Grassmann odd and so are the vertex
operators multiplying them. Furthermore, all states appearing in \refb{ephiRrep} and
\refb{efg1rep} are GSO even as required.
The quadratic part of the string field theory action, given in the normalization
used in \cite{Alexandrov:2021dyl}, is
\be\label{ensraction}
S_{NSR}=4\, \[ -{1\over 2}\, \langle \tilde \phi_{NSR} | c_0^- \, \XX_0\,
(Q_B+\bar Q_B)\, |\tilde \phi_{NSR}\rangle
+ \langle \tilde \phi_{NSR} | c_0^- \, (Q_B+\bar Q_B)\, |\phi_{NSR}\rangle \],
\ee
where $c_0^-=(c_0-\bc_0)/2$,
$Q_B,\bar Q_B$ are the holomorphic and anti-holomorphic BRST operators, and
$\XX_0,\bar \XX_0$ are the zero modes of the holomorphic and anti-holomorphic
picture changing operators.
We have
\begin{subequations}
\be \label{ebrs1}
Q_B = \ointop dz \, \jmath_B(z) \, ,
\ee
\be \label{ebrstcurrent}
\jmath_B =c \, \bigl(T_{m} + T_{\beta,\gamma} \bigr)+ \gamma \, T_F
+ b\, c\, \p c
-{1\over 4}\, \gamma^2 \, b\, ,
\ee
\be \label{epicture}
\XX(z) = 2\, \{Q_B, \xi(z)\} =2\,  c \, \partial \xi +
2\, e^\phi T_F - {1\over 2}\, \p \eta \, e^{2\phi} \, b
- {1\over 2}\, \p\left(\eta \, e^{2\phi} \, b\right),
\ee
\be
\XX_0 = \ointop z^{-1} dz \, \XX(z) \, .
\ee
\end{subequations}
Here $T_F=-\psi_\mu\p X^\mu + T_F^++T_F^-$.
Similar expressions exist in the anti-holomorphic sector.

Using the operator products given in \refb{e1.9a}, \refb{e1.10a}, \refb{ec}, and \refb{ec.12},
the action \refb{ensraction} evaluates to
\be\label{ec.21}
S_{NSR} =  \int{d^{4}p\over (2\pi)^{4}} \[-{1\over 2}\,
p^2  \wt\psi_m^\dalpha(-p)\! \sp_{\dalpha\beta} \wt \cfi_m^\beta(p)
- p^2 \ve_{\dalpha\dbeta} \wt\psi_m^\dalpha(-p) \cfi_m^\dbeta(p) -
p^2 \ve_{\alpha\beta} \wt\cfi_m^\alpha(-p) \psi_m^\beta(p) \] .
\ee
The fields $\wt\psi_m^\dalpha$ and $\wt \cfi_m^\beta$ do not actually represent independent physical
degrees of freedom --- they are part of the extra free fields that we need to add in order
to write the action of superstring field theory. Therefore, they can be eliminated using their
equations of motion:
\be
\sp_{\dalpha\beta} \wt \cfi_m^\beta(p)=-2\,  \ve_{\dalpha\dbeta}  \cfi_m^\dbeta(p),
\qquad
\sp_{\beta\dalpha} \wt\psi_m^\dalpha(p) = - 2\, \ve_{\beta\alpha}  \psi_m^\alpha(p)\, .
\ee
Substituting these relations into \refb{ec.21}, we get
\be
S_{NSR} = 2 \int{d^{4}p\over (2\pi)^{4}}\, \psi_m^\alpha(-p) \!\sp_{\alpha\dbeta} \cfi_m^\dbeta(p)\, .
\ee

A similar analysis can be performed in the RNS sector, leading to the total action:
\be
S_{NSR+RNS} = 2 \int{d^{4}p\over (2\pi)^{4}}\, \[
\psi_m^\alpha(-p) \!\sp_{\alpha\dbeta} \cfi_m^\dbeta(p) +
\psi_m^{\prime\dalpha}(-p)\! \sp_{\dalpha\beta} \cfi_m^{\prime\beta}(p)\] .
\ee
These results can be expressed in terms of fields that are even and odd under \oo,
with the even fields given by $\xxi_m^\alpha$, $\xxi_m^\dalpha$ introduced in \refb{einvfield}.
Keeping only the terms involving the even  fields, we can write
the kinetic term as
\be\label{ec.25}
2 \int{d^{4}p\over (2\pi)^{4}}\,
\xxi_m^\dalpha(-p) \sp_{\dalpha\beta} \xxi_m^\beta(p) \, .
\ee

In the construction of string field
theory one begins with all fields complex and then imposes an appropriate reality condition that
eliminates half of the degrees of freedom. In this spirit, before imposing the reality condition,
the fields $\xxi_m^\dalpha$ and
$\xxi_m^\beta$
appearing in \refb{ec.25} can be regarded as independent complex variables. We can
then impose the reality condition
\be \label{ereality}
\xxi_m^\dalpha(-p)= -\I\, (\xxi_m^\beta(p))^* (\gamma^0)_\beta^{~\dalpha}\equiv
- \I \, \(\overline{\xxi_m(p)}\)^\dalpha\, ,
\ee
which turns the action to
\be\label{ec.26}
-2 \I \int{d^{4}p\over (2\pi)^{4}}\,
\(\overline{\xxi_m(p)}\)^\dalpha \!\sp_{\dalpha\beta} \xxi_m^\beta(p)
= -2 \int d^4 x \, \bar{\xxi}_m(x)\! \not\!\partial \,  \xxi_m(x)\, ,
\ee
with the understanding that $\xxi(x)$ represents a chiral spinor.

For completeness we shall also determine the normalization of the scalar superpartners of
these fermions. They appear as components of the string field in the NSNS sector as
\be\label{eensns1}
|\phi_{NSNS}\rangle =
\int{d^{4}p\over (2\pi)^{4}}\, \[
\cpi_m(p) \, c\, \bar c\, e^{-\phi} e^{-\bar\phi} W^m
+ \cpi_m^*(p) \, c\, \bar c\, e^{-\phi} e^{-\bar\phi} Y^m \] e^{\I p.X}(0)|0\rangle.
\ee
The action is given by a formula similar to \refb{ensraction} except that we can take
$\XX_0=1$ and set
$|\tilde\phi_{NSNS}\rangle=|\phi_{NSNS}\rangle$ from the beginning. This gives
\be\label{ensns2}
S_{NSNS} = 4\times {1\over 2} \,  \langle
\phi_{NSNS} | c_0^- \, (Q_B+\bar Q_B)\, |\phi_{NSNS}\rangle\, .
\ee
Substituting \refb{eensns1} into \refb{ensns2} and using the operator product expansion
\refb{e1.8}, we get
\be\label{ensns3}
S_{NSNS} = -\int{d^{4}p\over (2\pi)^{4}}\,  \cpi_m^*(-p) \, p^2 \, \cpi_m(p)
= -\int d^4 x \, \p_\mu \cpi_m^* \, \p^\mu \cpi_m\, .
\ee
If one keeps only the part $\hat\phi_m$ that is even under the orientifold projection,
as defined in \refb{einvfield}, the
kinetic term takes the form:
\be\label{ensns4}
-{1\over 2} \int d^4 x \, \p_\mu \hat\phi_m \, \p^\mu \hat\phi_m\, .
\ee

\section{Regularization of the zero modes} \label{sdeform}

During our analysis we have fixed the normalization of the
integration measure over the modes of open
string field theory on the D-instanton by comparing the result of the
path integral with
the annulus partition function.
This procedure breaks down for the zero modes since the annulus amplitude
diverges. We resolved this by adding
a small constant to the $L_0$ eigenvalues of  the states in the Siegel gauge
so as to make the annulus amplitude finite, used this to fix the integration measure over
the zero modes, and then set the $L_0$ eigenvalues back to zero.
As indicated in the main
text and elsewhere \cite{Sen:2021tpp,Alexandrov:2021shf,Alexandrov:2021dyl},
the result does not depend on how we shift the $L_0$
eigenvalues of the zero modes,
but one concrete procedure involves putting slightly different boundary conditions on the two
boundaries of the annulus. In this appendix we shall elaborate on this procedure.

Putting slightly different boundary conditions on the two boundaries of the annulus may
be regarded as having a pair of identical instantons slightly separated in space-time,
and considering the partition function of an open string with two boundaries lying
on two different instantons. This corresponds to considering open string states with
off-diagonal Chan-Paton factors. On the other hand, open string with both ends on the same
instanton are described by diagonal Chan-Paton factors. Now let us take some mode
$\zeta^{(1)}$ living on the first open string and let
$a\,d\zeta^{(1)}$ be the integration measure
over that mode. Gauge invariance of string field theory requires that $a$ must be
independent of all the modes of string field theory since by construction the string field
theory is gauge invariant with flat integration measure over the string fields. Since the separation between the
two instantons is also a mode of open string field theory, $a$ must be independent of the
separation between the two instantons. Taking the separation to infinity and using cluster
property, we see that
$a\,d\zeta^{(1)}$ must also be the integration measure over this mode in the absence
of the second instanton. If $\zeta^{(2)}$ denotes the same mode of the open string living
on the second instanton, then the integration measure over
$\zeta^{(2)}$ will be $ad\zeta^{(2)}$ with the same constant $a$.
Then, since $\zeta^{(\pm)} = (\zeta^{(1)}\pm
\zeta^{(2)})/\sqrt 2$ are normalized in the same way as $\zeta^{(1)}$ and $\zeta^{(2)}$, the
integration measure over these modes must also be $a\,d\zeta^{(+)}$ and $a\,d\zeta^{(-)}$.
One the other hand, if we denote by $\zeta_1$ and $\zeta_2$ the same open string modes
on the D-instanton pair associated with Chan-Paton factors $\sigma_1/\sqrt 2$ and
$\sigma_2/\sqrt 2$, then in the limit of zero separation between the instantons
$\zeta_1$ and $\zeta_2$ are related to $\zeta^{(-)}$ by $SU(2)$
rotation and so must have the same integration measures $a\,d\zeta_1$ and $a\,d\zeta_2$.
Conversely, by determining the measure of integration over the modes
$\zeta_1$ and $\zeta_2$,
we can find the constant $a$, which in turn gives the measure of integration over the zero
modes of a single instanton.

Note
that for a separated pair of instantons there are two sets
of degenerate modes, involving strings connecting the first instanton to the second one and
the reverse, and the kinetic term of string field theory
will couple them. So effectively our procedure amounts
to computing the partition function of both sets of modes and then taking a square root.
This is related to the problem of defining a path integral over
chiral fermions, which
also requires doubling the degrees of freedom and then taking a
square root. This of course leaves the
phase of the partition function ambiguous. For some amplitudes this ambiguity can be
absorbed in a shift of the RR scalar fields, and for others we can in principle use the cluster
property to determine the phase. This has been discussed in
\cite{Alexandrov:2021shf}, and we shall not discuss it any further here.

\section{Theta functions and their properties} \label{sc}

In this appendix we shall review some of the properties of Jacobi theta functions that are
used in appendices \ref{saE} and \ref{sF}. We define for $\alpha,\beta=0,1$
\be
\vt_{\alpha\beta}(z|\tau)=\sum_{n\in \IZ+\frac{\alpha}{2}}e^{i\pi n\beta} q^{n^2/2} y^{n},
\qquad q=e^{2\pi\I\tau} ,\quad y=e^{2\pi \I z}.
\ee
If the first argument is zero, we will simply write $\vt_{\alpha\beta}(\tau)\equiv\vt_{\alpha\beta}(0|\tau)$.
These functions have the following product representations
\begin{subequations}
\bea
\vt_{00}(z|\tau)&=&\prod_{n=1}^\infty(1-q^n)\(1+(y+y^{-1})q^{n-\hf}+q^{2n-1}\),
\\
\vt_{01}(z|\tau)&=&\prod_{n=1}^\infty(1-q^n)\(1-(y+y^{-1})q^{n-\hf}+q^{2n-1}\),
\\
\vt_{10}(z|\tau)&=&q^{\frac18}(y^{\hf}+y^{-\hf})\prod_{n=1}^\infty(1-q^n)\(1+(y+y^{-1})q^{n}+q^{2n}\),
\\
\vt_{11}(z|\tau)&=& \I q^{\frac18}(y^{\hf}-y^{-\hf})\prod_{n=1}^\infty(1-q^n)\(1-(y+y^{-1})q^{n}+q^{2n}\).
\eea
\end{subequations}
They satisfy the relations
\begin{subequations}
\bea
\vt_{\alpha\beta}\(z+\hf\Big|\tau\)&=&\sum_{n\in \IZ+\frac{\alpha}{2}}e^{i\pi n(\beta+1)} q^{n^2/2}y^n
=\vt_{\alpha,\beta+1}\(z|\tau\),
\label{shiftzbyhalf}
\\
\vt_{\alpha\beta}\(z+\frac{\tau}{2}\Big|\tau\)&=&q^{-\frac18}\sum_{n\in \IZ+\frac{\alpha}{2}}e^{i\pi n\beta} q^{\hf(n+\hf)^2}y^n
=e^{-i\pi\beta/2} q^{-\frac18}y^{-\hf}\vt_{\alpha+1,\beta}\(z|\tau\),
\label{shiftzbytau}
\eea
\end{subequations}
where we understood that the dependence on $\alpha$ and $\beta$ is mod 2 up to a phase, i.e.\
$\vt_{(\alpha+2)\beta}=\vt_{\alpha\beta}$ and $\vt_{\alpha(\beta+2)}=e^{i\alpha\pi}\vt_{\alpha\beta}$.

The Jacobi identity reads
\be
\vt_{00}(\tau)^4=\vt_{01}(\tau)^4+\vt_{10}(\tau)^4.
\ee
The relations to the Dedekind eta function
\be
\eta(\tau)=q^{\frac{1}{24}}\prod_{n=1}^\infty(1-q^n) \, ,
\ee
are the following
\begin{subequations}
\be
\vt_{11}'(0|\tau)=-2\pi \eta(\tau)^3, \quad \vt_{\alpha\beta}'(z|\tau)\equiv \p_z\vt_{\alpha\beta}(z|\tau),
\label{dervt11}
\ee
\be
\vt_{10}(\tau)=\frac{2\eta(2\tau)^2}{\eta(\tau)},
\quad
\vt_{01}(\tau)=\frac{\eta(\hf\tau)^2}{\eta(\tau)},
\quad
\vt_{00}(\tau)=\frac{\eta(\hf(\tau+1))^2}{\eta(\tau+1)}
={\eta(\tau)^5 \over \eta(\hf \tau)^2 \eta(2 \tau )^2},
\label{thetavseta}
\ee
\be
\vt_{00}(\tau)\vt_{10}(\tau)\vt_{01}(\tau)=2\eta(\tau)^3.
\ee
\end{subequations}
An important identity is\cite{Abel:2006yk,Akerblom:2006hx}
\be\label{eE.2}
{\vt_{\alpha\beta}\(z|\tau\)^2
\vt'_{11}(0|\tau)^2\over \vt_{11}\(z|\tau\)^2 \vt_{\alpha\beta}(0|\tau)^2}= {\vt''_{\alpha\beta}(0|\tau)
\over \vt_{\alpha\beta}(0|\tau)} -\p_z^2 \log \vt_{11}(z|\tau)\, .
\ee
Another important set of identities are (see \cite[(8.199)]{GradRyzh}):
\begin{subequations}
\bea
\vt_{00}(z_1|\tau)\vt_{00}(z_2|\tau) =\vt_{00}(z_1+z_2|2\tau)\vt_{00}(z_1-z_2|2\tau)+\vt_{10}(z_1+z_2|2\tau)\vt_{10}(z_1-z_2|2\tau),  \hskip .4in ~
\label{quadtheta0000}\\
\vt_{00}(z_1|\tau)\vt_{01}(z_2|\tau)=\vt_{01}(z_1+z_2|2\tau)\vt_{01}(z_1-z_2|2\tau)-\vt_{11}(z_1+z_2|2\tau)\vt_{11}(z_1-z_2|2\tau), \hskip .4in ~
\label{quadtheta0001}\\
\vt_{01}(z_1|\tau)\vt_{01}(z_2|\tau)=\vt_{00}(z_1+z_2|2\tau)\vt_{00}(z_1-z_2|2\tau)-\vt_{10}(z_1+z_2|2\tau)\vt_{10}(z_1-z_2|2\tau),  \hskip .4in ~
\label{quadtheta0101}\\
\vt_{10}(z_1|\tau)\vt_{10}(z_2|\tau)=\vt_{10}(z_1+z_2|2\tau)\vt_{00}(z_1-z_2|2\tau)+\vt_{00}(z_1+z_2|2\tau)\vt_{10}(z_1-z_2|2\tau),  \hskip .4in ~
\label{quadtheta1010}\\
\vt_{11}(z_1|\tau)\vt_{11}(z_2|\tau)=\vt_{00}(z_1+z_2|2\tau)\vt_{10}(z_1-z_2|2\tau)-\vt_{10}(z_1+z_2|2\tau)\vt_{00}(z_1-z_2|2\tau).  \hskip .4in ~
\label{quadtheta1111}
\eea
\label{quadtheta}
\end{subequations}

We further need the modular transformations of the functions introduced above.
The modular transformation of the Dedekind eta function is given by
\begin{align} \label{modularEta}
\eta \( -\frac{1}{\tau}\)
=
(-\I \tau)^{1/2} \eta(\tau) \, ,
\qquad
\eta \( \tau+1 \)
=
e^{\frac{\pi\I}{12}}\, \eta(\tau) \, ,
\end{align}
while the modular transformation of the theta functions are given by
\begin{subequations} \label{modularTheta}
\begin{align}
&\vt_{00} \({z \over \tau} \bigg| -\!\frac{1}{\tau} \)
=
(-i \tau)^{1/2}\, e^{\frac{\pi \I z^2}{\tau}}\, \vt_{00} (z| \tau) \, ,
\qquad
\vt_{00} \(z | \tau + 1\) =  \vt_{01} (z| \tau) \, ,
\\
&\vt_{01} \({z \over \tau} \bigg| -\!\frac{1}{\tau} \)
=
(-i \tau)^{1/2}\, e^{\frac{\pi \I z^2}{\tau}}\, \vt_{10} (z| \tau) \, ,
\qquad
\vt_{01} \(z | \tau + 1 \)
=
\vt_{00} (z| \tau) \, ,
\\
&\vt_{10} \({z \over \tau} \bigg| -\!\frac{1}{\tau} \)
=
(-i \tau)^{1/2} \, e^{\frac{\pi \I z^2}{\tau}}\, \vt_{01} (z| \tau) \, ,
\qquad
\vt_{10} \(z | \tau + 1 \)
=
e^{\frac{ \pi \I }{4}}\,\vt_{10} (z| \tau) ,
\\
&\vt_{11} \({z \over \tau} \bigg|-\!\frac{1}{\tau} \)
=
-i(-i \tau)^{1/2} \, e^{\frac{\pi \I z^2}{\tau}}\, \vt_{11} (z| \tau) \, ,
\qquad
\vt_{11} \(z | \tau + 1 \)
=
e^{\frac{ \pi \I }{4}}\, \vt_{11} (z| \tau).
\end{align}
\end{subequations}

We also define
\be
\label{edeffkq}
\begin{split}
f_{k,Q} (z,\tau) = &\, {1\over \eta(\tau)} \sum_{m\in\IZ+\frac{Q}{k}} q^{{k\over 2} m^2}y^{km}
\\
=&\,  {1\over \eta(\tau)} \,  e^{i\pi\tau Q^2/k} e^{2\pi i Q z} \vt_{00}(kz + Q\tau|k\tau),
\qquad
y\equiv e^{2\pi i z}\, .
\end{split}
\ee
In particular, we have
\be
f_{1,0}(z,\tau)=\frac{\vt_{00}(z|\tau)}{\eta(\tau)}\, ,
\qquad
f_{2,0}(\tau) = \frac{\vt_{00}(2z|2\tau)}{\eta(\tau)}\, ,
\qquad
f_{2,1}(z,\tau)= \frac{\vt_{10}(2z|2\tau)}{\eta(\tau)}\, .
\label{evalf1}
\ee

\section{Evaluation of the partition function} \label{saE}

For computing the normalization of the instanton contribution to the amplitude,
we need to evaluate the quantity \refb{edefamintro}. If we have a solvable CFT for which
the spectrum of D-branes and open string states is known, we can evaluate it explicitly.
However, in most cases the model is not solvable and the best we can hope for
is to regard the model as a deformation of
a solvable model, and then use conformal perturbation theory to compute the spectrum in the
deformed theory. In such cases we can simplify the analysis by organizing the infinite sums over
states arising in the expressions for the partition function by summing over different representations
of the underlying symmetry algebra and identify the contribution from a given representation
as the character of that representation.
In this appendix we shall discuss this procedure in some detail.

\subsection{Open string channel}
\label{sE-open}

We define $q\equiv e^{-2\pi t}$, $\tau\equiv it$ and express the integrands $Z_A$ and $Z_M$ in
\refb{edefam} as sum of four sectors: NS with $(-1)^F$ insertion, NS with $(-1)^F(-1)^f$ insertion,
R with $(-1)^F$ insertion and R with $(-1)^F(-1)^f $ insertion. Here $F$ stands for space-time
fermion number and $(-1)^f$ stands for the GSO parity
of the state.
In each sector the integrand
may be expressed as the product of the contribution from free fields $X^\mu$, $\psi^\mu$,
$b,c$, $\beta,\gamma$ and the contribution from the internal part of the CFT associated
with Calabi-Yau compactification. We shall denote
the free field contributions by\footnote{The factors of 2 in front of the R sector contributions are introduced for convenience.
Their role is to cancel the factor of 1/2 appearing in \eqref{egenZA} due to the
R sector kinetic operator $G_0$ being the square root of $L_0$.}
$f_{00}$, $f_{01}$, $2f_{10}$ and $2f_{11}$, respectively, and the internal conformal field theory contributions by
$g_{00}$, $g_{01}$, $g_{10}$ and $g_{11}$, respectively. In the definitions of these quantities the
ground state contributions to $(-1)^F$ and $(-1)^f$ will be included in the definitions of
the free field contributions since they typically arise from the ghost sector.
NS sector ground state $c e^{-\phi}(0)|0\rangle$ has odd GSO parity and even space-time
fermion number while R sector ground state has even GSO parity and odd space-time
fermion number.\footnote{The assignment of GSO parity to R sector ground state is
convention dependent. This affects the sign of $f_{11}$
but this contribution vanishes anyway due to having equal number
of GSO even and GSO odd states at each level.}
We also replace $L_0$ in the $e^{-2\pi t L_0}$ factor by $L_0-{c\over 24}$ for convenience, even though after adding
the contribution from the free field part and the internal part these factors cancel out
since the total central charge vanishes. Finally, the minus sign appearing in the action of
\oo\ on the $SL(2,\IR)$ invariant vacuum, as discussed below \refb{etranslationa},
will be included in the contribution from the internal conformal field theory.

\subsubsection{Annulus with both ends on the D-instanton} \label{sE.1}

We first compute the free field contribution. We have
\begin{subequations}
\bea
\label{edd.1}
\hspace{-1cm}
&& f_{00} =
q^{\frac{9}{24}} q^{-\hf} \prod_{n=1}^\infty (1-q^n)^{-4} \prod_{n=1}^\infty (1+q^{n-{1\over 2}})^4 \prod_{n=1}^\infty (1-q^n)^{2}
\prod_{n=1}^\infty (1+q^{n-{1\over 2}})^{-2} =  \frac{\vt_{00}(\tau)}{\eta(\tau)^3}\, ,
\\
\label{edd.2}
\hspace{-1cm}
&& f_{01} =
-q^{\frac{9}{24}} q^{-\hf} \prod_{n=1}^\infty (1-q^n)^{-4} \prod_{n=1}^\infty (1-q^{n-{1\over 2}})^4 \prod_{n=1}^\infty (1-q^n)^{2}
\prod_{n=1}^\infty (1-q^{n-{1\over 2}})^{-2}= -\frac{\vt_{01}(\tau)}{\eta(\tau)^3} , \\
\label{edd.3}
\hspace{-1cm}
&& 2f_{10}=
- 4\, \prod_{n=1}^\infty (1-q^n)^{-4} \prod_{n=1}^\infty (1+q^n)^4 \prod_{n=1}^\infty (1-q^n)^{2}
\prod_{n=1}^\infty (1+q^n)^{-2}=-\frac{2 \vt_{10}(\tau)}{\eta(\tau)^3}\, ,
\\
\label{edd.4}
\hspace{-1cm}
&& 2f_{11}=
- (2-2)\, \prod_{n=1}^\infty (1-q^n)^{-4} \prod_{n=1}^\infty (1-q^n)^4 \prod_{n=1}^\infty (1-q^n)^{2}
\prod_{n=1}^\infty (1-q^n)^{-2}=0\, ,
\eea
\label{fDDall}
\end{subequations}
where the four factors come from $X^\mu$, $\psi^\mu$, $b,c$ and $\beta,\gamma$, respectively.
The factors $q^{9/24}$ in the NS sector contributions reflect that the free system carries total central charge $-9$,
and the factors $q^{-1/2}$ reflect that the NS sector ground state $ce^{-\phi}(0)|0\rangle$ has $L_0=-1/2$.
These results are nicely summarized by
\be
f_{\alpha\beta}=(-1)^{\alpha+\beta}\, \frac{\vt_{\alpha\beta}(\tau)}{\eta(\tau)^3}\, .
\label{fDD}
\ee

As mentioned before,
since the spectrum of states in the internal CFT is not known in general, the best we can do is
to organize the states into representations of the symmetry algebra and then evaluate the
contribution from a given
representation \cite{Eguchi:1988vra,Odake:1988bh,Odake:1989dm,Odake:1989ev}.
The relevant symmetry algebra in this case is the
$\NN=2$ superconformal algebra extended by the spectral flow.
For convenience we shall define, for the SCFT associated with Calabi-Yau compactification and for any representation $r$,
\be\label{edefgandf}
\begin{split}
\gr_{00}=&\, \Tr_{NS,r} \[q^{L_0} (-1)^F\],
\qquad
\gr_{01}=\Tr_{NS,r} \[q^{L_0} (-1)^F(-1)^f\] ,
\\
\gr_{10}= &\, \Tr_{R,r} \[ q^{L_0} (-1)^F\],
\qquad\ \
\gr_{11}= \Tr_{R,r} \[q^{L_0} (-1)^F (-1)^f\].
\end{split}
\ee

The representations of the extended $\NN=2$ superconformal algebra can be divided into two broad
classes: the massive representations and the massless
representations~\cite{Odake:1989dm,Odake:1989ev}. The massive
representations are characterized by a pair of numbers $(h,Q)$,
with $Q$ taking values 0,1 and $h$
taking any real value larger than $Q/2$.\footnote{It is also possible to have $Q=-1$ and $h>1/2$ representations.
Their characters are same as those of $Q=1$ and $h>1/2$, so we won't consider them below explicitly.}
On the other hand, there are three independent
massless representations, that we shall denote by $(vac)$, $(+)$ and $(-)$.
The $SL(2,\RRR)$ invariant vacuum state $|0\rangle$ belongs to the representation $(vac)$.

We first consider massive representations.  The results for all sectors take the form \cite{Odake:1989ev}:
\be
\ghQ_{\alpha\beta}=q^{\frac{3\alpha}{8}}\, g\(\frac{\alpha\tau+\beta}{2}\, ,\tau, h,Q\),
\label{eda.5a}
\ee
where
\be\label{eda.5b}
g(z,\tau; h,Q) = q^{h-{1\over 4} -{Q^2\over 4}} \eta(\tau)^{-1} f_{1,0}(z,\tau) f_{2,Q}(z,\tau)\, ,
\ee
and $f_{k,Q}$ are defined in \eqref{edeffkq}.
Substituting \eqref{evalf1} into \refb{eda.5b}, we get for $Q=0,1$,
\ben
g(z,\tau;h,Q)
&=&   \frac{q^{h-\frac{1+Q}{4}}  }{\eta(\tau)^3}\, \vt_{00}(z|\tau)\, \vt_{Q0}(2z|2\tau)\, ,
\label{edd.11new}
\een
which results in
\be
\ghQ_{\alpha\beta}=(-1)^{\beta Q}\, \frac{q^{h-\frac{1+Q}{4}} }{\eta(\tau)^3}\,
\vt_{\alpha\beta}(\tau)\, \vt_{\alpha+Q,0}(2\tau).
\label{gab}
\ee
Note in particular that $\ghQ_{11}=0$ due to $\vt_{11}(0|\tau)=0$.

We can now put these results together to compute the annulus partition function
for a massive representation with both boundaries lying on the instanton:
\be \label{edefzaii}
\ZhQ_{A,DD} = {1\over 4} \sum_{\alpha,\beta=0,1} f_{\alpha\beta}(\tau)\, \ghQ_{\alpha\beta}(\tau) .
\ee
Here the overall factor of $1/4$ is the result of the GSO and orientifold projections
that act as $(1+(-1)^f)/2$ and $(1+\hbox{\oo})/2$, respectively.
Note that the annulus partition function only captures the 1 part of $(1+\hbox{\oo})$, the \oo\ part being
represented by the M\"{o}bius strip that will be discussed in \S\ref{smobius}.
The factor of 2 in front of $f_{1\beta}$ cancelled $1/2$ appearing for the Ramond sector contribution due to
a similar factor in \refb{egenZA}. The subscript $DD$ denotes the fact that we have the Dirichlet boundary
condition along the non-compact directions along both boundaries of the annulus. Using
\refb{fDD} and \refb{gab}, we get
\be\label{ezaii}
\ZhQ_{A,DD}
= \frac{1}{4}\, \frac{q^{h-\frac{1+Q}{4}} }{\eta(\tau)^6}\,
\sum_{\alpha,\beta}(-1)^{\alpha+\beta(1+Q)}\, \vt_{\alpha\beta}(\tau)^2\, \vt_{\alpha+Q,0}(2\tau).
\ee
We can now use \eqref{quadtheta} to rewrite the sum as
\begin{align}
&\Bigl(\vt_{00}(2\tau)^2+\vt_{10}(2\tau)^2+(-1)^{1+Q}(\vt_{00}(2\tau)^2-\vt_{10}(2\tau)^2)\Bigr)\vt_{Q0}(2\tau) \non\\
&\hspace{2.5in} -2 \vt_{00}(2\tau)\vt_{10}(2\tau)\vt_{1-Q,0}(2\tau)=0 \, .
\end{align}
showing that $\ZhQ_{A,DD}$ vanishes.
This reproduces the result of \cite{Eguchi:1988vra}.

The analysis for massless multiplets proceeds along similar lines.
The characters are given
by formul\ae\ similar to that in \refb{eda.5a}, but with different functions $g(z,\tau)$.
We have \cite{Odake:1989dm}
\be
\label{echiralrep}
\begin{split}
g^{(vac)}(z,\tau) =&\, g(z,\tau;0,0) - g\(z,\tau;{1\over 2}, 1\),
\\
g^{(\pm)}(z,\tau) =&\,  \pm {1\over 2} \(f_{3,1}(z,\tau) - f_{3,-1}(z,\tau)\) + {1\over 2}\, g\(z,\tau;{1\over 2},1\) ,
\end{split}
\ee
with $f_{k,Q}$ defined in \refb{edeffkq}.
It is easy to see that $\Zi{vac}_{A,DD}$ and $\Zi{\pm}_{A,DD}$ also both vanish, which
follows from our previous analysis together with the identities:
\be\label{efidentity}
f_{3,1}(z,\tau) - f_{3,-1}(z,\tau)=0
\qquad
\hbox{for $z=0,{1\over 2}, {\tau\over 2}$}\, ,
\ee
so that this term does not contribute to any $\gi{\pm}_{\alpha\beta}$ except the one with
$(\alpha\beta)=(11)$. Since $f_{11}$ vanishes, the net contribution of the
$(\alpha\beta)=(11)$ term
to $\Zi{\pm}_{A,DD}$ also vanishes.
The identities \refb{efidentity} follow from \refb{edeffkq} and the standard periodicity and evenness
properties of $\vt_{00}(z|3\tau)$.

The vanishing of $\Zr_{A,DD}$ agrees with the general result of \cite{Eguchi:1988vra}.
This may be restated as the identity
\be \label{egidentity}
\frac{1}{4\, \eta(\tau)^3}\sum_{\alpha,\beta=0,1} (-1)^{\alpha+\beta}
\vt_{\alpha\beta}(\tau) \, \gr_{\alpha\beta}(\tau)=0\, .
\ee
This identity holds for all representations. Furthermore, we can exclude the $(\alpha\beta)=(11)$
term from the sum due to $\vt_{11}(\tau)=0$. One can view \refb{egidentity} as a consequence
of $\NN=2$ supersymmetry in four dimensions, since $\Zr_{A,DD}$ by itself does not have
any knowledge of the breaking of $\NN=2$ supersymmetry to $\NN=1$ due to the
orientifold projection.

Even though $\gr_{11}$ does not contribute to \refb{egidentity}, for later
use we shall now quote the values of $\gr_{11}$ for different representations.
Using \refb{gab} and \refb{echiralrep} we see that $\gr_{11}$ vanishes for the massive representations
and the $(vac)$ representation. On the other hand, for the $(\pm)$ representations we get
\be \label{eg11pm}
\gi{\pm}_{11} = \pm {1\over 2}\, q^{3/8}\, \( f_{3,1} \({1+\tau\over 2},\tau\)- f_{3,-1} \({1+\tau\over 2},\tau\)\)
=\pm 1\, .
\ee
Since the internal components of the operators
$S_\alpha(z)$ and $S_\dalpha(z)$ are related to the $SL(2,\IR)$
invariant vacuum $|0\rangle$ via spectral flow, the fermion zero modes
associated with broken supersymmetry come from the $(vac)$ representation that contains
the state $|0\rangle$. Therefore, they give vanishing contribution to $g_{11}$. Since we have
assumed that these are the only fermion zero modes on the D-instanton, we conclude that the
spectrum of open strings does not contain any $(\pm)$ representation and therefore $g_{11}$
vanishes for all open string states.

\subsubsection{Annulus between D-instanton and space-filling D-brane}
\label{sE.3}

In this case the contributions from a given representation in the internal CFT remain
unchanged although the specific representations and their multiplicities that appear in the
open string spactrum will be different from those appearing on the open string with
both ends on the D-instanton. The contributions from
the $X^\mu$ and $\psi^\mu$ fields change since they have Dirichlet-Neumann boundary conditions
and they become
\begin{subequations}
\bea
\label{edd.32}
\hspace{-1cm}
f_{00}&=& 4 q^{\frac{9}{24}} \prod_{n=1}^\infty (1-q^{n-{1\over 2}})^{-4} \prod_{n=1}^\infty (1+q^{n})^4 \prod_{n=1}^\infty (1-q^n)^{2}
\prod_{n=1}^\infty (1+q^{n-{1\over 2}})^{-2}
= {\vt_{10}(\tau)^2\, \eta(\tau)^3\over \vt_{01}(\tau)^2\, \vt_{00}(\tau)}\, ,
\non\\
\\
\label{edd.33}
\hspace{-1cm}
f_{01} &=& - (2-2) \, q^{\frac{9}{24}} \prod_{n=1}^\infty (1-q^{n-{1\over 2}})^{-4}
\prod_{n=1}^\infty (1-q^{n})^4 \prod_{n=1}^\infty (1-q^n)^{2}
\prod_{n=1}^\infty (1-q^{n-{1\over 2}})^{-2}= 0\, ,
\\
\label{edd.34}
\hspace{-1cm}
2f_{10}&=&
-  \prod_{n=1}^\infty (1-q^{n-{1\over 2}})^{-4}
\prod_{n=1}^\infty (1+q^{n-{1\over 2}})^4 \prod_{n=1}^\infty (1-q^n)^{2}
\prod_{n=1}^\infty (1+q^n)^{-2}
=-2\, {\vt_{00}(\tau)^2 \,\eta(\tau)^3\over \vt_{01}(\tau)^2\, \vt_{10}(\tau)}\, ,
\non\\ \\
\label{edd.35}
\hspace{-1cm}
2 f_{11}&=&
-\prod_{n=1}^\infty (1-q^{n-{1\over 2}})^{-4}
\prod_{n=1}^\infty (1-q^{n-{1\over 2}})^4 \prod_{n=1}^\infty (1-q^n)^{2}
\prod_{n=1}^\infty (1-q^n)^{-2}=-1\, .
\eea
\label{fDNall}
\end{subequations}
The contribution to the annulus partition function from a given representation
of the extended $\NN=2$ superconformal algebra is now given by
\be\label{edd.36}
\Zr_{A,DN} ={1\over 2}\sum_{\alpha,\beta} f_{\alpha\beta}(\tau)
\, \gr_{\alpha\beta}(\tau).
\ee
Note that the overall factor is $1/2$ instead of $1/4$ since we have two possible
orientations of the open string connecting the D-instanton to the space-filling D-brane
giving identical results.
Using \refb{gab}, the contribution from a massive representation is found to be
\ben
\ZhQ_{A,DN} &=&\hf\, \frac{q^{h-\frac{1+Q}{4}}}{\vt_{01}(\tau)^2}
\(\vt_{Q0}(2\tau)\,\vt_{10}(\tau)^2-\vt_{1-Q,0}(2\tau) \,\vt_{00}(\tau)^2\)
\non\\
&=& \hf\, q^{h-\frac{1+Q}{4}} \, \frac{2\vt_{Q0}(2\tau)\,\vt_{00}(2\tau)\, \vt_{10}(2\tau)
-\vt_{1-Q,0}(2\tau)\(\vt_{00}(2\tau)^2+\vt_{10}(2\tau)^2\)}{\vt_{00}(2\tau)^2-\vt_{10}(2\tau)^2}
\non\\
&=&  \frac{(-1)^{Q}}{2} \, q^{h -\frac{1+Q}{4} } \vt_{1-Q,0} ( 2 \tau),
\label{ZADN}
\een
where we used \eqref{quadtheta}.

The contribution due to the massless representations can be computed similarly using the formula
for the characters given in \refb{echiralrep}.  In particular, since we have argued that the
representations $(\pm)$ do not arise in the spectrum, the contribution only involves
$g(z,\tau;h,Q)$ and can be read out from \refb{ZADN}.
At the end we need to sum over all representations.
In particular, there is an infinite number of massive multiplets
labelled by $h$, both for $Q=0$ and $Q=1$. As can be seen from \refb{ZADN}, for either value of $Q$
the dependence on $h$ appears only through the overall factor of $q^h$.
Thus, the annulus partition function can be organized as
\be\label{endfinal}
Z_{A,DN} = \sum_{Q=0,1}  \frac{(-1)^{Q}}{2}  \vt_{1-Q,0}(2\tau)
\sum_{h\ge |Q|/2}  n_{h,Q} \,
q^{h -\frac{1+Q}{4} },
\qquad
q=e^{2\pi i\tau}=e^{-2\pi t}\, ,
\ee
where $n_{h,Q}$ is the number of representations of type $(h,Q)$ appearing in the
open string spectrum with one end on the instanton and the other end on the
space-filling brane, with the understanding that the
representation $(vac)$ contributes $1$ to $n_{0,0}$ and $-1$ to $n_{1,1/2}$.

\subsubsection{M\"{o}bius strip} \label{smobius}

For computing the M\"{o}bius strip contribution with D-instanton boundary condition,
we need to insert an orientifold operator inside
all the traces in the D-instanton partition functions. This has the following effects:
\begin{enumerate}
\item
For a holomorphic field $\phi(z)$ of conformal weight $h$,
\oo\ produces $(-1)^h \bar\phi(-\bar z)$. Therefore, if we use the
expansions
\be
\phi(z) =\sum_n \phi_n z^{-n-h}, \qquad \bar\phi(\bz) =\sum_n \bar\phi_n \bz^{-n-h},
\ee
then \oo\ takes $\phi_n$ to $(-1)^n \bar\phi_n$.

\item
This shows that if $\phi$ satisfies the Neumann boundary condition $\phi_n=\bar\phi_n$,
then \oo\ acts as $\phi_n\to (-1)^n \phi_n$. This is the case for the
generators of the extended superconformal algebra and the ghost fields.
On the other hand, if $\phi$ satisfies
the Dirichlet boundary condition $\phi_n=-\bar\phi_n$,
as $\p X^\mu$ and $\psi^\mu$, then \oo\
acts as $\phi_n\to -(-1)^n \phi_n$.

\item
Acting on any representation $r$, \oo\ produces an overall phase. This
phase, which
we shall denote by $\alpha_r$, must be such that acting on GSO even states
\oo\ has eigenvalues $\pm 1$, and after we combine this with $Z_{A,DD}$, only the
states with eigenvalue 1 are kept in the
spectrum.
\end{enumerate}

We shall now study how this affects these on the contribution to the partition function from a
given representation of the extended superconformal algebra. First of all, $\alpha_r$ will appear
as an overall multiplicative factor in the partition function. The effect of $\phi_n\to (-1)^n\phi_n$
can be represented as the $(-1)^{L_0}$ operation. This effectively replaces $q$ by $\hq=-q$ in
all the expressions provided we choose the overall phase appropriately by adjusting
$\alpha_r$.
We shall define $\htau=\tau+{1\over 2}$ so that $\hq=e^{2\pi i\htau}$.
Finally, the effect of the extra minus sign in the transformation laws of $X^\mu$ and $\psi^\mu$
due to the Dirichlet boundary condition is to  replace \refb{fDDall} by
\begin{subequations}
\bea
\label{edd.1a}
\hat f_{00}(\htau) &=& \hq^{\frac{9}{24}} \hq^{-\hf} \prod_{n=1}^\infty (1+\hq^n)^{-4}
\prod_{n=1}^\infty (1-\hq^{n-{1\over 2}})^4 \prod_{n=1}^\infty (1-\hq^n)^{2}
\prod_{n=1}^\infty (1+\hq^{n-{1\over 2}})^{-2}
\non\\
&=& 4\, {\vt_{01}(\htau)^2\,\eta(\htau)^3 \over \vt_{10}(\htau)^2 \,\vt_{00}(\htau)}
\, ,
\\
\label{edd.2a}
\hat f_{01}(\htau)  &=& -\hq^{\frac{9}{24}}  \hq^{-\hf} \prod_{n=1}^\infty
(1+\hq^n)^{-4} \prod_{n=1}^\infty (1+\hq^{n-{1\over 2}})^4
\prod_{n=1}^\infty (1-\hq^n)^{2}
\prod_{n=1}^\infty (1-\hq^{n-{1\over 2}})^{-2}
\non\\
&=& -4\, {\vt_{00}(\htau)^2\,\eta(\htau)^3 \over \vt_{10}(\htau)^2\, \vt_{01}(\htau)}
\, ,
\\
\label{edd.3a}
2 \hat f_{10}(\htau) &=&
- (2-2)\, \prod_{n=1}^\infty (1+\hq^n)^{-4} \prod_{n=1}^\infty (1-\hq^n)^4
\prod_{n=1}^\infty (1-\hq^n)^{2}
\prod_{n=1}^\infty (1+\hq^n)^{-2}=0\, ,
\\
\label{edd.4a}
2 \hat f_{11}(\htau)&=&
- 4\, \prod_{n=1}^\infty (1+\hq^n)^{-4} \prod_{n=1}^\infty (1+\hq^n)^4
\prod_{n=1}^\infty (1-\hq^n)^{2}
\prod_{n=1}^\infty (1-\hq^n)^{-2}=-4\, .
\eea
\label{fMall}
\end{subequations}
Combining these results with \refb{gab} with $q$ replaced by $\hq$ and an
overall multiplicative factor of $\alpha_r$, we get the contribution to the M\"{o}bius partition function
from a massive representation\footnote{We do not put on the M\"{o}bius partition function
any additional index indicating boundary conditions because the boundary always lies on the instanton.}
\ben\label{ed.29a}
\ZhQ_{M} &=&
{\alpha_r\over 4} \sum_{\alpha,\beta} \hat f_{\alpha\beta}(\hat\tau)
\ghQ_{\alpha\beta}(\hat\tau)
=
\alpha_r\, \hq^{h-\frac{1+Q}{4}} \, \frac{\vt_{Q0}(2\htau)}{\vt_{10}(\htau)^2}\(\vt_{01}(\htau)^2-(-1)^Q \vt_{00}(\htau)^2\)
\non\\
&=& (-1)^{1-Q}\alpha_r\, \hq^{h-\frac{1+Q}{4}} \, \frac{\vt_{Q0}(2\htau)\,\vt_{1-Q,0}(2\htau)^2}{\vt_{00}(2\htau)\,\vt_{10}(2\htau)}
\non\\
&=&  (-1)^{1-Q} \, \alpha_r\, \hq^{h -\frac{1+Q}{4} } \vt_{1-Q,0} ( 2 \htau),
\label{eE.31}
\een
where we again  used \eqref{quadtheta}.

In order to fix the allowed form of $\alpha_r$, let us consider the limit $\tau\to i\infty$.
In this limit \refb{eE.31} reduces to:
\be\label{elimit}
(-1)^{Q-1} 2^{1-Q} \,   \hq^{h -{Q\over 2}} \alpha_r, \qquad \hbox{for $Q=0,1$} \, .
\ee
In order that $\ZhQ_{M}$ has the interpretation as half of
a sum over states weighted by $(-1)^F\hbox{\oo}$,
\refb{elimit} must be half of an integer. Writing $\hq = e^{-2\pi t+i\pi}$,
we see that $\alpha_r$ must be of the form $e^{-i\pi (h-{Q\over 2})} \sigma_r$ where $\sigma_r=\pm 1$.

We can now write a formula for the M\"{o}bius partition function
analogous to \refb{endfinal} by summing over the contributions
from all representations:
\be
\begin{split}
Z_{M} =&\,  \sum_{Q=0,1} (-1)^{1-Q} \vt_{1-Q,0}(2\htau)
\!\!\sum_{h\ge |Q|/2} \sum_{\sigma=\pm 1} \sigma\,\tn_{h,Q,\sigma}\,
e^{-i\pi (h-{Q\over 2})} \hq^{h-{1+Q\over 4}},
\\
\hq =&\,  e^{2\pi i\htau}= - e^{-2\pi t} ,
\end{split}
\label{emobfinal}
\ee
where $\tn_{h,Q,\sigma}$ is the number of representations of the type $(h,Q)$ that pick up phase
$\alpha=\sigma e^{-i \pi (h-{Q\over 2})}$ under \oo. As in \refb{endfinal}, due to \refb{echiralrep}, the
representation $(vac)$ contributes $1$ to $\tn_{0,0,\sigma_{vac}}$ and $-1$ to
$\tn_{1/2,1,\sigma_{vac}}$. Here $\sigma_{vac}$ represents the value of $\sigma$ for the $(vac)$ representation.

We shall illustrate the procedure for determining $\sigma_r$ by computing it for the
$(vac)$ representation. The first equation in \refb{echiralrep} implies that the contribution
from the $(vac)$ representation is given by the difference between the contributions from the
representations $(h=0,Q=0)$ and $(h={1\over 2},Q=1)$. Therefore, \refb{elimit} gives the
leading contribution in the $\tau\to i\infty$
limit to be $(-2-1)\alpha_{vac}=-3\alpha_{vac}$. On the other hand,
this is to be identified as the contribution to $Z_{M}$ from the zero modes in the spectrum.
After the GSO projection, we have four orientifold even and $(-1)^F$ even
states \refb{etranslationa} and a pair of orientifold odd and $(-1)^F$ odd
states \refb{eghost} in the NS sector. Therefore, the sum over these states, weighted by
$(-1)^F$ times \oo, gives 6, and after division by the factor of 2 coming from the orientifold projection operator we get 3.
On the other hand, in the Ramond sector half of the zero modes are even and half are odd under \oo\ and therefore give
vanishing contribution. Thus, the expected contribution to $Z_{M}$ from the vacuum
representation is 3. This fixes $\alpha_{vac}=-1$ and hence  $\sigma_{vac}=-1$, leading to
$\tn_{0,0,-1}=1$, $\tn_{1/2,1,-1}=-1$ and $\tn_{0,0,1}=\tn_{1/2,1,1}=0$.

\subsection{Closed string channel} \label{sE4}

So far we have analyzed the contribution to the partition function from the
open string viewpoint by expressing the partition function as  integer linear
combination of characters in the open string channel.
This approach is useful for generating the large $t$ / small $q$ expansion of the partition function, {\it e.g.}
if we are computing the partition function near a solvable theory using perturbation theory.
Under a deformation of the background,
the conformal weights $h$ labelling the representations get deformed, but the integer
coefficients $n_{h,Q}$ and $\tn_{h,Q,\sigma}$
multiplying the characters are expected to remain unchanged.
For large $t$ we may get a reliable estimate by working out the deformation of the conformal
weights of a few low lying open string states.
However, there is a complementary approach in  which
we compute the partition function in the closed string
channel \cite{Ooguri:1996ck,Recknagel:1997sb} --- as inner products between boundary
states associated with D-instantons, D-branes and orientifold planes.

\subsubsection{Annulus between D-instanton and space-filling D-brane}
\label{sE-closed-DN}

First, let us consider the annulus amplitude $Z_{A,DN}$ --- we shall not consider $Z_{A,DD}$
since it vanishes anyway. As in \refb{eccd1},
this will be given by $\int_0^\infty d\ell \,\langle I|e^{-\pi\ell(L_0+\bar L_0)}c_0^- b_0^+ |S\rangle$
where $|I\rangle$ and $|S\rangle$ are the boundary states of the instanton and the space-filling brane, respectively,
as in \S\ref{subsec-effectorient}, $\ell=1/t$ and
$\tilde q=e^{-2\pi \ell}$ is the modular parameter associated with the closed string channel.
If $K_n$ denotes some generator of the extended $\NN=2$ superconformal algebra, then the
boundary state $|S\rangle$ satisfies
\be \label{eboundarystate}
(K_n- (-1)^h \bar K_{-n})|S\rangle=0\, ,
\ee
for all $n$. Here $h$ is the conformal weight of the operator whose modes are represented by
$K_n$ and the $(-1)^h$ factor arises from the conformal transformation $z = (1+iy)/(1-iy)$
that relates the disk coordinate $z$ to the upper half plane coordinate $y$.
Therefore, one copy of the extended superconformal algebra is preserved,
which we can take to be the holomorphic part of the algebra.

We can now express the internal SCFT part of the boundary states as
linear combinations of Ishibashi states \cite{Ishibashi:1988kg} of the
extended $\NN=2$ superconformal algebra --- with different boundary states being given by different linear
combinations. The Ishibashi states are obtained by starting with a given left-right
symmetric primary in the closed string sector and then taking a linear combination
of states obtained by acting with the generators of the left and right extended superconformal
algebra on this primary so that it satisfies the condition \eqref{eboundarystate}.
Therefore, the Ishibashi states are linear combinations of states in some particular
representation of the extended $\NN=2$ superconformal algebra, and the inner product between the Ishibashi states
associated with different representations vanishes. Furthermore, it follows from the general
analysis that the matrix element of $\tilde q^{L_0}$ between the Ishibashi states in the same
representation is again given by the characters of the corresponding
representation \cite{Recknagel:1997sb}, with the argument $q$ replaced by $\tilde q$.
Therefore, the integrand of the annulus amplitudes, given by the matrix element of
$c_0^-b_0^+e^{-\pi\ell (L_0+\bar L_0)}=c_0^-b_0^+ e^{-2\pi\ell L_0}$
between the boundary states, can be expressed as a linear combination of the characters of the extended superconformal
algebra with argument $\tilde q$. The four sectors,
labelled by NSNS sector with no $(-1)^f$ insertion, NSNS with $(-1)^f$ insertion, RR with
no $(-1)^f$ insertion and RR with $(-1)^f$ insertion will correspond to $g_{00}(\ttau)$,
$g_{01}(\ttau)$, $g_{10}(\ttau)$ and $g_{11}(\ttau)$, respectively, where
$\ttau=i\ell$.\footnote{Note that the effect
of $(-1)^f$ operation on a boundary state is to change the sign of the $(-1)^h$ factor in
\refb{eboundarystate}. In a more conventional description, {\it e.g.} in \cite{Quiroz:2001xz},
one denotes the boundary states
that differ by a change in sign of the $(-1)^h$ term for half integral $h$ by $+$ and $-$,
so that the action of $(-1)^f$ relates these states.
Note that in \cite{Quiroz:2001xz} the symbols $q$ and $\tilde q$ are opposite of what we have used here.}

One point to note is that while we have argued
that under our assumption on fermion zero modes the open string channel does not have
the $(\pm)$ representations, this does not preclude their existence in the closed string channel.
Indeed, the massless chiral multiplet fields discussed in \S\ref{sworld}
come from exactly these representations.
Nevertheless, using the result that $g^{(+)}-g^{(-)}= f_{3,1}-f_{3,-1}$ does not mix with any
other character under modular transformation \cite{Odake:1989dm},
one can argue that the $(+)$ and $(-)$ representations appear in equal numbers in the closed string channel and give
vanishing  contribution to $g_{11}$ due to \refb{eg11pm}.

The contribution from the free field sector  can be computed explicitly for each of
the four sectors separately and will be related by the modular
transformation $\tilde\tau=-1/\tau$ to $f_{\alpha\beta}$ computed in \refb{fDNall}.
More specifically, if we denote it by $\tilde f_{\alpha\beta}(\ttau)$,
then we have $\tilde f_{\alpha\beta}(\ttau) = f_{\beta\alpha} (\tau)$,
with the exchange of $\alpha,\beta$ being due to the fact that the modular transformation
$\tilde\tau=-1/\tau$ exchanges the boundary conditions on the fermions along the
width and the circumference of the annulus.
Then
the contribution from a particular character to a given amplitude may be expressed as\footnote{$\cZr_{A,DN}$
is to be distinguished from $\Zr_{A,DN}$ introduced in \S\ref{sE.3} in that  the former represents the contribution from
a particular representation in the closed string channel, while the latter represents the
contribution from a particular representation in the open string channel.}
\begin{align} \label{e43pre}
\cZr_{A,DN}={N_r\over 2} \sum_{\alpha,\beta}  \tilde f_{\alpha\beta}(\ttau)\, \gr_{\alpha\beta}(\ttau)
\,  .
\end{align}
Here $N_r$ is a normalization factor that reflects
the product of the coefficients of the Ishibashi states in
representation $r$ in the two boundary states whose inner product we are computing and
the factor of $1/2$ is due to the GSO projection.
The functions $\tilde f_{\alpha\beta}(\tilde q)$ are computed from \refb{fDNall} as follows:
\begin{subequations}
\bea
\tilde f_{00}(\ttau)
&=&
f_{00} \( -\frac{1}{\ttau}\)
=
{\vt_{10} \(-\frac{1}{\ttau} \)^2 \,\eta \( -\frac{1}{\ttau}\)^3
\over \vt_{01} \(-\frac{1}{\ttau} \)^2\, \vt_{00}\(-\frac{1}{\ttau}\)}
=
{(-i \ttau) \vt_{01}(\ttau)^2\eta(\ttau)^3 \over \vt_{10}(\ttau)^2 \vt_{00}(\ttau)}
=
-{\I \ttau \over 4} \hat f_{00} (\ttau) \, ,
\\
\tilde f_{01}(\ttau)
&=&
f_{10} \( -\frac{1}{\ttau}\)
=
{-\vt_{00} \( -\frac{1}{\ttau}\)^2 \eta \(  -\frac{1}{\ttau}\)^3
\over \vt_{10} \( -\frac{1}{\ttau}\)^2 \vt_{10}\( -\frac{1}{\ttau}\)}
=
 {\I \ttau\,  \vt_{00}(\ttau)^2\,\eta(\ttau)^3 \over \vt_{10}(\ttau)^2\, \vt_{01}(\ttau)}
=
-{\I \ttau \over 4} \hat f_{01} (\ttau) \, ,
\\
\tilde f_{10}(\ttau)
&=&
f_{01} \( -\frac{1}{\ttau} \) = 0  \, ,
\eea
\end{subequations}
where $\hat f_{\alpha\beta}$ have been defined in \refb{fMall} and
we have used the modular transformations given
in~\eqref{modularEta}, \eqref{modularTheta} to arrive at this result. We do not need the
expression for $\tilde f_{11}$ since $g_{11}$ contributions would vanish at the end.
This shows that the evaluation of \eqref{e43pre} is analogous to that of \eqref{ed.29a},
the only differences being the replacement $\htau \to \ttau$
in arguments, $\alpha_r\to N_r$ in the normalization
and extra multiplication by $-\I \ttau/2$. This gives
\begin{align} \label{e43}
{\mathcal Z}^{(h,Q)}_{A,DN}
={N_r\over 2}\, (- \I  \ttau) (-1)^{1-Q} \tilde q^{h - {1 +Q \over 4} }  \vt_{1-Q,0} ( 2 \ttau)\,  ,
\end{align}
for massive representations.

After summing over all representations we can express
the partition function as
\begin{align} \label{e43sum}
\cZ_{A,DN}
=-\frac{ \I  \ttau}{2}  \sum_{Q=0,1} (-1)^{1-Q}  \vt_{1-Q,0} ( 2 \ttau) \sum_{h\ge |Q|/2} N_{h,Q}\,
\tilde q^{h - {1 +Q \over 4} },
\qquad
\tilde q=e^{2\pi i\ttau} = e^{-2\pi/t}\,  ,
\end{align}
where as before, $N_{h,Q}$ is the normalization constant that multiplies the Ishibashi state
associated to the representation $(h,Q)$.
Using \refb{echiralrep}, we also see that the contribution from the representation $(vac)$
gets added to $N_{0,0}$ with weight 1 and to $N_{1,1/2}$ with weight $-1$,
whereas the contribution from the $(\pm)$ representations are added to $N_{1,1/2}$ with weight $1/2$. Note that even though for individual representations the open string channel
result $Z^{(h,Q)}_{A,DN}$ differs from the closed string channel result $\cZ^{(h,Q)}_{A,DN}$,
the total contributions $Z_{A,DN}$ and $\cZ_{A,DN}$ are the same.

Since we derived the form of $\tilde f_{\alpha\beta}$'s by rewriting the expressions in the
open string channel in terms of the new variable $\ttau$,
we need to use the integration measure induced from the open string
channel which is given by $dt/(2t)=d\ell/(2\ell)$.
The factor $\ell$ in the denominator cancels the factor $-i\ttau=\ell$ in the
expression for $Z_{A,DN}$ in \refb{e43sum}. The resulting integrand has a power
series expansion in $\tilde q$ and therefore can be interpreted as the matrix element of
$c_0^- b_0^+ \tilde q^{L_0}$ between the boundary states of the D-instanton and the
space-filling D-brane.

\subsubsection{M\"{o}bius strip}
\label{sE-open-mobius}

For the M\"{o}bius strip, given by
$\int_0^\infty d\ell \,\langle I|e^{-\pi \ell (L_0+\bar L_0)} c_0^- b_0^+ |C\rangle$
where $|C\rangle$ is the crosscap state, the analysis is similar but with a few differences.
First of all, the relation between $\ell$ and $t$ changes to
$\ell=1/(4t)$~\cite{Polchinski:1998rq}. Second,
a crosscap state $|C\rangle$ satisfies
\begin{align}
	(K_n - (-1)^{n+h} \bar K_{-n}) \, |C\rangle=0 \, .
\end{align}
Therefore, it is given by a linear combination of $(-1)^{L_0}$ acting on the Ishibashi
states up to an overall phase. Hence, we need to calculate the matrix element of
$-e^{-2\pi \ell L_0} c_0^- b_0^+=\check q^{L_0} c_0^- b_0^+$ between Ishibashi states where
\begin{align}
	\check q=e^{2\pi i\ctau} \quad \text{with} \quad \ctau={1\over 2}+{i\over 4t} = {\htau-1\over 2\htau -1} \, .
\end{align}
The matrix element in the $(\alpha\beta)$ sector will now be given by linear
combinations of $g_{\alpha\beta}(\check q)$.
The contribution from the free field sector can be computed explicitly
and is related by the modular transformation $\ctau= {\htau-1\over 2\htau -1}$
to $\hat f_{\alpha\beta}$ computed in \refb{fMall}.
Denoting the new functions by $\check f_{\alpha\beta}(\ctau)$,
we can write the equations relating the two sets of functions as:
\be
\check f_{\alpha'\beta'}(\ctau) =\hat f_{\alpha\beta}(\htau),
\qquad \begin{pmatrix} \beta'+1\\ \alpha'+1 \end{pmatrix}
= \begin{pmatrix} 1 & & -1\\ 2 && -1 \end{pmatrix} \begin{pmatrix} \beta+1\\ \alpha+1 \end{pmatrix} \quad \hbox{mod 2}\, ,
\ee
up to relative phases which we will determine below.
The modular transformation matrix acting on $\scriptsize\begin{pmatrix} \beta+1\\ \alpha+1 \end{pmatrix}$
is the same one relating $\ctau$ and $\hat\tau$
and the shift of $\alpha,\beta$ by 1 in the modular transformation law is a reflection
of the fact that NS sector corresponds to anti-periodic boundary condition and the R sector
corresponds to periodic boundary condition. This gives
\be \label{frelation}
\check f_{00}(\ctau)= \hat f_{01}(\hat\tau),
\qquad
\check f_{01}(\ctau)= \hat f_{00}(\hat\tau),
\qquad
\check f_{10}(\ctau)= \hat f_{10}(\hat\tau)\, .
\ee

We can find $\check f_{\alpha \beta} (\ctau)$ explicitly using modular transformations
of $f_{\alpha \beta} (\htau)$. In order to do that, first notice
\begin{align}
\htau= {\ctau-1\over 2\ctau -1} = T S T^2 S (\ctau)
\quad \text{where} \quad
T(\ctau) = \ctau + 1,
\quad
S(\ctau) = - {1 \over \ctau}\, .
\end{align}
Upon using~\eqref{modularEta}-\eqref{modularTheta} accordingly, we obtain
\begin{subequations} \label{Mobiusf}
\begin{align}
\check f_{00}(\ctau)
&=
\hat f_{01} \(T S T^2 S (\ctau)  \) \non
\\
&=
-4 \,e^{\I \pi/4} (2 \ctau-1)  {\vt_{01}(\ctau)^2\,\eta(\ctau)^3 \over \vt_{10}(\ctau)^2 \,\vt_{00}(\ctau)}
=
-e^{\I \pi/ 4} (2 \ctau -1) \hat f_{00} (\ctau) \, ,
\\
\check f_{01}(\ctau)
&=
\hat f_{00} \(T S T^2 S (\ctau)  \)  \non
\\
&=
-4 \,e^{- \I \pi/4} (2 \ctau-1)  {\vt_{00}(\ctau)^2\,\eta(\ctau)^3 \over \vt_{10}(\ctau)^2 \,\vt_{01}(\ctau)}
=
e^{-\I \pi/ 4} (2 \ctau -1) \hat f_{01} (\ctau) \, ,
\\
\check f_{10}(\ctau)
&=
\hat f_{10} \(T S T^2 S (\ctau)  \) = 0  \, .
\end{align}
\end{subequations}

\newcommand{\ZZhQ}{{\mathcal Z}^{(h,Q)}}

The contribution from a particular character associated with a massive representation
to a given amplitude may now be expressed as:
\be
\begin{split}
\ZZhQ_{M} =&\, {N'_r \over 2}
\sum_{\alpha,\beta}  \check f_{\alpha\beta}(\check \tau)\, \ghQ_{\alpha\beta}(\check \tau)
\\
= &\, {N'_r \over 2}\, (2 \ctau -1) \[-e^{\I \pi/4}
\hat f_{00} (\ctau) \ghQ_{00} (\ctau) + e^{2 \pi i\delta} e^{-\I \pi/4} \hat f_{01} (\ctau) \ghQ_{01} (\ctau) \].
\end{split}
\label{ClosedMobius}
\ee
Here $1/2$ is from the GSO projection and $N'_r$ is a (complex) normalization factor
that reflects the product of the coefficient of the Ishibashi state in
representation $r$ in the crosscap state and  a similar coefficient
in the D-instanton boundary state.
We added the relative phase $e^{2 \pi i\delta}$ between the two terms and this arises as follows.
Since $\hq=-q$ is a negative real number, the definition of $\hat f_{\alpha\beta}(\hq)$ in \refb{fMall} contains
certain phases which eventually combine with similar phases of $\ghQ_{\alpha\beta}(\hq)$ to give
a real result. Since we have defined the functions $\check f_{\alpha\beta}(\check q)$ just by
reexpressing the $\hat f_{\alpha\beta}(\hq)$'s in terms $\check q$, the original phases
continue to be present in their definition. However, $\ghQ_{\alpha\beta}(\check q)$
have been written down directly by using the character formula
and not by rewriting the open string channel results in terms of closed string channel results.
Therefore, while combining $\ghQ_{\alpha\beta}(\hq)$ with $\check f_{\alpha\beta}(\check q)$,
we may need to multiply by possible phases. Since the overall phase can be absorbed into
$N'_r$, only the relative phase is important, which we call $\delta$. This relative phase will be
determined below explicitly.

Substituting \eqref{Mobiusf} and \eqref{gab}, and using \refb{quadtheta},
one can rewrite \eqref{ClosedMobius} as
\ben
\hspace{-0.5in}
\ZZhQ_{M}  &=& -2N'_r \, (2 \ctau -1)\, \cq^{h - {1 +Q \over 4} }
\Bigl(e^{\I \pi/4}\vt_{01} (\ctau)^2 + (-1)^Q e^{2 \pi i\delta-\I \pi/4}
\vt_{00} (\ctau)^2 \Bigr)\,\frac{\vt_{Q0} (2\ctau)}{\vt_{10} (\ctau)^2} \, .
\een
More explicitly this is
\ben \label{eE.49}
\ZZhQ_{M} &=& -N'_r \, (2 \ctau -1)\,\cq^{h - {1 +Q \over 4} }
\[\Bigl((-1)^Qe^{2 \pi i\delta-\I \pi/4}+e^{\I \pi/4}\Bigr)\, \frac{\vt_{00} (2\ctau)^2}{\vt_{1-Q,0} (2\ctau)}
\right.
\non\\
&&\left.\qquad
+\Bigl((-1)^Qe^{2 \pi i\delta-\I \pi/4}-e^{\I \pi/4}\Bigr)\, \frac{\vt_{10} (2\ctau)^2}{\vt_{1-Q,0} (2\ctau)}\]
\non\\
&=& (-1)^{1-Q} N'_r \, (2 \ctau -1)\,\cq^{h - {1 +Q \over 4} }
\[\Bigl(e^{2 \pi i\delta-\I \pi/4}-e^{\I \pi/4}\Bigr)\, \vt_{1-Q,0} (2\ctau)
\right.
\non\\
&&\left.\qquad
+\Bigl(e^{2 \pi i\delta-\I \pi/4}+e^{\I \pi/4}\Bigr)\, \frac{\vt_{Q0} (2\ctau)^2}{\vt_{1-Q,0} (2\ctau)}\].
\een

In order to fix $\delta$, we first observe that $\ZZhQ_{M}$ receives contributions only from NSNS sector, as is clear from
\refb{ClosedMobius}. Since $\cZ_{A,DD}$ can be regarded as the matrix element of $\tilde q^{L_0}$ between a pair of
D-instanton boundary states, while $\cZ_{M}$ can be regarded as the matrix element of $(-\check q)^{L_0}$
between a D-instanton boundary state and a crosscap state, the powers of $\check q$ appearing in $\cZ_{M}$ must be a subset
of the powers of $\tilde q$ appearing in $\cZ_{A,DD}$, the latter being the $L_0$ eigenvalues
appearing in the expression for the boundary state of the instanton, leaving out the momentum contribution.
The NSNS sector contribution to $\cZ_{A,DD}$ from a given representation
may be computed in the same way as $\cZ_{A,DN}$, and can be written, using \refb{edd.1}, \refb{edd.3}, as:
\be
\begin{split}
& {1\over \eta\(-\frac{1}{\tilde\tau}\)^3} \,
\[ \vt_{00}\(-\frac{1}{\tilde\tau}\) g^{(h,Q)}_{00}(\tilde \tau) -\vt_{10}\(\!-{1\over \tilde \tau}\) g^{(h,Q)}_{01}(\tilde \tau)\]
\\
& \hspace{1in} =\,  {(-i\tilde\tau)^{-1}\over \eta\(\tilde\tau\)^3}\,
\Bigl[ \vt_{00}( \tilde\tau)\, g^{(h,Q)}_{00}(\tilde \tau) -\vt_{01}(\tilde \tau)\,
g^{(h,Q)}_{01}(\tilde \tau)\Bigr]\, .
\end{split}
\ee
Using \refb{egidentity}, $\vt_{11}(\tilde\tau)=0$ and \refb{gab}, for massive representations
we can rewrite this as
\be
{(-i\tilde\tau)^{-1}\over \eta\(\tilde\tau\)^3} \, \vt_{10}(\tilde\tau)\, \ghQ_{10}(\tilde \tau)
= {(-i\tilde\tau)^{-1}\over \eta\(\tilde\tau\)^6} \, \tilde q^{h - {1+Q\over 4}}
\vt_{10}(\tilde\tau)^2\, \vt_{1+Q,0}(2\tilde\tau)\, .
\ee
For large imaginary
$\tilde\tau$ or small $\tilde q$, this expression has an expansion of the form:
\be\label{elargeq}
(-i\tilde\tau)^{-1} \, \tilde q^{h - {Q\over 2}} \sum_{n=0\atop n\in\ZZZ}^\infty a_n \tilde q^n\, .
\ee
We now compare this result with the small $\check q$ expansion of
\refb{eE.49} which for the two possible values of $Q$ is given by
\begin{subequations}  \label{eQ0Mtot}
\begin{align}
\label{eQ0M}
\cZi{h,0}_{M}
&=  {1 \over 2} \, N'_r\,  (2 \check \tau -1) e^{3 \pi i/4} \check q^h \Bigl[(\I + e^{2 \pi i \delta})
\, \check q^{-1/2} + 4 \, (-\I + e^{2 \pi i \delta}) + \cdots \Bigr],
\\
\label{eQ1M}
\cZi{h,1}_{M}
&= {1 \over 2}  \, N'_r\,  (2 \check \tau -1) e^{3 \pi i/4} \check q^{h-1/2}
\Bigl[-2 \, (-\I + e^{2 \pi i \delta}) \, - 8 \, (\I + 	e^{2 \pi i \delta}) \check q^{1/2}  + \cdots \Bigr].
\end{align}
\end{subequations}
Demanding that the powers of $\check q$ appearing in these expansions are a subset
of the powers of $\tilde q$ that appear in the expansion \refb{elargeq}, where the sum
over $n$ runs over integers, we see that the first term inside the square bracket in \refb{eQ0M}
and the second term inside the square bracket in \refb{eQ1M} must vanish. This gives\footnote{Note
that the power of $\check \tau -{1\over 2}$ in the expression for $\cZ_{M}$ does not match
that of $\tilde \tau$ in the expression for $\cZ_{A,DD}$. This can be traced to the fact that
$\cZ_{A,DD}$ receives an additional contribution from integration over the momenta of closed strings
exchanged between the two boundary states.}
\be
e^{2\pi i\delta}=-i \quad \text{for} \quad Q=0,1 \, .
\ee

Substituting this into \refb{eE.49}, and absorbing an overall phase
$-e^{i\pi/4}$ into the definition of $N'_r$, we find
\begin{align}
\ZZhQ_{M} = 2 N'_r  \, (2 \check \tau -1) \, (-1)^{1-Q} \check q^{h
-{1 \over 4} - {Q \over 4}} \vt_{1-Q,0} (2 \check \tau) \, .
\end{align}
In order that $\ZZhQ_{M}$ to be real, $i  N'_r e^{i\pi (h-{Q\over 2})}$ must be real.
After summing over all representations we get
\be
\begin{split}
\label{emobclosed}
\cZ_{M}  = &\, 2 \, (2 \check \tau -1) \, \sum_{Q=0,1}  (-1)^{1-Q}  \vt_{1-Q,0} (2 \check \tau)
\sum_{h\ge |Q|/2} N'_{h,Q}  \, e^{-i\pi (h-{Q-1\over 2})}  \check q^{h-{1 \over 4} - {Q \over 4}},
\\
\check q =&\,  e^{2\pi i\check\tau}=- e^{-\pi / (2t)} \, ,
\end{split}
\ee
for real constants $N'_{h,Q}$.
As before,
the contribution from the representation $(vac)$
gets added to $N'_{0,0}$ with weight 1 and to $N'_{1,1/2}$ with weight $-1$ and the
contribution from the representations $(\pm )$ are added to $N'_{1,1/2}$ with weight $1/2$.
As in the case of $\cZ_{A,DN}$, the integration measure is given by $dt/(2t)=d\ell/(2\ell)$.
The factor of $\ell$ coming from the $(2 \check \tau -1)$ factor in \refb{emobclosed}
cancels the $\ell$ in the denominator of the integration measure and the
resulting expression has the interpretation of the matrix element of
$c_0^-b_0^+\check q^{L_0}$ between the boundary state and the crosscap state. Finally,
as in the case of annulus partition function, the total contribution $\cZ_M$ from the closed
string channel is equal to its counterpart $Z_M$ computed using the open string
channel.

Under a general
deformation of the background, starting from a solvable model,
the coefficients $N_r$, $N_r'$ as well as the conformal weights of the left-right symmetric
closed string primaries
labelling the Ishibashi states are expected to change. We could calculate these
shifts using perturbation theory, {\it e.g.} the changes in $N_r$ and $N_r'$ can be computed
by analyzing the change in the one-point function on the disk or $\mathbb{RP}^2$ under
deformation of the background.
This approach will be useful for computing the
behaviour of the partition function in the small $t$ limit since only a few low lying
closed string primary states will contribute in this limit. Once we analyze both the small $t$
and large $t$ expansions of the partition function, we can find an approximate interpolating
function that gives the partition function over the entire range of $t$ and then integrate it
to compute the normalization constant $K_0$ via \refb{edefam}.

If we work with a Calabi-Yau threefold of large volume then we expect many light closed string
states and the expansion described above may not be efficient. In that case there is an
alternate approach where we regard the system of D-branes, orientifold planes and the instanton
as sources for ten-dimensional supergravity fields. We can then solve for the fields in terms
of the sources using standard Green's function technique and then substitute the solution
into the action in terms
of the sources to get the partition function. This has been illustrated in \cite{Baumann:2006th}.
Since this effectively allows us to resum
the contribution from many light closed string states, it
should be possible to augment the analysis given above with this approach to get a better
approximation of the contribution from the closed string channel.
However, this procedure does not account for light or string scale open string states living at
the intersection of the D-instanton with D-branes and O-planes, and these must be
accounted for separately by working in the open string channel.

\section{Relation between instanton correction and threshold correction} \label{sF}

In this appendix we shall derive the relation between the instanton
partition function and threshold correction to gauge coupling on a space-filling
D-brane \cite{Abel:2006yk,Akerblom:2006hx}. We begin by discussing
a few points  that require special attention.
\begin{enumerate}
\item
Our strategy will be to show that if we replace the D-instanton by a space-filling D-brane,
then the threshold correction to the gauge coupling of the space-filling brane is proportional
to the instanton partition function. We must note however that if the original configuration
satisfies the tadpole cancellation condition then the addition of a new space-filling brane
generally violates the tadpole cancellation condition. For this reason we shall regard the space-filling
brane as a `probe brane' that does not affect the background.
Formally, we shall achieve this by taking the number of probe branes to zero at the end of the computation.
To distinguish the probe brane from other space-filling branes that
were already present earlier, we shall label the former by $P$ and the latter by $N$.

\item
When we replace the Dirichlet boundary condition on the D-instanton by a Neumann
boundary condition, we need to take even number $(2k)$ of branes and use a different action
of \oo\ on the Chan-Paton factors. This leads to an $Sp(k)$ gauge group
on the brane \cite{Gimon:1996rq}. The replacement of $O(1)$ by $Sp(k)$ will produce extra
factors from traces over Chan-Paton factors that need to be taken into account in the
intermediate steps of analysis. We shall ignore this issue and various other overall normalization
factors in the analysis of \S\ref{sF.1}, \S\ref{sF.2} and return to the discussion of normalization
in \S\ref{sF.4}. There we shall also discuss the effect of the formal $k\to 0$ limit that is needed
to get the desired answer.

\item
As we have argued earlier, our assumption that the only zero modes on the D-instanton
are those associated with broken translations and supersymmetry implies that $g_{11}$
vanishes. Therefore, in our analysis we shall  focus only on the terms proportional
to $g_{\alpha\beta}$ with $(\alpha\beta)\ne (11)$.
\end{enumerate}

\subsection{Annulus partition function with DN boundary condition} \label{sF.1}

We begin by analyzing the annulus partition function with one boundary on the instanton and
the other boundary on a space-filling D-brane.
By focussing on the contribution from a given representation in the internal SCFT, we can
express the partition function $\Zr_{A,DN}$ given in \refb{edd.36} as:
\be \label{eE.1}
\Zr_{A,DN}=-{1\over 2}\,
{\sum_{\alpha,\beta}}' (-1)^{\alpha+\beta} {\vt_{\alpha\beta}\({\tau\over 2}|\tau\)^2
\eta(\tau)^3\over \vt_{11}\({\tau\over 2}|\tau\)^2 \vt_{\alpha\beta}(\tau)} \,\gr_{\alpha\beta} \, ,
\ee
where we have used \eqref{fDNall} and \eqref{shiftzbytau},
$\gr_{\alpha\beta}$ have been defined in \refb{edefgandf}, and $\sum'$
denotes the sum over $(\alpha\beta)\ne (11)$. We could
simplify this using \refb{ZADN}, but instead we use
the identities \eqref{eE.2} and \eqref{dervt11} to express \refb{eE.1} as
\be
\Zr_{A,DN}=-{1\over 8\pi^2 \eta(\tau)^3} \[
{\sum_{\alpha,\beta}}' (-1)^{\alpha+\beta} \vt''_{\alpha\beta}(0|\tau) \gr_{\alpha\beta}
- \p_z^2 \log \vt_{11}(z|\tau)|_{z=\frac{\tau}{2}} {\sum_{\alpha,\beta}}' (-1)^{\alpha+\beta}
\vt_{\alpha\beta}(\tau) \gr_{\alpha\beta}\].
\ee
The second term vanishes as a consequence of \refb{egidentity}. Therefore, we get
\be\label{ecomp}
\Zr_{A,DN}=-{1\over 8\pi^2 \eta(\tau)^3}
{\sum_{\alpha,\beta}}' (-1)^{\alpha+\beta} \vt''_{\alpha\beta}(\tau)\, \gr_{\alpha\beta}\, .
\ee

We shall now compare this with the threshold correction to the gauge coupling when the
D-instanton is replaced by a space-filling D-brane $P$. On the partition function this will essentially
have the effect of changing the boundary condition on the $X^\mu$ and $\psi^\mu$ fields from DN to NN, leaving the
internal part unchanged. This is almost the same as
the partition function computed with DD boundary condition on $X^\mu$ and $\psi^\mu$,
giving the results \refb{fDDall}. However, the open strings now carry momentum
along the non-compact directions, producing an extra factor
\be\label{eE.6}
\VV\, \int {d^4 k\over (2\pi)^4} e^{-2\pi t k^2} =\frac{\VV}{2^{6}\pi^{4} t^{2}}\, ,
\ee
where $\VV = \int{d^4x} $ is the volume of the four dimensional
space-time. After summing over all four sectors, the resulting
partition function in a representation $r$ will be given by an expression similar to the left hand side
of \refb{egidentity}
\be \label{eidrep}
\Zr_{A,NN}=\frac{\VV}{2^{6}\pi^{4} t^{2}} \, \frac{1}{2\, \eta(\tau)^3}\sum_{\alpha,\beta=0}^1 (-1)^{\alpha+\beta}
\vt_{\alpha\beta}(\tau) \, \gr_{\alpha\beta}(\tau) \, ,
\ee
and vanishes by the same identity \refb{egidentity} that ensured
the vanishing of the annulus amplitude with DD boundary conditions. The $4\, \eta(\tau)^3$
factor in the denominator of the left hand side of \refb{egidentity}
has been replaced by $2\, \eta(\tau)^3$ in order to
account for the two different orientations of the open string living on the $PN$ system.

Our goal however is to compute the threshold correction to the gauge coupling.
This requires switching on a background self-dual electromagnetic field strength and
setting the field to zero after taking two derivatives of the partition function with respect
to the background field \cite{Lust:2003ky}. The effect of switching on the background self-dual electromagnetic
field is to twist the boundary condition on half of the $X^\mu$'s and $\psi^\mu$'s by an
angle $2\pi\phi$ and the other half of these fields by an angle
$-2\pi\phi$ where $\phi$ is determined by the background field.
We denote the strength of the background electromagnetic field by $B$, normalized so that
(see \refb{eprecisenorm} for the precise normalization)
\be \label{ephiB}
\phi = {1\over \pi} \, \tan^{-1}(\pi B)\, .
\ee
This will replace the first two products in \refb{edd.1} by
\be\label{eF.6a}
{\pi^2 t^2 B^2\over \sinh^2(\pi t\phi)} \,
\prod_{n=1}^\infty (1-q^{n+\phi})^{-2}  (1-q^{n-\phi})^{-2}
\prod_{n=1}^\infty (1+q^{n+\phi-{1\over 2}})^2 (1+q^{n-\phi-{1\over 2}})^2.
\ee
The factor $t^2$ cancels the factor $t^{-2}$
coming from the momentum integration in  \refb{eE.6}, consistently with the fact that for non-zero $B$, the continuum of states
labelled by momenta are organized into discrete Landau levels
created by the oscillators $\alpha^\mu_{-\phi}$
for a pair of $\mu$'s. In the limit $B\to 0$, \refb{eF.6a} reduces to the first two products in \refb{edd.1}, as it should be.
The net effect of \refb{eF.6a} is to multiply \refb{edd.1} by
\be
-4 \pi^2 t^2 B^2 \,
{\eta(\tau)^6\over \vt_{11}(i t\phi|\tau)^2}\, {\vt_{00}(it\phi|\tau)^2\over \vt_{00}(\tau)^2}\, .
\ee
Repeating the analysis for the other cases, we find
that \refb{edd.1}-\refb{edd.3} are multiplied by:
\be\label{eE.9}
-4 \pi^2 t^2 B^2 \,
{\eta(\tau)^6\over \vt_{11}(i t\phi|\tau)^2} \,
{\vt_{\alpha\beta}(it\phi|\tau)^2\over \vt_{\alpha\beta}(\tau)^2}\, .
\ee
We shall not worry about the corrected form of
\refb{edd.4} since it multiplies $g_{11}$ which vanishes.
The twisted partition function is obtained by multiplying the summand in \refb{eidrep} by
\refb{eE.9}:
\be\label{eE99pre}
\Zr_{A,PN}=\frac{\VV\,\eta(\tau)^{-3}}{2^{7} \pi^{4} t^{2}}\, {\sum_{\alpha,\beta}}' (-1)^{\alpha+\beta}
(-4 \pi^2 t^2 B^2) \,
{\eta(\tau)^6\over \vt_{11}(i t\phi|\tau)^2} \,
{\vt_{\alpha\beta}(it\phi|\tau)^2\over \vt_{\alpha\beta}(\tau)^2}\,
\vt_{\alpha\beta}(\tau) \,\gr_{\alpha\beta}(\tau)\, .
\ee
We can simplify this result using \refb{eE.2} for $z=it\phi$ and express it as
\ben\label{eE99}
\Zr_{A,PN}&=& -\frac{\VV B^2 }{ 2^{5} \pi^{2}} \,
{\eta(\tau)^{3}\over \vt'_{11}(0|\tau)^2}
{\sum_{\alpha,\beta}}' (-1)^{\alpha+\beta}
\(\vt''_{\alpha\beta}(0|\tau) -\vt_{\alpha\beta}(\tau) \p_z^2 \log \vt_{11}(z|\tau)|_{z=it\phi}\)
\gr_{\alpha\beta}(\tau) \non\\ &=&
-\frac{\VV B^2}{2^{7} \pi^{4}\eta(\tau)^{3}} \,
{\sum_{\alpha,\beta}}' (-1)^{\alpha+\beta}
\vt''_{\alpha\beta}(0|\tau)\,  \gr_{\alpha\beta}(\tau)\, ,
\een
where we have used \eqref{dervt11} and \refb{egidentity} in the last step.

A consistency test of this formula is that in the strong field limit this should be proportional
to the annulus amplitude with DN boundary condition with positive constant of proportionality, since
in this limit the space-filling brane $P$
develops a large D-instanton charge density and we expect the
result to be proportional to the amplitude where we replace the space-filling brane carrying
a self-dual background field by a collection of D-instantons. Since the strong field limit corresponds to $B\to\infty$,
$\phi\to{1/2}$ we see that indeed in this limit \refb{eE99} is proportional to the result \refb{ecomp} for
D-instantons with positive constant of proportionality. In fact, the individual terms in the
original sum \refb{eE99pre} can be shown to be proportional to the individual terms in
the sum \refb{eE.1} in this limit.

To compute the threshold correction, we are supposed to take two derivatives of
\refb{eE99} with respect to the strength $B$ of the background gauge field and
then set $B=0$. This gives
\be
\label{eE.11}
\p_B^2 \Zr_{A,PN}|_{B=0}=-\frac{\VV}{64\pi^{4}\eta(\tau)^3} \,
{\sum_{\alpha,\beta}}' (-1)^{\alpha+\beta} \vt''_{\alpha\beta}(0|\tau)\, \gr_{\alpha\beta}
= {\VV\over 8\pi^2}\, \Zr_{A,DN}\, .
\ee
where in the last step we used \refb{ecomp}. This is indeed proportional to $\Zr_{A,DN}$.

\subsection{M\"{o}bius strip contribution} \label{sF.2}

We now turn to the M\"{o}bius strip contribution $Z_{M}$. As discussed in \S\ref{smobius},
the first effect is to replace $\tau$ by $\htau=\tau+{1\over 2}$ in the arguments of the functions on the
left hand side of  \refb{egidentity}. The other effect is the replace the factors
\refb{fDDall} in the summand in \refb{egidentity} by
\refb{fMall} and multiply the result by an overall phase
factor $\alpha_r$ that determines the parity of the representation of internal SCFT under
\oo. This gives
\be \label{ezmd1}
\Zr_{M}= \alpha_r {\eta(\htau)^3\over \vt_{10}(\htau)^2}
{\sum_{\alpha,\beta}}' (-1)^{\alpha+\beta} {\vt_{\alpha\beta}\({1\over 2}|\htau\)^2
\over \vt_{\alpha\beta}(\htau)}\, \gr_{\alpha\beta}(\htau)\, ,
\ee
where as above the sum is restricted to $(\alpha\beta)\ne (11)$ due to
vanishing of $g_{11}$. The identity \refb{eE.2} with $z=1/2$ and $\tau$ replaced by $\htau$
gives
\be\label{eE.21}
{\vt_{\alpha\beta}\({1\over 2}|\htau\)^2
\over  \vt_{\alpha\beta}(\htau)}= {\vt_{11}\({1\over 2} |\htau\)^2\over
\vt'_{11}(0|\htau)^2}
\[{\vt''_{\alpha\beta}(0|\htau)
} - \vt_{\alpha\beta}(\htau) \p_z^2 \log \vt_{11}(z|\htau)_{z={1\over 2}}\] .
\ee
When we substitute this into \refb{ezmd1}, the contribution from the second
term in \refb{eE.21} vanishes due to
\refb{egidentity}. Using $\vt'_{11}(0|\htau)=-2\pi \eta(\htau)^3$ and $\vt_{11}\({1\over 2}|\htau\)=-\vt_{10}(\htau)$,
this gives
\be
\label{eE.13}
\begin{split}
\Zr_{M} & = {\alpha_r \over 4\pi^2 \eta(\htau)^3}\,
{\vt_{11}\({1\over 2}|\htau\)^2\over \vt_{10}(\htau)^2}\,
{\sum_{\alpha,\beta}}' (-1)^{\alpha+\beta} \vt''_{\alpha\beta}(0|\htau)\,
\gr_{\alpha\beta}(\htau)
\\
& = {\alpha_r\over 4\pi^2 \eta(\htau)^3}\,
{\sum_{\alpha,\beta}}' (-1)^{\alpha+\beta} \vt''_{\alpha\beta}(0|\htau)\,
\gr_{\alpha\beta}(\htau)\, .
\end{split}
\ee

On the other hand, the M\"{o}bius strip contribution with
space-filling D-brane boundary condition in the presence of background gauge
field can be analyzed in the same way as in the case of the annulus contribution with NN boundary
conditon, and is given by the  following set of operations on \refb{eE99}:
\begin{enumerate}
\item
Replace $\tau$ by $\htau$ following the discussion at the beginning of \S\ref{smobius}.

\item
Include an overall multiplicative factor of $\alpha_r$.

\item
Include a factor of $1/2$ since we have only one open string instead of
a pair of open strings with different orientations.

\item
Replace $\phi$ by $2\phi$ since both ends of the open string lie
on the same brane (after orientation reversal) and therefore feel the effect of the
background electromagnetic field, where $\phi$ continues to be given by \refb{ephiB}.
This also requires that we multiply the result by a factor of 4 so that \refb{eF.6a}
continues to approach the product of the first two factors of \refb{edd.1} in the
limit $B\to 0$ (with $q$ replaced by $\hat q$).

\item
Include an extra factor of $-1$ due to the fact that if we replace the D-instanton
boundary condition by a space-filling brane boundary condition along the four non-compact directions keeping the
internal part unchanged, a $O(k)$ type instanton is replaced by an $Sp(k)$ type
instanton \cite{Gimon:1996rq}.
This effectively changes the sign of \oo\ on the states in the sector
that contains the gauge field. In order to see this, note that changing the boundary
condition from Dirichlet to Neumann will make the states \refb{etranslationa},
representing gauge fields, odd under \oo\ and project them out unless we include an extra minus sign in the
transformation law of the vacuum or equivalently the Chan-Paton factor.
The need for the extra minus sign will be
verified independently below by taking the strong field limit.
\end{enumerate}
This gives a net multiplicative factor of $-2\,\alpha_r$ and the replacement of $\tau$ by
$\hat\tau$ in \refb{eE99}, leading to,
\be\label{emobelec}
\Zr_{M,P}= \frac{\alpha_r\,\VV B^2 }{2^{6} \pi^{4}\eta(\htau)^{3}} \,
{\sum_{\alpha,\beta}}' (-1)^{\alpha+\beta}
\vt''_{\alpha\beta}(0|\htau)  \,\gr_{\alpha\beta}(\htau)\, .
\ee
One can check that in the strong coupling limit $B\to\infty$,
$\phi\to 1/2$, this is proportional to the M\"{o}bius strip
amplitude \refb{eE.13} with Dirichlet boundary condition with positive
constant of proportionality. Furthermore, the individual terms in the sum
\refb{eE99pre} with $\phi$ replaced by $2\phi$, $\tau$ replaced by
$\htau$ and carrying an extra factor of $-2\,\alpha_r$,
approach the individual terms in the sum \refb{ezmd1} up to a positive constant
of proportionality in this limit.

To compute the threshold correction, we need to take two derivatives of
\refb{emobelec} with respect to the background field $B$ and set $B=0$. This gives:
\be
\label{eF.16}
\p_B^2 \Zr_{M,P}|_{B=0}=\frac{\alpha_r\,\VV}{ 32\pi^{4} \eta(\htau)^3}\,
{\sum_{\alpha,\beta}}' (-1)^{\alpha+\beta} \vt''_{\alpha\beta}(0|\htau)\,
\gr_{\alpha\beta}(\htau) = {\VV\over 8\pi^2} \, \Zr_{M}\, ,
\ee
where in the last step we used
\refb{eE.13}. This is indeed proportional to $\Zr_{M}$.
We now see that the constant of proportionality $\VV/(8\pi^2)$
is exactly the same in \refb{eE.11} and \refb{eF.16}.
This establishes that the total threshold
correction to the gauge coupling of a space-filling D-brane is proportional to the partition
function on a D-instanton obtained by replacing the Neumann boundary condition on the
space-filling brane by the Dirichlet boundary condition.

\subsection{Fixing the normalization} \label{sF.4}

The analysis of \S\ref{sF.1}, \S\ref{sF.2} establishes the relation between the D-instanton
partition function and the threshold correction to the gauge coupling on a space-filling
probe D-brane up to a constant of proportionality.
Our goal in this section will be to compute this constant.

Let us consider a single space-filling probe D-brane and switch on a gauge field strength
$F_{12}=-F_{21}=f$ along the 12 plane. We consider an open string whose $\sigma=0$ end is on this probe D-brane and
the $\sigma=\pi$ end is on some other space-filling D-brane on which no gauge field has
been switched on. We shall normalize $F_{\mu\nu}$ such that
the boundary conditions for open strings moving in such a
background Abelian gauge field is given by (in $\alpha^\prime = 1$ unit)
\be
\begin{split} \label{bc}
\partial_\sigma X^\mu + i \, F^{\mu}_{\; \nu} \, \partial_\tau X^\nu =&\,  0
\quad  \text{at} \quad \sigma =0\, ,
\\
\partial_\sigma X^\mu=&\,  0 \quad  \text{at} \quad \sigma =\pi\, .
\end{split}
\ee
In~\cite{Abouelsaood:1986gd}, it has been shown that with this normalization the
Euclidean action on
the D-brane is given by the Dirac-Born-Infeld action
\be \label{eDBI}
-T_P\, \int d^4 x \sqrt{\det ({\mathds 1}+F)} = -T_P \int d^4 x \(1+{1\over 2} f^2 + \cdots\) ,
\ee
where $T_P$ is the tension of the probe brane and
in the second expression we evaluated the action in the chosen background.

Let us define
\begin{align}
X^\pm \equiv {1 \over \sqrt{2} } \left(X^1 \pm i X^2 \right) ,
\end{align}
so that the boundary conditions~\eqref{bc}  take the form,
\be
\begin{split}
\partial_\sigma X^\pm \pm f \,  \partial_\tau X^\pm  = &\, 0\quad  \text{at} \quad \sigma =0\, ,
\\
\partial_\sigma X^\pm= &\, 0 \quad  \text{at} \quad \sigma =\pi \, .
\end{split}
\ee
We will take the mode expansion for $X^{\pm}$ to be
\begin{align}
X^\pm = a^\pm \, e^{-p_\pm\tau} \cos(p^\pm \sigma + \pi \phi_\pm)\, ,
\end{align}
where $p_\pm$ are real numbers and $a^\pm$ are constants of integration.
Applying the boundary conditions at $\sigma = 0$ we get
\begin{align}
\mp \tan \pi \phi_\pm = f \, .
\end{align}
Therefore, we can take
\be\label{ephiNEW}
\phi_-=-\phi_+=\phi={1\over \pi} \tan^{-1}f\, .
\ee
Furthermore, applying the Neumann boundary conditions at $\sigma = \pi$ gives
\begin{align}
\sin(p_\pm \pi \mp \pi \phi) = 0 \implies p_\pm =n\pm \phi, \quad n \in \mathbb{Z} \, .
\end{align}
Comparing this with \refb{eF.6a}, we can identify
$\phi$ with the twist $\phi$ appearing there. Furthermore, comparison of
\refb{ephiB} and \refb{ephiNEW} gives,
\be \label{eprecisenorm}
f = \pi B\, .
\ee
Substituting this into the action \refb{eDBI}, we get the  part of the
classical action quadratic in the field to be
\be\label{eactEV}
-{\pi^2\over 2} \, T_P\int B^2 \, d^4 x= -{\VV\over 4 g_O^2}  B^2 \, ,
\ee
where $g_O$ is the coupling constant
of the open string field theory on the probe D-brane,
related to the tension via $T_P=(2\pi^2 g_O^2)^{-1}$ \cite{Sen:1999xm}.
Note that even though we have used the PN system to find the relation between $f$ and
$B$, once the relation has been found, it also holds for the M\"{o}bius strip with boundary
lying on $P$.

The actual system of interest
differs from the one discussed above in two ways.
First, as in \S\ref{sF.1}, \S\ref{sF.2}, the electromagnetic field has to be switched
on in two planes 12 and 34 at the same time by self-duality. This produces a factor of 2 in \refb{eactEV}. Second,
as discussed at the beginning of this appendix, we
need to have $(2k)$  D-branes for consistency of the orientifold projection.
Since \oo\ exchanges the branes, we cannot switch on gauge fields
inside only one of the branes. However, we can switch on a background gauge
field inside an $Sp(1)$ subgroup, i.e. equal and opposite fields on a pair of branes.
This produces another factor of $2$. Therefore, the
net classical action is given by
\be \label{eCLASS}
-{\VV\over g_O^2} \,  B^2 \, .
\ee

The threshold correction results of \S\ref{sF.1}, \S\ref{sF.2} can be regarded as one-loop corrections to the
second derivative of this expression with respect to $B$.
This was found to be $\VV/(8\pi^2)$ times the instanton partition function
$Z_{\mbox{\scriptsize D-inst}}$. This includes contributions from the annulus and
the M\"{o}bius strip. Taking into
account the fact that we have switched on gauge fields on
a pair of D-branes gives an extra factor of 2, while integrating this
twice to produce the term proportional to $B^2$ gives a factor of 1/2. Therefore, the
one-loop correction to \refb{eCLASS} from the annulus with $PN$ boundary condition
and the M\"{o}bius strip with $P$ boundary condition can be written as:
\be
\frac{\VV\, B^2}{8\pi^2} \, Z_{\mbox{\scriptsize D-inst}}\, ,
\ee
so that
\be \label{efinthr}
\delta\,{ \left( 8\pi^2\over g_O^2 \right) } = - Z_{\mbox{\scriptsize D-inst}}\, .
\ee
This is consistent with the results of~\cite{Akerblom:2007uc} after identifying $g_O$
with the Yang-Mills coupling constant $g_{YM}$ on the brane where the latter is defined by normalizing the action as:
\be
S_{YM} =-{1\over 2}\,  \Tr(F_{\mu\nu} F^{\mu\nu}), \qquad
F_{\mu\nu} =\p_\mu A_\nu - \p_\nu A_\mu + g_{YM} [A_\mu, A_\nu]\, ,
\ee
with the trace taken over the fundamental representation of the $Sp(k)$ group.
The identification of $g_O$
with $g_{YM}$ follows from the comparison of the triple gluon vertices computed from
string field theory and the Yang-Mills action given above.

The right hand side of \refb{efinthr} captures the contribution to the threshold
correction due to the $PN$ annulus and the M\"{o}bius strip with the boundary on $P$.
However, the full threshold correction to the gauge coupling on the probe brane also involves a contribution from
annulus with both boundaries lying on the probe brane.
This contribution  will have insertion of background gauge
field on one of the boundaries, producing a $k$-independent factor from the trace
over Chan-Paton factor on that boundary. However, the other boundary will produce
a factor of $k$ from a similar trace. Thus, the net contribution from these
open strings will be proportional to $k$.
Since this contribution to the threshold correction is not related
to any contribution to the D-instanton partition function, we have to remove this. This can
be done by taking the formal limit $k\to 0$ in the final result.

Let us now review what is known about such threshold corrections. Our assumption that
the only fermion zero modes on the D-instanton are those associated with broken
supersymmetry translates to the statement that on the probe brane there is no
massless chiral
multiplet transforming in a non-trivial representation of the gauge group $Sp(k)$.
In this case the formula for the threshold correction to $8\pi^2/g_O^2$
takes the form (see {\it e.g.}
\cite{Akerblom:2007uc}, eqs.(5.1), (5.2)):
\be\label{e3n.44thr}
 8\pi^2 \Re(f^{(1)}) -{3\over 2} \, T(G) \log{M_p^2\over \mu^2} -{1\over 2}\,  T(G) \cK
+ T(G) \, \log {1\over g_O^2}\, ,
\ee
where most of the quantities on the right hand side have already been defined below \refb{e3n.44},
$\mu$ is an infrared cut-off, giving the energy scale at which the threshold correction is being
computed, and $T(G)$ is defined via the relation $\Tr_{adj}(L^a L^b)=T(G) \delta^{ab}$ where
$L^a$ are the generators of $Sp(k)$ normalized so that $\Tr_{fund}(L^a L^b)=\delta^{ab}/2$.
For $Sp(k)$ we have $T(G)=k+1$. Using \refb{efinthr}, \refb{e3n.44thr},
and after adding infrared cutoff $\mu^2$ to
$Z_{\mbox{\scriptsize D-inst}}$ like in~\refb{e3n.44}, we have
\be\label{e3n.44thrnew}
Z_{\mbox{\scriptsize D-inst}}(\mu^2)
= \lim_{k\to 0} \[
-8\pi^2 \Re(f^{(1)}) + {3\over 2}\, (k+1) \log{M_p^2\over \mu^2} + {1\over 2} \,(k+1) \cK
- (k+1) \log {1\over g_O^2}\].
\ee
If we now use $g_O^2\propto g_s/V_\Gamma$,
we recover \refb{e3n.44}. All the additive constant ambiguities that arise due to the
precise definition of $\mu$ and the precise relation between $g_O$ and $g_s$
can be absorbed into the definition of $f^{(1)}$. Various other subtleties involving this
equation have been discussed below  \refb{e3n.44}.


\providecommand{\href}[2]{#2}\begingroup\raggedright\endgroup

\end{document}